\documentclass[a4paper,12pt]{article}            
\usepackage{jheppub}
\pdfoutput=1

\usepackage[utf8]{inputenc}
\usepackage{amsmath,amssymb}
\usepackage{xcolor}
\usepackage{tensor}
\usepackage{graphicx}
\usepackage{subcaption}
\usepackage{xurl} 
\usepackage{hyperref}

\newcommand{\highlight}[1]{\textcolor{red}{#1}}
\title{Linear response beyond hydrodynamic poles}
\author[1,2]{Andrea Amoretti,}
\author[1,2]{Daniel K. Brattan,} 
\author[1,2]{Jonas Rongen.}

\emailAdd{andrea.amoretti@ge.infn.it}
\emailAdd{danny.brattan@gmail.com}
\emailAdd{jonas.ludovico.rongen@edu.unige.it}

\affiliation[1]{Dipartimento di Fisica, Universit\`a di Genova,
via Dodecaneso 33, I-16146, Genova, Italy}
\affiliation[2]{I.N.F.N. - Sezione di Genova, via Dodecaneso 33, I-16146, Genova, Italy}

\begin{abstract}
{\ We consider the problem of writing an effective, linearised theory in small derivatives that reproduces the Mittag-Leffler expansion of a charge current correlator with an arbitrary number of simple poles. We demonstrate how such a framework: can be compatible with hydrostaticity without modification of thermodynamics, properly accounts for the differing notions of smallness in time and space derivatives including setting the lowest order effective equation of motion, and corrects the effective equations in derivatives. As an application, we apply the results to charge fluctuations of the D3/D5 probe brane and quantify how the transport coefficients behave when quasihydrodynamics emerges at large charge density.}
\end{abstract}
\begin{document}

\maketitle

\section{Introduction  }

Hydrodynamics provides a universal long-wavelength description of many-body systems near equilibrium, encoding conservation laws and symmetry principles in an expansion in derivatives. In recent years it has become increasingly clear that this framework admits systematic extensions that incorporate additional long-lived excitations beyond the strictly hydrodynamic sector \cite{Amoretti:2014zha,Amoretti:2014ola,Davison:2014lua,Davison:2015bea,Davison:2015taa} \cite{Amoretti:2016bxs,Amoretti:2017xto,Delacretaz:2017zxd,Amoretti:2017axe,Donos:2018kkm} \cite{Gouteraux:2018wfe,Grozdanov:2018fic,Amoretti:2018tzw,Delacretaz:2019wzh,Amoretti:2019cef} \cite{Donos:2019txg,Amoretti:2019kuf,Armas:2020bmo,Amoretti:2020ica,Armas:2021vku,Delacretaz:2021qqu,Gouteraux:2023uff,Armas:2024iuy}. These extended theories, collectively referred to as relaxed or quasihydrodynamics, describe situations in which gapped modes remain parametrically slow and significantly affect transport and late-time dynamics.

A common feature of many quasihydrodynamic systems is the appearance, in retarded correlation functions, of several isolated poles close to the origin of the complex frequency plane \cite{Withers_2018,Grozdanov:2018fic,Amoretti:2020ica,Amoretti_2022}. Such poles coexist with the diffusive hydrodynamic pole and reorganise the structure of linear response \cite{Davison:2014lua,Davison:2015taa,Gouteraux:2023uff}. While this phenomenon has been observed in a wide range of microscopic models, a sharp and general effective description remains subtle. In particular, many existing approaches treat the gaps of non-hydrodynamic modes as small parameters in a derivative expansion, implicitly assuming that poles can be generated order by order in frequency.

In this work we argue that this viewpoint is fundamentally misleading. Poles in Green's functions are intrinsically non-perturbative in frequency, irrespective of whether they correspond to hydrodynamic or gapped modes. This includes the diffusive pole itself: although its dispersion relation vanishes as $|\vec{k}| \to 0$, its very existence cannot be recovered at any finite order in a small-frequency expansion. This fact becomes manifest when retarded correlators are viewed through their Mittag-Leffler representation \cite{Conway}, in which a meromorphic function is decomposed uniquely into a sum over its isolated poles plus a holomorphic remainder.

From this perspective, quasihydrodynamics is not characterised by small gaps, but rather by the necessity of treating a finite set of poles as fundamental, non-perturbative input data for the effective theory. Once this set is specified, the remaining freedom lies entirely in the holomorphic part of the correlator, which admits a controlled derivative expansion with a finite radius of convergence.

The goal of this paper is to construct an effective, linearised theory that realises this philosophy explicitly. We develop a systematic framework that reproduces the full Mittag-Leffler expansion of a conserved $U(1)$ current correlator, including both its complete pole structure and its holomorphic part, within a well-defined disc of analyticity in the complex frequency plane. Crucially, this is achieved while preserving standard hydrostaticity conditions \cite{Jensen:2012jh,Haehl_2015} and without modifying thermodynamics. Time and space derivatives are treated on different footings: spatial derivatives are expanded perturbatively, while time derivatives associated with pole locations are resummed exactly.

In particular, our construction proceeds by promoting the spatial current to an independent dynamical variable whose equation of motion involves a finite product of first-order differential operators, each corresponding to a prescribed pole. In this way, an arbitrary but finite number of simple poles can be incorporated from the outset. The resulting effective theory naturally distinguishes between non-perturbative structures, namely the poles, and perturbative data, namely the momentum dependence of residues and the holomorphic sector, allowing for a precise matching to microscopic Green's functions.

As an application, we analyse charge transport in the D3/D5 probe brane system at finite density \cite{Karch:2002sh,Brattan:2012nb,Brattan:2013wya,Brattan:2014moa,Jokela_2015}. This model provides a clean holographic setting with finite DC conductivity and no coupling to momentum, making it an ideal testbed for our framework. We show that the emergence of quasihydrodynamic behaviour at large charge density can be understood quantitatively as a truncation effect in the Mittag-Leffler representation: as additional poles approach the origin, transport coefficients reorganise in a way that is - accurately and precisely - captured by our effective theory.

The structure of the paper is as follows: in section two we review the hydrodynamics of a single conserved $U(1)$ current and emphasise the non-perturbative nature of the diffusive pole. In section three we construct the effective linearised theory of many poles, ensuring compatibility with hydrostaticity and deriving the resulting Green's functions. Section four applies the formalism to the D3/D5 probe brane system and analyses the onset of quasihydrodynamics. We conclude in section five with a discussion and outlook.

\section{\texorpdfstring{Hydrodynamics of a single $U(1)$ conserved current}{Hydrodynamics of a single U(1) conserved current}} \label{section:single_u1}

{\noindent To introduce the ideas we wish to pursue more concretely, consider the effective hydrodynamic description of a linearised $U(1)$ charge current decoupled from momentum degrees of freedom. The framework for describing the behaviour of such a system at finite temperature includes the charge conservation equation
	\begin{eqnarray}
		\label{Eq:U1conservation}
		\partial_{t} \delta \rho + \partial_{i} \delta J^{i} &=& 0 \; , 
	\end{eqnarray}
describing time evolution, where $\delta \rho$ is a fluctuation of the charge density and $\delta J^{i}$ the fluctuation of the spatial current. The conservation equation is supplemented by a constitutive relation expressing this spatial charge current in terms of fluctuations of an applied external electromagnetic field $\delta E_{i}$ and the chemical potential $\delta \mu$,
	\begin{subequations}
	\begin{eqnarray}
		\label{Eq:HydroSpatialCharge}
		\delta J^{i} &=& \sigma_{(0)} \left( \delta E_{i} - \partial_{i} \delta \mu \right) + \ldots \; , \\
		\delta E_{i} &=& \partial_{i} \delta a_{t} - \partial_{t} \delta a_{i}  \; , 
	\end{eqnarray}
	\end{subequations}
where the ellipsis in \eqref{Eq:HydroSpatialCharge} corresponds to higher derivative terms. The spatial current-current correlator is readily obtained by solving the equation of motion \eqref{Eq:U1conservation} in the presence of the background gauge field \cite{Kovtun_2012, Hartnoll:2016apf}. With a slight abuse of terminology, we can call the following quantity,
	\begin{subequations}
	\begin{eqnarray}
		\label{Eq:DefConductivity}
		\sigma^{ij}(\omega,\vec{k}) &=&	 \frac{1}{-i \omega} \left[ \langle J^{i} J^{j} \rangle_{\mathrm{R}}(\omega,\vec{k}) -\langle J^{i} J^{j} \rangle_{\mathrm{R}}(0,\vec{k})  \right] \; ,  \\
		\label{Eq:Generalisedconductivity}
							  &=& \left(  \sigma_{(0)} - \frac{i \frac{\sigma_{(0)}^2}{\chi_{\rho \rho}} \vec{k}^2}{\omega + i \frac{\sigma_{(0)}}{\chi_{\rho \rho}} \vec{k}^2} \right) \frac{k^{i} k^{j}}{\vec{k}^2} + \sigma_{(0)} \left( \delta^{ij} - \frac{k^{i} k^{j}}{\vec{k}^2} \right) \; , 
	\end{eqnarray}
	\end{subequations}
the conductivity at non-zero wavevector $\vec{k}$ \footnote{Our convention is that
    \begin{eqnarray}
    	\label{Eq:FourierConvention}
        f(t,\vec{x}) = \int \frac{d^{d+1}k}{(2\pi)^{d+1}} f(\omega,\vec{k}) e^{-i( \omega t - \vec{k} \cdot \vec{x})} \; . 
    \end{eqnarray}
}. Notice that it is a complex quantity.}

{\ Focusing on the longitudinal part of the correlator in \eqref{Eq:Generalisedconductivity}, we see the sum of a pole and a trivial holomorphic function (i.e. a constant). Importantly, the residue of the diffusive pole is zero as $\vec{k} \rightarrow \vec{0}$, which ensures that the DC conductivity is finite when $\vec{k} \rightarrow \vec{0}$. This should be compared to theories with unbroken translation invariance and a non-trivial overlap between the charge and momentum sectors; in such theories the $\vec{k} \rightarrow 0$ limit of the theory has a pole at $\omega=0$ and thus the zero frequency limit is divergent \cite{Hartnoll:2007ih, Hartnoll:2012rj}. We do not consider momentum degrees of freedom and neatly side-step such issues. However, there is no reason that our approach and attitude cannot also be extended to these cases.}

{\ What features of the conductivity \eqref{Eq:Generalisedconductivity} can we expect as we proceed to progressively higher derivative corrections in the constitutive relation \eqref{Eq:HydroSpatialCharge}? A trick that is often employed is to use the lowest order equation of motion
	\begin{eqnarray}
		\partial_{t} \delta \mu &=& 0 + \mathcal{O}(\partial^2)
	\end{eqnarray}
and its derivatives, to avoid introducing time derivatives of the chemical potential. However, we have no reason not to include arbitrary time derivatives of the electric field. It is not hard to see that such time derivatives correct the trivial holomorphic term $\sigma_{(0)}$, replacing it with a function of frequency and wavevector. Similarly, the residue and pole get corrections in the wavevector as we introduce higher spatial derivatives, with the result that the conductivity looks like
	\begin{eqnarray}
\label{Eq:Infiniteseries}
\sigma^{ij}(\omega,\vec{k}) &=& 
\left( 
   \sum_{n=0}^{\infty} f_{(\mathrm{L}),n}(\vec{k}^2) (i\omega)^{n} 
   - 
   \frac{i \vec{k}^2 R_{\mathfrak{D}}(\vec{k}^2)}{\omega + i \mathfrak{D}(\vec{k}^2) \vec{k}^2}
\right) 
\frac{k^{i} k^{j}}{\vec{k}^2} 
\nonumber \\
&\;& 
+ 
\sum_{n=0}^{\infty} f_{(\mathrm{T}),n}(\vec{k}^2) (i\omega)^{n} 
\left( \delta^{ij} - \frac{k^{i} k^{j}}{\vec{k}^2} \right)
\end{eqnarray}
where we have acknowledged that longitudinal and transverse parts of the conductivity can be distinct while allowing ourselves to sum all possible corrections in frequency and wavevector.}

{\ Again, we emphasise the important role of an overall $\vec{k}^2$ in the residue of the diffusive pole in \eqref{Eq:Infiniteseries}. Working at $\vec{k} = \vec{0}$, there is no evidence of a diffusive pole in the correlator, instead $\sigma^{ij}(\omega,0)$ is a holomorphic function.  Similarly, the small frequency expansion of $\sigma^{ij}(\omega,\vec{k} \neq 0)$ has the form
	\begin{eqnarray}
		\sigma^{ij}(\omega,\vec{k}) &=& \left(  \sum_{n=0}^{\infty} \left( f_{(\mathrm{L}),n}(\vec{k}^2) - \frac{\vec{k}^2 R_{\mathfrak{D}}(\vec{k}^2)}{(\mathfrak{D}(\vec{k}^2) \vec{k}^2)^{n+1}} \right) (i\omega)^{n} \right) \frac{k^{i} k^{j}}{\vec{k}^2} \nonumber \\
		&\;& + \sum_{n=0}^{\infty} f_{(\mathrm{T}),n}(\vec{k}^2) (i\omega)^{n} \left( \delta^{ij} - \frac{k^{i} k^{j}}{\vec{k}^2} \right) \; .  
	\end{eqnarray}
In other words, it too is a holomorphic function with no apparent pole in the longitudinal term for all $|\vec{k}|>0$. Notice that as $f_{(\mathrm{L})}(\vec{k}^2)$ is finite as $\vec{k} \rightarrow 0$, the radius of convergence of the longitudinal series in small frequency goes to zero as $\vec{k} \rightarrow \vec{0}$ i.e. it is dictated by the position of the diffusive pole. This behaviour of the small frequency series is the smoking gun of a diffusive pole.} 

{\ To summarise the point we are trying to make here: the diffusive pole is in fact a non-perturbative object in small frequency; it's presence can only be detected from $\sigma^{ij}(\omega,\vec{k})$ at small frequency by summing the entire series. For systems with finite DC conductivities, there is no evidence of a diffusive pole in $\sigma^{ij}(\omega,\vec{0})$. For systems with infinite DC conductivities, the situation is more subtle for $\vec{k}=0$ with the small frequency expansion of $\sigma^{ij}(\omega,\vec{0})$ introducing a non-perturbative $1/\omega$ term.}

\begin{figure}[t!]
    \centering
    \includegraphics[width=9cm]{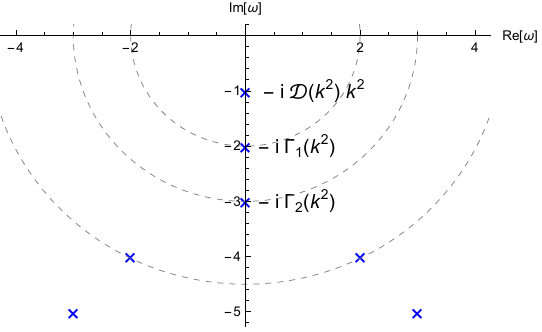}
    \caption{A schematic diagram showing the distribution of poles (blue crosses) in the longitudinal conductivity at some charge density. The labels indicate the diffusive pole $\omega_{\mathfrak{D}}= - i \mathfrak{D}(\vec{k}^2) \vec{k}^2$, the first gapped pole $\omega = - i \Gamma_{1}(\vec{k}^2)$ and the second gapped pole $\omega = - i \Gamma_{2}(\vec{k}^2)$. The dashed circles show discs of convergence for the series expansion in frequency of the holomorphic parts of the charge conductivity. The smallest dashed circle gives the convergence radius of the usual hydrodynamic charge conductivity \eqref{Eq:Infiniteseries} - i.e. up to the first non-hydrodynamic pole. Subsequent dashed circles indicate the effect of including higher poles in the effective description. In particular, we can correctly match our effective Green's function to the full Green's function on a larger segment of the real frequency axis, as displayed, when we include more poles.}
    \label{fig:christmastree}
\end{figure}

{\ Continuing in this vein, we have written our holomorphic functions in \eqref{Eq:Infiniteseries} as series in small $\omega$ with intent. The reason is that generally these will have finite radii of convergence (even at $\vec{k}^2=0$). For the holographic theories we will be interested in, this finite radius indicates the appearance of a new pole (see figure \ref{fig:christmastree}) \cite{Withers_2018,Grozdanov:2019kge,Jansen:2020hfd,Abbasi:2020ykq}. To continue the conductivity to frequencies beyond this new pole, and thus cover a larger part of the real frequency axis with our approximation, we must resum the series to produce
\begin{eqnarray}
\label{Eq:Resummed2}
\sigma^{ij}(\omega,\vec{k}) &=& 
\left(
  \sum_{n=0}^{\infty} \tilde{f}_{(\mathrm{L}),n}(\vec{k}^2) (i\omega)^{n}
  + \frac{i R_{(\mathrm{L})}(\vec{k}^2)}{\omega + i \Gamma_{(\mathrm{L})}(\vec{k}^2)}
  - \frac{i \vec{k}^2 R_{\mathfrak{D}}(\vec{k}^2)}{\omega + i \mathfrak{D}(\vec{k}^2) \vec{k}^2}
\right) \frac{k^{i} k^{j}}{\vec{k}^2}  
\nonumber \\
&\;& + 
\left(
  \sum_{n=0}^{\infty} \tilde{f}_{(\mathrm{T}),n}(\vec{k}^2) (i\omega)^{n}
  + \frac{i R_{(\mathrm{T})}(\vec{k}^2)}{\omega + i \Gamma_{(\mathrm{T})}(\vec{k}^2)}
\right)
\left( \delta^{ij} - \frac{k^{i} k^{j}}{\vec{k}^2} \right) \; .
\end{eqnarray}
We note that spatial rotation invariance requires that
	\begin{subequations}
	\begin{align}
		& \lim_{\vec{k}^2 \rightarrow 0} R_{(\mathrm{L})}(\vec{k}^2) = \lim_{\vec{k}^2 \rightarrow 0} R_{(\mathrm{T})}(\vec{k}^2) \; , \qquad \lim_{\vec{k}^2 \rightarrow 0} \Gamma_{(\mathrm{L})}(\vec{k}^2) = \lim_{\vec{k}^2 \rightarrow 0} \Gamma_{(\mathrm{T})}(\vec{k}^2) \; ,  \\
		& \lim_{\vec{k}^2 \rightarrow 0} \tilde{f}_{(\mathrm{L}),n}(\vec{k}^2) = \lim_{\vec{k}^2 \rightarrow 0} \tilde{f}_{(\mathrm{T}),n}(\vec{k}^2)  \; , 
	\end{align}
	\end{subequations}
but at non-zero $\vec{k}$ the residues, poles and holomorphic terms can evolve differently in the longitudinal and transverse sectors.}

{\ The structure of the expression given in \eqref{Eq:Resummed2} is a Mittag-Leffler representation of $\sigma^{ij}(\omega,\vec{k})$ \cite{Conway}. The Mittag-Leffler form of a meromorphic function on an open disc of the complex frequency plane ($D_{\mathrm{R}}$) is the representation of that function as the sum over its isolated poles lying in $D_{\mathrm{R}}$ plus a holomorphic function. Most importantly to our analysis - the Mittag-Leffler expansion of a given meromorphic function is both unique and exact. If we can reproduce the Mittag-Leffler expansion inside a disc, we know everything it is possible to know about the function on that disc.} 

{\ Unfortunately, in the resummation to produce \eqref{Eq:Resummed2} we have lost the link to the effective theory we began with which had only a single pole.  In this paper our remit is to write down the effective, linearised theory, in ``small derivatives'', that reproduces the Mittag-Leffler expansion of the $U(1)$ charge conductivity $\sigma^{ij}(\omega,\vec{k})$ at non-zero frequency and wavevector with an arbitrary number of simple poles. Importantly, we will not consider theories which have branch cut singularities.}

{\ To accomplish this goal, we must take seriously the non-perturbative nature of the diffusive pole and slightly shift our usual hydrodynamic intuition. In standard (relativistic) hydrodynamics one expands in small frequency and wavevector. However, poles are intrinsically non-perturbative in frequency, and this becomes evident when we examine how a Mittag–Leffler representation of the conductivity behaves as we enlarge its domain of definition.}

{\ Assume that we are given a Mittag-Leffler representation of $\sigma^{ij}(\omega,\vec{k})$ which is valid on a disc $D_\mathrm{R}$ in the complex $\omega$-plane for fixed $\vec{k}$. If we try to extend this disc, either the function is entire, we hit a branch point or we eventually encounter new poles. Suppose that there is a new pole as is typically the case for holographic theories. Such new poles can, in principle, be detected from the radius of convergence of the holomorphic part of the Mittag–Leffler series. Identifying a new pole from a series expansion requires knowledge of the series representing the holomorphic function to all orders in $\omega$. The appearance of a pole at the boundary of the disc is therefore a genuinely non-perturbative effect in frequency, even for the diffusive pole. For this reason, in our effective theory, we must specify from the outset how many poles we include as input data.}

{\ Treating poles as non-perturbative has important consequences. It constrains not only the homogeneous part of the effective equations but also the structure of the source terms describing couplings to background fields. In the literature on “hydrodynamic” models containing gapped modes, one often finds statements about treating the gap as small in derivatives. For the linearised theory this is misleading: one never expands in the size of the gap. Every pole, gapped or hydrodynamic, is non-perturbative in small frequency, regardless of its location in the complex plane. In this work we show that one can construct an effective theory, consistent with standard hydrostaticity conditions, that reproduces a Mittag–Leffler expansion with an arbitrary but finite number of isolated poles.}


{\ For legibility, let us constrain ourselves in the types of conductivity \eqref{Eq:DefConductivity} we will consider. We shall work with linearised fluctuations of a theory that has spatial rotational invariance around a background at non-zero temperature $T$ and charge density $\rho$, but only consider zero background electric and magnetic fields. The conductivities we shall be interested in will satisfy the following five constraints:
    \begin{enumerate}
    	\item on a disc $D_{\mathrm{R}}$ of radius $R$ centered at the origin of complex frequency plane, the conductivity $\sigma^{ij}(\omega, \vec{k})$ has a diffusive pole ($\omega \rightarrow 0$ as $\vec{k} \rightarrow \vec{0}$) plus a finite number of isolated gapped poles sitting in the lower half of the complex frequency plane,
        \item the conductivity is derived from a theory that respects spatial rotation and spatial parity invariance,
        \item the $U(1)$ gauge Ward identity is satisfied,
        \item at zero wave-vector $\vec{k}=\vec{0}$, there exists a disc around $\omega=0$ in which the AC conductivity $\sigma_{\mathrm{AC}}(\omega)$ is holomorphic,
        \item the model has finite charge ($\chi_{\rho \rho}$), polarisation ($\chi_{\mathrm{EE}}(\vec{k})$) and magnetisation ($\chi_{\mathrm{BB}}(\vec{k})$) susceptibilities.
    \end{enumerate}
Let us comment upon these constraints: the necessity of point one should be clear from the discussion of the previous section.  The second and third of our constraints are nothing more than imposition of symmetries with the former allowing us to break up the tensor structures of our conductivity into rotationally invariant terms for simplicity. Finally, the fourth and fifth are purely phenomenological. The fourth constraint excludes from our analysis theories coupled to momentum degrees of freedom which enjoy translational invariance.  Regarding point five, in holographic models we generally have ready access to the quantities $\sigma_{\mathrm{AC}}(\omega)$, $\chi_{\mathrm{EE}}(\vec{k})$, $\chi_{\mathrm{BB}}(\vec{k})$ detailed in these constraints, as we are only required to evaluate the holographic Green's functions as a series expansion in their respective parameters near the origin in complex frequency or wavevector respectively. These expansions can readily be determined using the holographic approximants discussed in section \ref{section:holographicapproximant}. }

{\ With this background taken as read, one finds that certain components of the conductivity are independent of our listed constraints above, while others must satisfy particular relationships as a consequence of these restrictions. Let us then consider the maximal disc $D_{\mathrm{R}}$, centered at the origin in complex frequency space, that contains $N-1$ gapped poles in addition to the diffusive pole when $\vec{k} \neq \vec{0}$. In general the radius $R$ will be a function of $\vec{k}$. With our constraints, one finds that the conductivities have the following form:
        \begin{subequations}
        \label{Eq:TotalChargeCorrelator}
	\begin{align}
		\sigma^{ij}(\omega,\vec{k}) &= \sigma_{(\mathrm{L})}(\omega,\vec{k}^2) \left(  \frac{k^{i} k^{j}}{\vec{k}^2} \right)  + \sigma_{(\mathrm{T})}(\omega,\vec{k}^2) \left( \delta^{ij} - \frac{k^{i} k^{j}}{\vec{k}^2} \right) \; , \\
	   \label{Eq:CompleteChargeCorrelatorL}
            \sigma_{(\mathrm{L})}(\omega,\vec{k}^2) 
        &= - \frac{i \vec{k}^2 \left( R_{\mathfrak{D}}(0) + R_{\mathfrak{D}}'(0) \vec{k}^2 + \highlight{\tilde{R}_{\mathfrak{D}}(\vec{k}^2)} \vec{k}^4 \right)}{\omega + i \left( \mathfrak{D}  + \highlight{\tilde{\mathfrak{D}}(\vec{k}^2)} \vec{k}^2 \right) \vec{k}^2} \nonumber \\
        &\;  + \sum_{m=1}^{N-1} \frac{i \left( \highlight{R_{(0),m}} - \vec{k}^2 \highlight{R_{(\mathrm{L}),m}(\vec{k}^2)} \right)}{\omega + i (\highlight{\tau_{(0),m}^{-1}} - \vec{k}^2 \highlight{\tau_{(\mathrm{L}),m}^{-1}(\vec{k}^2)})} \nonumber \\
        &\; + \sum_{n=0}^{N_{D}} \frac{(i\omega)^n}{n!}  \left[  \highlight{\sigma_{\mathrm{AC}}^{(n)}(0) }  - n! \left( \sum_{m=1}^{N-1}  \frac{\highlight{R_{(0),m}}}{( \highlight{\tau_{(0),m}^{-1}} )^{n+1}} \right)  \right] \nonumber \\
        &\;  - \vec{k}^2 \left( f_{(\mathrm{L}),0}(\vec{k}^2) + i \omega f_{(\mathrm{L}),1}(\vec{k}^2)  + \sum_{n=2}^{\infty} \frac{(i\omega)^n}{n!} \highlight{f_{(\mathrm{L}),n}(\vec{k}^2)}  \right) + \mathcal{O}(\omega^{N_{D}+1})    \; ,  \\
        	  \label{Eq:CompleteChargeCorrelatorT}
            \sigma_{(\mathrm{T})}(\omega,\vec{k}^2) 
        &= \sum_{m=1}^{N-1} \frac{i \left( \highlight{R_{(0),m}} - \vec{k}^2  \highlight{R_{(\mathrm{T}),m}(\vec{k}^2)} \right)}{\omega + i (\highlight{\tau_{(0),m}^{-1}} - \vec{k}^2 \highlight{\tau_{(\mathrm{T}),m}^{-1}(\vec{k}^2)})}  \nonumber  \\
        &\; + \sum_{n=0}^{N_{D}} \frac{(i\omega)^n}{n!}  \left[  \highlight{\sigma_{\mathrm{AC}}^{(n)}(0)}   - n! \left( \sum_{m=1}^{N-1}  \frac{\highlight{R_{(0),m}}}{( \highlight{\tau_{(0),m}^{-1}} )^{n+1}} \right) - \vec{k}^2  \highlight{f_{(\mathrm{T}),n}(\vec{k}^2)}  \right] + \mathcal{O}(\omega^{N_{D}+1})  \; ,
	\end{align} 
    \end{subequations}
where $\sigma_{\mathrm{AC}}^{(n)}(0)$ is the $n^{\mathrm{th}}$ derivative of the AC conductivity and we have introduced a high frequency cut-off $N_{D}\geq1$ which counts how many terms in frequency we wish to include in the holomorphic part of our approximation. An exact match within the disc of convergence is given when $N_{D} \rightarrow \infty$.}

{\ Terms in red in \eqref{Eq:TotalChargeCorrelator} are system dependent, being unfixed by our constraints. Consequently one can specify any reasonable value for them. Terms in black are fixed by the existence of finite AC conductivity\footnote{We have chosen to write $\sigma_{\mathrm{AC}}$ as a function of $i \omega$ to emphasis the complex value of this function and to make certain formulae cleaner.} $\sigma_{\mathrm{AC}}(i \omega)$, charge susceptibility $\chi_{\rho \rho}$ and, polarisation $\chi_{\mathrm{EE}}(\vec{k})$ and magnetisation $\chi_{\mathrm{BB}}(\vec{k})$ susceptibilities or by the terms in red. The detailed derivation of the relations can be found in Appendix \ref{appendix:refining}.  The expressions for the black terms can be quite complicated, with the exception of the Einstein relation, 
    \begin{eqnarray}
        \label{Eq:EinsteinRelation}
        \mathfrak{D} &=& \frac{\sigma_{\mathrm{DC}}}{\chi_{\rho \rho}}
    \end{eqnarray}
and they are thus un-illuminating. Hence we relegate details of their derivation to appendix \ref{appendix:refining}. It is sufficient to know for our purposes that our effective theory satisfies the constraints on the conductivity and has precisely the freedom necessary to match the red terms. Moreover, as explained in section \ref{section:holographicapproximant}, the holographic approximant yields the three constraining quantities: $\sigma_{\mathrm{AC}}(i\omega)$, $\chi_{\mathrm{EE}}(\vec{k})$ and  $\chi_{\mathrm{BB}}(\vec{k})$ as series about zero in their respective arguments.}

{\ Before continuing, we remind the reader of the important point that the expressions \eqref{Eq:TotalChargeCorrelator} are exact and unique in their disc of definition, which is typically up to the next pole out from the origin of complex frequency space that is not included in the summation. Reproducing them from our effective theory is the same as knowing the full Green's function in the disc.}

\section{An effective theory of many poles} \label{sec:manypoles}

\subsection{Hydrostaticity and the effective action   }\label{sec:Hydrostatic}

{\noindent We begin by examining what form the effective equations must take if there exists a hydrostatic generating functional \cite{Jensen:2012jh,Haehl_2015}. We shall use this as a basis for our effective theory and for simplicity stick to $(2+1)$-dimensions where the magnetic field is a pseudo-scalar. The set of stationary states described by such a generating functional are constrained to vanish under the action of a time-like Killing vector which acts via Lie derivatives on the fields. Let $V^{\mu}$ be this Killing vector and define the fluid velocity $u^{\mu}$ and temperature $T$ in terms of the Killing vector to be
    \begin{subequations}
    \label{Eq:ThermodynamicFrameQuantities}
	\begin{eqnarray}
		u^{\mu} &=& \frac{V^{\mu}}{\sqrt{-V^2}} \; , \qquad T = \frac{1}{ \beta_0 \sqrt{-V^2} }\; ,	
	\end{eqnarray}
where $\beta_0$ specifies the normalization of the temperature. As we decouple momentum and temperature fluctuations, these terms will be constant in the effective theory we consider. On the other hand, the integral around the thermal circle of the gauge field defines a (potentially space and time dependent) chemical potential which we can identify with \cite{Jensen:2012jh,Kovtun_2016}
	\begin{eqnarray}
		\mu &=& \frac{V^{\mu} A_{\mu}	+ \Lambda}{\sqrt{-V^2}}
	\end{eqnarray}
    \end{subequations}
where $\Lambda$ is a gauge parameter present to ensure that $\mu$ is gauge invariant. We can also define a gauge field strength $F_{\mu \nu}$ which, for $d = 2$, one can decompose with respect to the fluid velocity to define a (vector) electric field $E_{\mu}$ and (pseudo-scalar) magnetic field $B$. These are given by
\begin{subequations}
 \begin{eqnarray}
    F_{\mu \nu} &=& u_\mu E_\nu - u_\nu E_\mu - \Sigma_{\mu \nu} B \; , \qquad
    E_\mu = F_{\mu \nu} u^{\nu}\; , \qquad B = -\frac{1}{2} \epsilon^{\mu \alpha \beta} u_{\mu} F_{\alpha \beta} \; , \qquad
 \end{eqnarray}
 where
 \begin{equation}
    \Sigma^{\mu \nu} = \epsilon^{\mu \nu \rho} u_{\rho} \;  , \qquad \epsilon^{\mu \nu \rho} = u^{\mu} \Sigma^{\nu \rho} + u^{\nu} \Sigma^{\rho \mu} + u^{\rho} \Sigma^{\mu \nu} \; , \qquad \Sigma^{\mu \nu} \Sigma_{\nu \rho} = \Pi\indices{^{\mu}_{\rho}} \; .
\end{equation}
\end{subequations}
Imposing vanishing of the Lie derivative acting on the temperature and chemical potential \eqref{Eq:ThermodynamicFrameQuantities} leads to the following conditions 
\begin{subequations}
\begin{align}
    \label{Eq:ThermodynamicConstraints}
    & \nabla_\nu^{\perp} T = - T a_\nu, \quad \quad \nabla_\nu^{\perp} \mu = E_\nu - \mu a_\nu \; , \\
    				&\; \nabla_{\mu} = - u_{\mu} D +  \nabla^{\perp}_{\mu} \; ,  \qquad \nabla_{\mu}^{\perp} = \Pi\indices{_{\mu}^{ \nu}} \nabla_{\nu} \; , \qquad D = u^\mu \nabla_\mu \; ,
\end{align}
\end{subequations}
where $a^\mu \equiv u^\nu \nabla_{\nu} u^\mu$ is the acceleration vector satisfying $u_\mu a^\mu = 0$. In what follows we shall take a flat spacetime with $V^{\mu} = (1,\vec{0})$ and as such $a^{\mu}=0$. The constraints \eqref{Eq:ThermodynamicConstraints}, and those derived by acting with $\mathcal{L}_{V}$ on $F_{\mu \nu}$, will define hydrostatic states of our system.} 

{\ Following \cite{Kovtun_2016} one proceeds order by order in derivatives to construct a generating functional with effective action $W$ whose variations produce the hydrostatic constitutive relations. To order two in fluctuations of the gauge field, the effective action for the generating functional must be given in terms of the effective fields by
    \begin{subequations}
	\begin{align}
        \label{Eq:GeneratingFunctional}
		&\; W_{(2)}[A,\delta A] = \frac{1}{2} \int d^{2+1}x \sqrt{-g} \; \left[ \chi_{\rho \rho} (\delta \mu)^2 + F_{1} \delta E_{\mu} \Pi^{\mu \nu} \delta E_{\nu} + \delta E_{\mu} \Pi_{(\mathrm{L})}^{\mu \nu} F_{2}(\nabla_{\perp}^2)\left[ \delta E_{\nu} \right] \right. \nonumber \\
				&\; \left. \hphantom{W_{(2)}[A,\delta A] = \ \frac{1}{2} \int d^{2+1}x \sqrt{-g} \; \left[ \right.}  + \delta E_{\mu} \Pi_{(\mathrm{T})}^{\mu \nu} F_{3}(\nabla_{\perp}^2)\left[ \delta E_{\nu} \right] +  (\delta B) F_{4}(\nabla_{\perp}^2)[\delta B] \right] \; , \nonumber \\
				&\; 
	\end{align}
where
	\begin{align}
				&\; \Pi_{(\mathrm{L})}^{\mu \nu} =  \nabla_{\perp}^{\mu} \nabla_{\perp}^{\nu}  \; , \qquad \Pi_{(\mathrm{T})}^{\mu \nu} = \Pi^{\mu \nu} \nabla_{\perp}^2  - \nabla_{\perp}^{\mu} \nabla_{\perp}^{\nu} \; , \qquad
	\end{align}
    \end{subequations}
and, $F_{2}$, $F_{3}$ and $F_{4}$ are arbitrary polynomials of $\nabla_{\perp}^2$, and all derivatives act to the right. In writing this effective action for the generating functional we have also assumed that the space is flat so that the Riemann tensor vanishes and consequently we can commute covariant derivatives without concern. The linearised fluctuation of the $U(1)$ charge current is related to the generating functional by
    \begin{subequations}
	\begin{eqnarray}
		\delta J^{\mu} &=& - \frac{\delta W_{(2)}[A,\delta A]}{\delta (\delta A_{\mu})} \; ,
	\end{eqnarray}
where 
	\begin{align}
		&\; \delta \mu = u^{\mu} \delta A_{\mu} \; , \qquad \delta E_{\mu} = u^{\nu} \left( \partial_{\mu} \delta A_{\nu} - \partial_{\nu} \delta A_{\mu} \right) \; , \qquad \\
		&\; \delta B = - \frac{1}{2} u_{\mu} \epsilon^{\mu \nu \rho} \left( \partial_{\nu} \delta A_{\rho} - \partial_{\rho} \delta A_{\nu} \right)  \; . 
	\end{align}
    \end{subequations}
With these variations to hand, one finds that a hydrostatic configuration for our system must have a charge current of the form \cite{Kovtun_2016}
	\begin{subequations}
    \label{Eq:HydrostaticConstitutive}
	\begin{eqnarray}
		\delta J^{\mu} &=&  (\chi_{\rho \rho} \delta \mu) u^{\mu} - \nabla_{\nu} \delta M^{\nu \mu} \; , \\
        \label{Eq:PolarisationTensor}
	       \delta M^{\mu \nu} &=& u^{\nu} \delta p^{\mu} - u^{\mu} \delta p^{\nu} - \epsilon^{\mu \nu \rho} u_{\rho} \delta m \; ,
	\end{eqnarray}
	\end{subequations}
where $\delta M^{\mu \nu}$ is the fluctuation of the electromagnetic polarisation tensor with the polarisation and magnetisation components given by
	\begin{subequations}
	        \label{Eq:PolarisationMagnetisation}
	\begin{eqnarray}
		\delta p^{\mu} &=&  F_{1} \delta E^{\mu}
					      + \left( F_{2}[\nabla_{\perp}^2] \Pi_{(\mathrm{L})}^{\mu \nu}  + F_{3}[\nabla_{\perp}^2] \Pi_{(\mathrm{T})}^{\mu \nu} \right) \delta E_{\nu} \; ,  \\
					       \delta m &=& F_{4}[\nabla_{\perp}^2] \delta B \; . \qquad
	\end{eqnarray}
	\end{subequations}
Consequently the polarisation and magnetisation susceptibilities are
	\begin{subequations}
			\label{Eq:PolarisationSusceptibility}
	\begin{eqnarray}
		\chi_{\mathrm{EE}}^{\mu \nu} &=& \frac{\delta p^{\mu}}{\delta E_{\nu}} = F_{1} \delta^{\mu \nu} + F_{2}[\nabla_{\perp}^2] \Pi_{(\mathrm{L})}^{\mu \nu} + F_{3}[\nabla_{\perp}^2] \Pi_{(\mathrm{T})}^{\mu \nu} \; , \\
		\chi_{\mathrm{BB}} &=& \frac{\delta m}{\delta B} = F_{4}[\nabla_{\perp}^2] \; , \qquad
	\end{eqnarray}
	\end{subequations}
respectively. Note that the polarisation is perpendicular to the fluid velocity and that by definition $\delta p^\mu = u_\rho \delta M^{\rho \mu}$. These currents form the basis about which we shall build our effective theory of generic linearised fluctuations. This, and the choice of operators we add to build our effective equations, is why our effective framework will be compatible with the standard hydrostaticity conditions of hydrodynamics, and thus thermodynamics, even though there are additional gapped modes.}

\subsection{The effective linearised theory of many poles   }

{\noindent We now come to the point where we can state our effective, linearised theory. We begin with the hydrostatic constitutive relations of \eqref{Eq:HydrostaticConstitutive} and choose to work in a frame\footnote{A brief discussion of potential transformations to alternate frames is given in appendix \ref{appendix:frametrans}.} where
	\begin{eqnarray}
        \label{Eq:FrameConditions}
		u_{\mu} \delta J^{\mu} := 	- \left( \chi_{\rho \rho} \delta \mu - \nabla_{\mu}^{\perp} \delta p^{\mu} \right)  \; . 
	\end{eqnarray}
Corrections to the expression for $\delta J^{\mu}$ are then entirely transverse to $u^{\mu}$ at all orders in derivatives. To accomplish our goal of incorporating gapped modes, we wish to promote the spatial part of $\delta J^{\mu}$ to an independent hydrodynamic variable. This concurs with previous approaches \cite{Grozdanov:2018fic} and bares more than a surface resemblance to the approach taken in deriving Muller-Israel-Stewart theory \cite{Baier:2007ix,ISRAEL1976310,ISRAEL1979341}. Thus we take our constitutive relation for $\delta J^{\mu}$ to be
    \begin{subequations}
    \label{Eq:EffectiveEquations}
    \begin{eqnarray}
        \label{Eq:ChargeCurrentConstitutive}
        \delta J^{\mu} &=& \left( \chi_{\rho \rho} \delta \mu - \nabla_{\nu}^{\perp} \delta p^{\nu} \right) u^{\mu} + \left( \delta \bar{J}^{\mu} - \Pi\indices{^{\mu}_{\nu}} \nabla_{\rho} \delta M^{\rho \nu} \right)
    \end{eqnarray}
where $p^{\mu}$ and $M^{\mu \nu}$ are given in \eqref{Eq:PolarisationTensor} and \eqref{Eq:PolarisationMagnetisation} and $\delta \bar{J}^{\mu}$ vanishes at hydrostaticity. Supplementing this constitutive equation are the following two equations of motion
	\begin{eqnarray}
        \label{Eq:ChargeConserv}
		\nabla_{\mu} \delta J^{\mu} &=& \chi_{\rho \rho} D \delta \mu + \nabla_{\mu}^{\perp} \delta \bar{J}^{\mu} = 0 \; , \\
        \label{Eq:Nonconserv}
		\left[ \Pi_{n=1}^{N-1} \left( \Pi^{\mu \nu} D + \Gamma_{n}^{\mu \nu}[\nabla_{\perp}^2] \right) \right] \delta \bar{J}_{\nu} 
		&=& \bar{\sigma}^{\mu \nu}\left[ \nabla_{\perp}^{2} \right] \left( \delta E_{\nu} - \nabla^{\perp}_{\nu} \delta \mu \right) \nonumber \\ 
		&\;& + r^{\mu \nu}[D,\nabla_{\perp}^{2}] (D \delta E_{\nu}) \; . \qquad
	\end{eqnarray}
	\end{subequations}
In the above we have employed the $(2+1)$-dimensional Bianchi identity
    \begin{eqnarray}
        D \delta B + \Sigma^{\mu \nu} \nabla_{\mu}^{\perp} \delta E_{\nu} &=& 0 \; ,
    \end{eqnarray}
to eliminate any potential time derivatives of the magnetic field which must appear with $\Sigma^{\mu \nu}$ to maintain spatial parity invariance. The first equation \eqref{Eq:ChargeConserv} is nothing more than charge conservation, while the second equation \eqref{Eq:Nonconserv} tells us about non-conservation of the spatial charge current.}

{\ In the hydrostatic limit the second equation \eqref{Eq:Nonconserv} reduces to 
	\begin{eqnarray}
		 \Gamma_{n}^{\mu \nu}[\nabla_{\perp}^2] \delta \bar{J}_{\mu} &=& 0 \; ,
	\end{eqnarray}
which is of course satisfied if the $\delta \bar{J}^{\mu}$ current vanishes at hydrostaticity, which is expected if it corresponds to the non-hydrostatic part of the current i.e. it is build only from the hydrostaticity conditions. Thus it follows that $U(1)$ charge conservation is automatically satisfied at hydrostaticity, as expected since it is a consequence of $U(1)$ gauge invariance of the generating functional \eqref{Eq:GeneratingFunctional}. Outside hydrostaticity, one can see that the operator acting on $\delta \bar{J}^{\mu}$ will introduce exactly $(N-1)$ poles in the complex frequency plane, while charge conservation \eqref{Eq:ChargeConserv} introduces one more in the longitudinal sector. What is perhaps less obvious is that there is sufficient room on the right hand side of \eqref{Eq:Nonconserv} to build the holomorphic part of any Green's function in the disc, allowing us to absolutely match the Green's function of the charge currents. We shall demonstrate this by obtaining the Green's functions for our effective theory in later sections and comparing to \eqref{Eq:TotalChargeCorrelator}.}

{\ At the risk of repeating ourselves, we remind the reader that an important part of this work is a realignment in thinking about the derivative expansion. Many approaches to the hydrodynamics of theories with gapped modes (i.e. quasihydrodynamics) try to treat the gap as small in derivatives. Our approach to the problem does not require this restriction. Instead, one must understand the role of the operator
    \begin{eqnarray}
        \label{Eq:ZerothOrderOp}
        \left[ \Pi_{n=1}^{N-1} \left( \Pi^{\mu \nu} D + \Gamma_{n}^{\mu \nu}[\nabla_{\perp}^2] \right) \right]
    \end{eqnarray}
as being the leading term, in its entirety, in the derivative expansion. Our procedure never makes time derivative corrections to this operator. Instead, order by order corrections in $D$ only appear on the right hand side of \eqref{Eq:Nonconserv} in the term
    \begin{eqnarray}
        \label{Eq:SourceTerm}
        \bar{\sigma}^{\mu \nu}\left[ \nabla_{\perp}^{2} \right](\delta E_{\nu} - \nabla_{\nu}^{\perp} \delta \mu) +
		     r^{\mu \nu}[D,\nabla_{\perp}^{2}] (D \delta E_{\nu}) \; . 
    \end{eqnarray}
 Quasihydrodynamic approaches then emerge when we treat one gap in \eqref{Eq:ZerothOrderOp} as parametrically smaller than all others i.e. one pole under tuning moves very close to the origin in the complex frequency plane and we truncate to a single gapped pole. Importantly, every quasihydrodynamic model can be embedded in our approach, which more general.

For example, in the case of the D3/D5 probe brane, at large charge density, one finds a diffusive $\omega_{\mathfrak{D}}$ and quasihydrodynamic pole that at leading order in large charge density \underline{and} small wavevector behave like
	\begin{subequations}
	\begin{eqnarray}
		\omega_{\mathfrak{D}} &=& - i \mathfrak{D} \vec{k}^2 + \mathcal{O}(\vec{k}^4) \; , \\
		\omega_{\mathrm{quasi}} &=& - i \left( \frac{0.539}{\sqrt{\tilde{\rho}}} - i \mathfrak{D} \vec{k}^2 + \mathcal{O}(\vec{k}^4) \right) \; . 
	\end{eqnarray}
	\end{subequations}
where again $\mathfrak{D}$ is the diffusion constant and $\tilde{\rho}$ is the temperature normalised charge density. As $\tilde{\rho} \rightarrow \infty$, treating $1/\sqrt{\tilde{\rho}} \sim |\vec{k}|$ in terms of small parameters, the gapped mode becomes longer and longer lived while all other modes tend to fixed positions further from the origin of the complex frequency plane; we compute more terms in these dispersion relations in section \ref{sec:omegandk} and one can see the regime begin to emerge in fig. \ref{fig:qnm_evolution}. As opposed to quasihydrodynamics, our formalism allows to us include $\omega_{\mathrm{quasi}}$ for arbitrary charge density.

{\ If one follows our framework, and chooses the correct basis for expansion of this term \eqref{Eq:SourceTerm}, then they are able to completely match the derived effective Green's functions and the measured Green's function order by order in derivatives. We shall elaborate further on the time derivative operator basis in section \ref{sec:NonzeroomegaGreen}. Notice however that here we are definitively treating derivative counting in $D$ as distinct from derivative counting in $\nabla_{\perp}$. In the latter case of derivative counting in $\nabla_{\perp}$, one counts as per usual in hydrodynamics and we expect the expansions of operators such as $\Gamma_{n}^{\mu \nu}[\nabla_{\perp}^2]$ in $\nabla_{\perp}$ to be valid up to the first collision of poles. However, if one were to try and expand \eqref{Eq:ZerothOrderOp} treating $D \sim \nabla_{\perp}$ then one would find erroneous results.}

 Given our distinct treatments of space and time, one may naturally worry about covariance in our theory. We first remind the reader that Lorentz invariance is broken at finite temperature. In particular boosts, which are the only non-trivial transformation linking space and time, are not symmetries of the theory. Time and space are already distinct in the usual formulation of hydrodynamics. The key difference between our approach and that typically found in hydrodynamics is that we choose to resum the frequency dependent terms, which otherwise has no effect on the covariance of the theory.

\subsubsection{Charge, polarisation and magnetisation susceptibilities  }\label{sec:Hydrostaticsub}

{\noindent Let us derive the correlators from our effective theory given in \eqref{Eq:EffectiveEquations} and see if we arrive at the desired result (i.e.~the expressions of \eqref{Eq:TotalChargeCorrelator}). We begin by considering zero frequency but non-zero $\vec{k}$, roughly corresponding to the hydrostatic limit.  If we choose our coordinates such that $u^{\mu}=(1,\vec{0})$, which we shall do henceforth, we can readily solve our effective equations for the charge current correlators. We make note of the following identifications at non-zero frequency and wavevector
    \begin{subequations}
    \label{Eq:Defs}
    \begin{align}
       & \epsilon_{txy} = 1 \; , \qquad
        \delta \vec{E} = i \left( k \delta a_{t} + \omega \delta a_{x}, \omega \delta a_{y} \right) \; , \qquad \delta B = - i k \delta a_{y} \; , \\
        & \delta \vec{p} = i \left( (F_{1} - k^2 F_{2}[-k^2]) (k \delta a_{t} + \omega \delta a_{x}) , \omega (F_{1} - k^2 F_{3}[-k^2]) \delta a_{y} \right) \; , \\
        & \delta m = - \delta M^{xy} = - i k F_{4}[-k^2] \delta a_{y} \; , \label{eq:magnetization}
    \end{align}
    \end{subequations}
where we have chosen $\vec{k}=(k,0)$ as these will be useful throughout the following sections.}

{\ With the definitions of \eqref{Eq:Defs} to hand, we find from the non-conservation equation \eqref{Eq:Nonconserv} at zero frequency that
\begin{subequations}
        \begin{eqnarray}
        \label{Eq:omega0eqns}
        \delta \bar{J}^{y} &=& 0 \; , \\ 
        \Pi_{n=1}^{N-1} \left( \Gamma_{(0),n} - k^2 \Gamma_{(\mathrm{L}),n}(-k^2) \right) \delta \bar{J}^{x}  &=& i k \left( \bar{\sigma}_{(0)} - k^2 \bar{\sigma}_{(\mathrm{L})}(-k^2) \right) \left( \delta a_{t} - \delta \mu \right) \; , \qquad 
    \end{eqnarray}
\end{subequations}
where we have used that spatial rotation invariance implies that a general spatial two-tensor $\Xi^{ij}(\nabla_\perp^2)$ can be decomposed as
\begin{equation} \label{Eq:SpatialTensorDecomposition}
    \Xi^{ij} (-\vec{k}^2) = \Xi_{(0)} \delta^{ij} - \Xi_{(\mathrm{L})}(-\vec{k}^2) k^i k^j - \Xi_{(\mathrm{T})}(-\vec{k}^2) \left( \vec{k}^2 \delta^{ij} - k^i k^j \right) \; 
\end{equation}
and which will be used in various parts of the remainder of the paper. The conservation equation \eqref{Eq:ChargeConserv} on the other hand tells us
    \begin{eqnarray}
        0 &=& i k \delta \bar{J}^{x} \; , 
    \end{eqnarray}
for all values of $k$ which, combined with \eqref{Eq:omega0eqns}, has a solution if we identify
    \begin{eqnarray}
        \delta a_{t} &=& \delta \mu \; . 
    \end{eqnarray}
Using \eqref{Eq:ChargeCurrentConstitutive} we then see that the Green's functions are given by
   \begin{subequations}
    \begin{eqnarray}
            \langle J^{t} J^{t} \rangle_{\mathrm{R}}(0,\vec{k}) 
            &=& - \left( \chi_{\rho \rho} + k^2 \left( \chi_{\mathrm{EE}}^{(0)} + k^2 \chi_{\mathrm{EE}}^{(\mathrm{L})}(k^2) \right) \right) \; , \\
                \langle J^{y} J^{y} \rangle_{\mathrm{R}}(0,\vec{k}) 
             &=&  k^2 \chi_{\mathrm{BB}}(k^2) \; , 
    \end{eqnarray}
    \end{subequations}
where we have used the identifications 
	    \begin{eqnarray}
        \label{Eq:SusceptibilityMatch}
       F_{1} = \chi_{\mathrm{EE}}^{(0)}  \; , \qquad F_{2}[-k^2] =  \chi_{\mathrm{EE}}^{(\mathrm{L})}(k^2) \; , \qquad F_{4}[-k^2] &=& \chi_{\mathrm{BB}}(k^2) \; , \qquad 
    \end{eqnarray}
following from \eqref{Eq:PolarisationSusceptibility} and decomposed $\chi_{\mathrm{EE}}^{\mu \nu}$ defined in \eqref{Eq:PolarisationSusceptibility}  into
	\begin{eqnarray}
			\chi_{\mathrm{EE}}^{\mu \nu} \partial_{\mu} \otimes \partial_{\nu} 
		&=& \left( \chi_{\mathrm{EE}}^{(0)} + k^2 \chi_{\mathrm{EE}}^{(\mathrm{L})}(k^2) \right)  \left( \frac{k^{i} k^{j}}{\vec{k}^2} \right) \partial_{i} \otimes \partial_{j} \nonumber \\
		&\;& + \left( \chi_{\mathrm{EE}}^{(0)} + k^2 \chi_{\mathrm{EE}}^{(\mathrm{T})}(k^2) \right)  \left( \delta^{ij} - \frac{k^{i} k^{j}}{\vec{k}^2} \right) \partial_{i} \otimes \partial_{j} \; . 
	\end{eqnarray}
Finite $\chi_{\rho \rho}$, $\chi_{\mathrm{EE}}(\vec{k})$ and $\chi_{\mathrm{BB}}(\vec{k})$ are constraints we imposed to arrive at \eqref{Eq:TotalChargeCorrelator}. Our effective theory has reproduced these constraints and we have identified three (we are missing $F_{3}[-k^2] \sim \chi_{(\mathrm{EE})}^{(\mathrm{T})}(k^2)$) of the terms that appear in the on-shell action \eqref{Eq:GeneratingFunctional} in terms of quantities we can evaluate using the holographic approximant. Most importantly, the number of poles we include in our effective theory has no impact on these ``thermodynamic susceptibilities'' - thermodynamics is safe from modification.}

\subsubsection{Dispersion relations  }\label{sec:modes}

{\noindent Having satisfied the phenomenological constraints used to derive \eqref{Eq:TotalChargeCorrelator}, let us see if the source-free versions of our effective equations \eqref{Eq:EffectiveEquations} can reproduce the poles in the conductivity. Those in the sector transverse to the wave-vector are easily determined and have the form
   \begin{subequations}
   \label{Eq:TransverseModes}
    \begin{eqnarray}
        \omega_{n=1,\ldots,N-1} &=& - i \left( \Gamma_{(0),n} - k^2 \Gamma_{(\mathrm{T}),n}(-k^2) \right) \; . 
    \end{eqnarray}
where the notation for the decomposition has been introduced in \eqref{Eq:SpatialTensorDecomposition}. We thus identify
	\begin{eqnarray}
		\tau_{(0),n}^{-1} =  \Gamma_{(0),n}  \; , \qquad  \Gamma_{(\mathrm{T}),n}(-k^2) = \tau_{(\mathrm{T}),n}^{-1}(k^2) \; , 
	\end{eqnarray}	
	\end{subequations}
in \eqref{Eq:TotalChargeCorrelator} in terms of quantities appearing in our effective theory \eqref{Eq:EffectiveEquations}. As long as the transverse pole motion is smooth in $\vec{k}$ we will always be able to tune $\Gamma_{(\mathrm{T}),n}(-k^2)$ order by order in small $k^2$ (i.e. spatial derivatives) to match the motion up to the first collision of any pair of the poles included in our Mittag-Leffler representation of the conductivity.}

{\ The modes in the longitudinal sector are a little more complicated because they are determined not only by dependence of the longitudinal  relaxation terms $\Gamma_{(\mathrm{L})}(-k^2)$ on $\vec{k}$ but also the terms necessary to yield a diffusive mode. To determine the modes we are instructed to find the zeroes of the following polynomial
    \begin{equation}
        \label{Eq:LongitudinalPolynomial}
        \omega \Pi_{n=1}^{N-1} \left( \omega + i ( \Gamma_{(0),n} - k^2 \Gamma_{(\mathrm{L}),n}(-k^2) ) \right) + \frac{i^N k^2}{\chi_{\rho \rho}}  ( \bar{\sigma}_{(0)} - k^2 \bar{\sigma}_{(\mathrm{L})}(-k^2) )\; . 
    \end{equation}
This is a polynomial of order $N$ in $\omega$ and we have $N$ free functions of $\vec{k}$ 
$\bigl(\bar{\sigma}_{(0,\mathrm{L})}(k^2)$ and $\Gamma_{(0,\mathrm{L}),\,n}(k^2)\bigr)$ that we can tune to match pole motions. Let us do this to leading order in $\vec{k}^2$ to illustrate our point. At $\vec{k}=\vec{0}$ one can check that the solutions are
    \begin{eqnarray}
        \omega_{\mathfrak{D}} = 0 \; , \qquad \omega_{n=1,\ldots,N-1} = - i \Gamma_{(0),n} 
    \end{eqnarray}
exactly matching the $\vec{k}=\vec{0}$ modes in \eqref{Eq:TransverseModes}, up to the presence of the additional diffusive pole $\omega_{\mathfrak{D}}$. The equivalence of longitudinal and transverse sector modes at $\vec{k}=\vec{0}$ is as one would expect for a spatially rotation invariant theory. At leading non-zero order in $\vec{k}$, these modes become
    \begin{subequations}
    \label{Eq:longitudinalmodes}
    \begin{eqnarray}
        \omega_{\mathfrak{D}} &=& - i \left( \frac{\bar{\sigma}_{(0)}}{\chi_{\rho \rho} \Pi_{m=1}^{N-1} \Gamma_{(0),m}} \right) k^2 + \mathcal{O}(k^4) \; , \\
        \omega_{n} &=& - i \Gamma_{(0),n} - i \left( \frac{ \bar{\sigma}_{(0)}}{\chi_{\rho \rho} \Gamma_{(0),n} \Pi_{m=1,i \neq n}^{N-1} \left( \Gamma_{(0),n} - \Gamma_{(0),m} \right)} - \Gamma_{(\mathrm{L}),n}(0) \right) k^2 \nonumber \\
        			&\;& + \mathcal{O}(k^4)  \; . \qquad 
    \end{eqnarray}
    \end{subequations}
Taking $n \geq 2$ derivatives with respect to $k^2$ of \eqref{Eq:LongitudinalPolynomial} and subsequently setting $k^2=0$, one can see at each new order in derivatives one has the free parameters $\frac{\partial^{n-2}}{\partial (k^2)^{n-2}} \sigma_{(\mathrm{L})}(0)$ and $\frac{\partial^{n-1}}{\partial (k^2)^{n-1}} \Gamma_{(\mathrm{L}),n}(0)$ which can be used to match any observed pole motion with absolute precision up to the radius of convergence of the respective series in small $k^2$. In particular, we identify
    \begin{subequations}
    \label{Eq:LongitudinalModeIDs}
    \begin{eqnarray}
    	\label{Eq:IdentificationofD}
        \mathfrak{D} &=&  \left( \frac{\bar{\sigma}_{(0)}}{\chi_{\rho \rho} \Pi_{m=1}^{N-1} \Gamma_{(0),m}} \right) \; , \\
        \label{Eq:LongitudinalModeIDstau0}
        \tau_{(0),m}^{-1} &=& \Gamma_{(0),m} \; , \\
        \tau_{(\mathrm{L}),m}^{-1}(k^2) &=& \left( \frac{\bar{\sigma}_{(0)}}{\chi_{\rho \rho} \Gamma_{(0),n} \Pi_{m=1,i \neq n}^{N-1} \left( \Gamma_{(0),n} - \Gamma_{(0),m} \right)} - \Gamma_{(\mathrm{L}),n}(0) \right) k^2 \nonumber \\
        &\;& + \mathcal{O}(k^4) \; .
    \end{eqnarray}
    \end{subequations}
Thus, whatever the pole motions, on the condition that they are smooth in $\| \vec{k} \|$ and $\| \vec{k} \|$ is sufficiently small so as to avoid pole collisions \cite{Withers_2018,Grozdanov:2019kge} then we can match the modes of our effective theory to the poles of the conductivity \eqref{Eq:TotalChargeCorrelator} precisely.}

\subsubsection{Spatially homogeneous Green's functions  }
\label{sec:NonzeroomegaGreen}

{\noindent Let us now work in the opposite extreme to section \ref{sec:Hydrostaticsub} and take $\vec{k} = \vec{0}$ with $\omega \neq 0$. Our conservation equation \eqref{Eq:ChargeConserv} tells us that
    \begin{eqnarray}\label{Eq:ChargeConservZeroK}
        - i \omega \chi_{\rho \rho} \delta \mu(\omega) &=& 0 \; ,  
    \end{eqnarray}
i.e. $\delta \mu(\omega) = 0$. Meanwhile, the non-conservation equation \eqref{Eq:Nonconserv} becomes
    \begin{eqnarray} \label{Eq:NonconservZerok}
        \delta \bar{J}^{i}
        &=& i \omega \left[ \frac{\bar{\sigma}_{(0)} - i \omega r_{(0)}(- i \omega)}{ \Pi_{n=1}^{N-1} \left(- i \omega + \Gamma_{(0),n} \right)} \right] \delta a^{i} \; .
    \end{eqnarray}
At $\vec{k}=\vec{0}$ there are also contributions from the polarisation vector to the total spatial charge current, $\delta J^{i} = \delta \bar{J}^{i} - i \omega \delta p^{i}$, recalling that $p^\mu = u_\rho M^{\rho \mu}$. Subsequently this means that
    \begin{eqnarray} \label{eq:currentCorrelatorZerok}
        \langle J^{i} J^{j} \rangle_{\mathrm{R}}(\omega,\vec{0}) &=& - i \omega \left( \frac{\bar{\sigma}_{(0)} - i \omega r_{(0)}(- i \omega)}{ \Pi_{n=1}^{N-1} \left(- i \omega + \Gamma_{(0),n} \right)} - i \omega \chi_{\mathrm{EE}}^{(0)}  \right) \delta^{ij}  \; ,
    \end{eqnarray}
where we have employed $F_{1}=\chi_{\mathrm{EE}}^{(0)}$ from \eqref{Eq:SusceptibilityMatch} and \eqref{Eq:Defs}. For this above expression to correspond to a Mittag-Leffler expansion involving $N-1$ simple poles (the diffusion pole has cancelled with a zero as expected), we see that the function $r_{(0)}(- i \omega)$ cannot be arbitrary, which acts as a constraint on our effective system. In particular, 
    \begin{subequations}
    \label{eq:ZerokNumerator}
    \begin{eqnarray}
        \label{Eq:Hydrozerokconstraint}
               r_{(0)}(- i \omega)
        &=& \frac{i}{\omega} \left[ \left( \sigma_{\mathrm{AC}}(i\omega) + i \omega \chi_{\mathrm{EE}}^{(0)}  - \sum_{n=0}^{\infty} \sum_{m=1}^{N-1} \frac{R_{(0),m}(i \omega)^n}{(\Gamma_{(0),m})^{n+1}} \right) Q(\omega) \right. \nonumber \\
        &\;& \left. \hphantom{ - \frac{i}{\omega} \left[ \right.} \vphantom{ \sum_{n=0}^{\infty} \sum_{m=1}^{N-1} \frac{R_{(0),m}(i \omega)^n}{(\Gamma_{(0),m})^{n+1}}} + \sum_{n=1}^{N-1} R_{(0),n} P_{n}(\omega) - \bar{\sigma}_{(0)} \right]  \; , \qquad \\
        P_{n}(\omega) &=& \Pi_{m=1, m \neq n}^{N-1} \left(- i \omega + \Gamma_{(0),m} \right)  \; , \; \; \; Q(\omega) = \Pi_{n=1}^{N-1} \left(- i \omega + \Gamma_{(0),n} \right) \; , \qquad
    \end{eqnarray}
    \end{subequations}

The coefficients $R_{(0),n}$ and the holomorphic function $\sigma_{\mathrm{AC}}(i \omega)$ have been named appropriately to match onto the form of the conductivity given in \eqref{Eq:TotalChargeCorrelator} once we use the identification $\Gamma_{(0),m} = \tau_{(0),m}^{-1}$ from \eqref{Eq:LongitudinalModeIDstau0}. With the identification \eqref{Eq:Hydrozerokconstraint} our effective theory reproduces the current-current correlator of \eqref{Eq:TotalChargeCorrelator} at $\vec{k} = \vec{0}$ i.e.
   \begin{subequations}
    \begin{eqnarray}\label{eq:currentCorrelatorZerok2}
        \langle J^{i} J^{j} \rangle_{\mathrm{R}}(\omega,\vec{0})
        &=& - i \omega \left[ \sum_{n=0}^{N_{D}} c_{n} (i \omega)^n + \sum_{n=1}^{N-1} \frac{i R_{(0),n}}{ \omega + i\Gamma_{(0),n}} \right] \delta^{ij} + \mathcal{O}(\omega^{N_{D}+1})  \; , \qquad \\
        c_{n} &=& \frac{1}{n!} \sigma_{\mathrm{AC}}^{(n)}(0) - \sum_{m=1}^{N-1} \frac{R_{(0),m}}{(\Gamma_{(0),m})^{n+1}}  \; , 
    \end{eqnarray}
    \end{subequations}
where $\sigma_{\mathrm{AC}}^{(n)}(0)$ is the $n^{\mathrm{th}}$ derivative of the AC conductivity with respect to frequency and we have introduced a frequency cut-off $N_{D}$ representing the order in time derivatives to which we wish to work.}

{\ We can fix the function $r_{(0)}(- i \omega)$ defined by \eqref{Eq:Hydrozerokconstraint} using the AC conductivity, the gapped pole positions and the gapped pole residues $R_{(0),n}$.  Generally, one might be worried about whether the term in brackets on the right hand side of \eqref{Eq:Hydrozerokconstraint} is at least order one in small frequency so that $r_{(0)}(- i \omega)$ is not divergent as we take $\omega \rightarrow 0$. This is where the Einstein relation appears as 
	\begin{eqnarray}
		r_{(0)}(- i\omega)
        &=& \frac{i}{\omega} \left[ \left(\sigma_{\mathrm{DC}} \left( \Pi_{n=1}^{N-1} \Gamma_{(0),n} \right) - \bar{\sigma}_{(0)}\right) + \mathcal{O}(\omega) \right] \nonumber \\
        &\overset{\eqref{Eq:IdentificationofD}}{=}& \frac{i}{\omega} \left[ \left( \Pi_{n=1}^{N-1} \Gamma_{(0),n} \right) \left(\sigma_{\mathrm{DC}}   -\mathfrak{D} \chi_{\rho \rho} \right) + \mathcal{O}(\omega) \right] \nonumber \\
        &\overset{\eqref{Eq:EinsteinRelation}}{=}& \mathcal{O}(\omega^{0}) \; . 
	\end{eqnarray}
The Einstein relation is thus important for the matching, but it was already a natural consequence of the constraints we imposed on our desired conductivities \eqref{Eq:TotalChargeCorrelator}.}

{\ Having matched the correlator at non-zero frequency (but zero wave-vector), we return now to the choice of a suitable basis for the source term \eqref{Eq:SourceTerm} in our effective non-conservation equation \eqref{Eq:Nonconserv} discussed in the paragraph containing \eqref{Eq:ZerothOrderOp}. It is clear that to have isolated order one poles, which is the typical situation we will encounter, we must at least have the right hand side of \eqref{Eq:Nonconserv} up to order $D^{N-1}$ in frequency. In particular, the source term must be expressible as a polynomial in frequency of the form \eqref{Eq:Hydrozerokconstraint}. Again, the lowest order equation of motion in our derivative expansion contains the operator $D^{N-1}$ in the source (coupling to the background field) term and in the source-free equation. Subsequently, we order by order correct in frequency $\sigma_{\mathrm{AC}}(i \omega)$, truncating to whatever order is sufficient for purposes of accuracy. These order by order corrections in frequency in our effective approach only affect the holomorphic part of the Mittag-Leffler representation; unlike corrections in $\vec{k}$ which will also affect pole positions and residues.}

\subsubsection{Complete Green's functions}

{\noindent Finally, let us sketch the Green's functions for a general $(\omega,\vec{k})$-dependent fluctuation. The system splits into transverse and longitudinal sectors with respect to the wavevector. The latter is easier to write and we find that the transverse fluctuation of the non-hydrostatic current has the form
    \begin{eqnarray}
            \delta \bar{J}^{y}
        &=& i \omega \frac{\left( \bar{\sigma}_{(0)} - k^2 \bar{\sigma}_{(\mathrm{T})}(-k^2) \right) - i \omega \left( r_{(0)}(- i \omega) - k^2 r_{(\mathrm{T})}(- i \omega,-k^2) \right)}{\Pi_{n=1}^{N-1} \left( - i \omega + ( \Gamma_{(0),n} - k^2 \Gamma_{(\mathrm{T}),n}(-k^2) ) \right)} \delta a_{y} \; . \qquad
    \end{eqnarray}
Using \eqref{Eq:ChargeCurrentConstitutive}, the fluctuation of the non-hydrostatic part of the spatial current is related to the fluctuation of the complete transverse current by
    \begin{eqnarray}
        \delta J^{y} &=& \delta \bar{J}^{y} - i \omega \delta p^{y} + i k \delta m 
    \end{eqnarray}
and thus
    \begin{eqnarray}
            \sigma_{(\mathrm{T})}(\omega,\vec{k})
        &=& \frac{i}{\omega} \left[ \langle J^{y} J^{y} \rangle_{\mathrm{R}}(\omega,\vec{k}) - k^2 \chi_{\mathrm{BB}}(k^2) \right] \nonumber \\
        &=& \frac{\left( \bar{\sigma}_{(0)} - k^2 \bar{\sigma}_{(\mathrm{T})}(-k^2) \right) - i \omega \left( r_{(0)}(\omega) - k^2 r_{(\mathrm{T})}(\omega,-k^2) \right)}{\Pi_{n=1}^{N-1} \left( - i \omega + ( \Gamma_{(0),n} - k^2 \Gamma_{(\mathrm{T}),n}(-k^2) ) \right)}  \nonumber \\
        &\;& - i \omega  \left( \chi_{\mathrm{EE}}^{(0)}  + k^2 \chi_{\mathrm{EE}}^{(\mathrm{T})}(k^2) \right) \; . \qquad
     \end{eqnarray}
Similarly to the $\vec{k} = \vec{0}$ case given in section \ref{sec:NonzeroomegaGreen}, we must write the right hand side in a suitable basis and thus
    \begin{subequations}
    \label{eq:FinitekNumerator}
    \begin{eqnarray}
        &\;& r_{(0)}(- i \omega) - k^2 r_{(\mathrm{T})}(- i \omega,-k^2)  \nonumber \\
        &=& \frac{i}{\omega} \left[  - \left( \bar{\sigma}_{(0)} - k^2 \bar{\sigma}_{(\mathrm{T})}(-k^2) \right) + \sum_{n=1}^{N-1} \left( R_{(0),n} - k^2 R_{(\mathrm{T}),n}(-k^2) \right) P_{n}(\omega;\vec{k})    \right. \nonumber \\ 
        &\;& \hphantom{\frac{i}{\omega} \left[ \right.} +  \sum_{n=0}^{N_{D}} \frac{(i\omega)^n}{n!}  \left(  \sigma_{\mathrm{AC}}^{(n)}(0)   - n! \left( \sum_{m=1}^{N-1}  \frac{R_{(0),m}}{(\tau_{(0),m}^{-1})^{n+1}} \right) - \vec{k}^2  f_{(\mathrm{T}),n}(\vec{k}^2) \right) Q^{(\mathrm{T})}(\omega;\vec{k})   \nonumber \\
        &\;& \left. \hphantom{\frac{i}{\omega} \left[ \right. +  \; } + i \omega \left( \chi_{\mathrm{EE}}^{(0)}  + k^2 \chi_{\mathrm{EE}}^{(\mathrm{T})}(k^2) \right)  Q^{(\mathrm{T})}(\omega;\vec{k})  \vphantom{ \sum_{n=1}^{N-1} \left( R_{(0),n} - k^2 R_{(\mathrm{T}),n}(-k^2) \right) P_{n}(\omega;\vec{k})} \right]  \; ,
          \end{eqnarray}
where we have defined
   \begin{eqnarray}
        &\;& P_{n}^{(\mathrm{T})}(\omega;\vec{k}) = \Pi_{m=1, m \neq n}^{N-1} \left(- i \omega + \left( \Gamma_{(0),m} - k^2 \Gamma_{(\mathrm{T}),m}(k^2) \right) \right) \; , \\
        &\;& Q^{(\mathrm{T})}(\omega;\vec{k})  = \Pi_{m=1}^{N-1} \left(- i \omega + ( \Gamma_{(0),m} - k^2 \Gamma_{(\mathrm{T}),m}(k^2) ) \right)
    \end{eqnarray}
    \end{subequations}
We note that $\chi_{\mathrm{EE}}^{(\mathrm{T})}(k^2)$ appears in this expression, allowing us in principle to determine its value.}

{\ The only part of our effective theory relevant to the expression \eqref{eq:FinitekNumerator} which has yet to be determined is the value of $r_{(\mathrm{T})}(-i \omega,-k^2)$. We proceed in the following manner: we construct a set of equations up to order $N_{D}$ by differentiating \eqref{eq:FinitekNumerator} with respect to $\omega$ and setting $\omega = 0$. For this to be sensible, \eqref{eq:FinitekNumerator} must have a smooth limit as $\omega \rightarrow 0$ for all values of $k^2$. To see this is the case, we remind ourselves that unlike the longitudinal correlator\footnote{In the longitudinal case $\bar{\sigma}_{(\mathrm{L})}(-k^2)$ is fixed by the modes - see section \ref{sec:modes}.}, $\bar{\sigma}_{(\mathrm{T})}(-k^2)$ has not yet been fixed. We find that if we fix
	\begin{eqnarray}
		\label{Eq:BarSigmaTrel}
		\bar{\sigma}_{(\mathrm{T})}(-k^2)
		&=& \frac{1}{k^2} \left[ - \bar{\sigma}_{(0)} + \sum_{n=1}^{N-1} \left( R_{(0),n} - k^2 R_{(\mathrm{T}),n}(-k^2) \right) P_{n}(0;\vec{k}) \right. \nonumber \\
		&\;& \left. \hphantom{ \frac{1}{k^2} \left[  \right.} +\left(  \sigma_{\mathrm{DC}}  - \left( \sum_{m=1}^{N-1}  \frac{R_{(0),m}}{\Gamma_{(0),m}} \right) - \vec{k}^2  f_{(\mathrm{T}),n}(\vec{k}^2) \right) Q^{(\mathrm{T})}(0;\vec{k})   \right] \; , \qquad
	\end{eqnarray}
then the limit is smooth. One can further check that this expression \eqref{Eq:BarSigmaTrel} is smooth as $k^2 \rightarrow 0$ as a consequence of the Einstein relation \eqref{Eq:EinsteinRelation}.}

{\ With this said, the coupled equations obtained by differentiating \eqref{eq:FinitekNumerator} with respect to frequency are then expanded in orders of $k^2$, with the $k^2=0$ case being given by \eqref{Eq:Hydrozerokconstraint}. At each order in frequency and wavevector we have an unfixed derivative of $r_{(\mathrm{T})}(-i \omega,-k^2)$ appearing linearly in the equation. We can solve the resultant linear equations for these derivatives without concern for the value of $f_{(\mathrm{T}),n}(k^2)$ appearing in \eqref{Eq:CompleteChargeCorrelatorT} as long as this latter quantity is smooth. This identifies the unknown term $r_{(\mathrm{T})}(-i \omega,-k^2)$ in our effective theory  to the desired derivative orders, while demonstrating that we can match the transverse conductivities of \eqref{Eq:CompleteChargeCorrelatorT} to arbitrary accuracy.}

{\ A similar procedure occurs for the longitudinal correlator, with the important caveat that the modes are not straightforwardly given in terms of the quantities appearing in our effective theory (see section \ref{sec:modes}). In particular, we have to solve the following matrix equation
    \begin{eqnarray}
        &\;& \left( \begin{array}{ccc}
                     - i \omega \chi_{\rho \rho} & \; & i k \\ i k \left( \bar{\sigma}_{(0)} + k^2 \bar{\sigma}_{(\mathrm{L})}(k^2) \right) & \; & \prod_{n=1}^{N-1} \left( - i \omega + ( \Gamma_{(0),n} + k^2 \Gamma_{(\mathrm{L}),n} ) \right)
                    \end{array} \right) 
        \left( \begin{array}{c} 
                    \delta \mu \\
                    \delta \bar{J}^{x}
                \end{array} \right) \nonumber \\
        &=& i \left( \begin{array}{c} 
                        0 \\
                        \left( \bar{\sigma}_{(0)} + k^2 \bar{\sigma}_{(\mathrm{L})}(k^2) \right) - i \omega ( r_{(0)}(\omega) + k^2 r_{(\mathrm{L})}(\omega, k^2) ) \\
                    \end{array} \right) \left( k \delta a_{t} + \omega \delta a_{x} \right) \; . \qquad
    \end{eqnarray}
Inverting, we find
    \begin{align}
            \left( \begin{array}{c} 
                    \delta \mu \\
                    \delta \bar{J}^{x}
                \end{array} \right)
        &= - \frac{i \left( k \delta a_{t} + \omega \delta a_{x} \right)}{\Xi} \times \nonumber \\
        &\hphantom{= \;} \left( \begin{array}{c} 
                        i k \left( \bar{\sigma}_{(0)} + k^2 \bar{\sigma}_{(\mathrm{L})}(k^2) \right) - k \omega ( r_{(0)}(\omega) + k^2 r_{(\mathrm{L})}(\omega, k^2) ) \\
                        i \omega \chi_{\rho \rho} \left( \bar{\sigma}_{(0)} + k^2 \bar{\sigma}_{(\mathrm{L})}(k^2) \right) + \chi_{\rho \rho} \omega^2 ( r_{(0)}(\omega) + k^2 r_{(\mathrm{L})}(\omega, k^2) ) \\
                    \end{array} \right) \; , \nonumber \\
            \Xi
        &= - i  \chi_{\rho \rho} \omega \prod_{n=1}^{N-1} \left( - i \omega + ( \Gamma_{(0),n} + k^2 \Gamma_{(\mathrm{L}),n} ) \right) + k^2 (\bar{\sigma}_{(0)} + k^2 \bar{\sigma}_{(\mathrm{L})}(k^2 ) )
    \end{align}
and consequently, the longitudinal conductivity takes the form
    \begin{eqnarray}
    		\label{Eq:GenericLongConduct}
        		\sigma_{(\mathrm{L})}(\omega,\vec{k})
        &=& \frac{(- i  \omega \chi_{\rho \rho}) \left( \bar{\sigma}_{(0)} + k^2 \bar{\sigma}_{(\mathrm{L})}(k^2) \right) - \chi_{\rho \rho} \omega^2 ( r_{(0)}(\omega) + k^2 r_{(\mathrm{L})}(\omega, k^2) )}{\Xi} \nonumber \\
        &\;& - i \omega \left( \chi_{\mathrm{EE}}^{(0)} + k^2 \chi_{\mathrm{EE}}^{(\mathrm{L})}(k^2) \right) \; , \qquad
    \end{eqnarray}
where the final term accounts for polarisation effects. It is straightforward to see by direct computation that
    \begin{eqnarray}
        \langle J^{x} J^{x} \rangle_{\mathrm{R}}(\omega,\vec{k}) 
        &=& \frac{k}{\omega} \langle J^{x} J^{t} \rangle_{\mathrm{R}}(\omega,\vec{k}) \; , 
    \end{eqnarray}
as required by the $U(1)$ Ward identity. By varying $\delta \mu$ instead and using the constitutive relation for the current, we soon find
    \begin{eqnarray}
                \langle J^{t} J^{t} \rangle_{\mathrm{R}}(\omega,\vec{k}) 
        &=& \frac{k}{\omega} \langle J^{t} J^{x} \rangle_{\mathrm{R}}(\omega,\vec{k}) = \frac{k}{\omega} \langle J^{x} J^{t} \rangle_{\mathrm{R}}(\omega,\vec{k}) \; . 
    \end{eqnarray}
The result is thus  time reversal covariant. Matching the coefficients between the generic conductivities \eqref{Eq:TotalChargeCorrelator}  and our expression above \eqref{Eq:GenericLongConduct} follows much the same procedure as for the transverse case with the notable caveat that the expression for the denominator is significantly more complicated. Moreover, rather than $\bar{\sigma}_{(\mathrm{T})}(-k^2)$ being constrained so that \eqref{eq:FinitekNumerator} is finite in the $\omega \rightarrow 0$ limit, $f_{(\mathrm{L}),0}(k^2)$ and $f_{(\mathrm{L}),1}(k^2)$ in \eqref{Eq:CompleteChargeCorrelatorT} are fixed so that the result is smooth (see appendix \ref{appendix:refining} for further details).}

\section{Applied to the D3-D5 probe brane; the emergence of quasihydrodynamics as a ``truncation error'' } \label{sec:D3D5}

\begin{table}[t!]
  \centering
  \begin{tabular}{|c||cccccccccc|} \hline
			 & $t$ & $x$ & $y$ & $z$ & $X^1$ & $X^2$ & $X^3$ & $X^4$ & $X^5$ & $X^6$\\ \hline \hline
    $N_c$ \,\,\, D$3$ & $\times$ & $\times$ & $\times$ & $\times$ & & &  &  & & \\
    $N_f$ \,\,\, D$5$ & $\times$ & $\times$ & $\times$ & & $\times$ & $\times$ & $\times$ & & & \\ \hline
  \end{tabular}
  \caption{The embeddings of the D3 and D5 branes in ten dimensional Minkowski space.}
  \label{tab:branembedding}
\end{table}

{\noindent We will now use particular fluctuations of the D3/D5 probe brane system, consisting of $N_{c}$ D3 branes intersecting with $N_{f}$ D5 branes in the probe limit ($N_{c} \gg N_{f}$), as a testbed for our framework. The schematic embedding of the branes in the full ten-dimensional Minkowski space is displayed in table \ref{tab:branembedding}. Back-reacting the D3-branes on the geometry leads to an AdS$_{5}\times S^{5}$ spacetime with a black hole. On the corresponding dual field theory side, the D3 branes give rise to $\mathcal{N}=4$ SYM, while the D5-branes introduce $\mathcal{N}=2$ flavour degrees of freedom, localized on a (2+1) dimensional intersection. The resultant strongly coupled dual theory is a $(2+1)$-dimensional defect CFT which we can tune to have finite charge density by placing suitable conditions on probe brane gauge fields living on the intersection. It is well known \cite{Karch:2007pd,Karch:2008fa,Brattan:2012nb,Brattan:2013wya,Brattan:2014moa,Chen:2017dsy} that these systems have finite DC conductivities at non-zero charge density without any coupling to momentum degrees of freedom and thus provide suitable testbeds for our formulation. 

\subsection{Charge fluctuations of the D3-D5 model }\label{section:D3D5fluctuation}


{\noindent In the following we review the D3-D5 probe brane system at finite charge density and temperature, and derive the equations of motion of linearised charge fluctuations. We will work with ingoing Eddington Finkelstein coordinates adapted to asymptotically AdS spacetimes, as this is convenient for developing the holographic approximant. Consequently, the geometry generated by back-reacting D3-branes has the metric
    \begin{subequations}
    \label{Eq:spacetime}
    \begin{eqnarray}
        ds^2 & = & \frac{L^2}{r^2} \left[- f(r) dv^2 - 2 dv dr + dx^2 + dy^2 + dz^2 \right] + L^2 ds^2_{S^5} \; , \\
        f(r) & = & 1-\frac{r^4}{r_{H}^4} \; , 
    \end{eqnarray}
    \end{subequations}
where $r_H=(\pi T)^{-1}$, with $T$ being the Hawking temperature and associated to the thermal bath, and $L$ the AdS radius which we set to be $L = 1$. The radial coordinate $r$ goes from the black hole horizon at $r = r_H = 1$ to the conformal boundary at $r=0$. We will only consider massless black hole embeddings of the D5-brane in this geometry, thus maintaining chiral symmetry of the flavour degrees of freedom at some non-zero $U(1)_B$ baryon number charge density \cite{Karch:2002sh,Brattan:2012nb,Brattan:2013wya,Brattan:2014moa,Jokela_2015}.}

{\ The probe limit is defined such that the $N_f$ D5-branes do not backreact on the blackhole geometry. Their embedding is described by the Dirac-Born-Infeld (DBI) action
  \begin{eqnarray}
   \label{Eq:D3D5probe}
   \mathcal{S}_{\mathrm{D5}} &=& - N_f T_{D5} \int d^{6} \xi \sqrt{-\det\left(g_{ab} + F_{ab} \right)} \;
  \end{eqnarray}
where $\xi$ are the embedding coordinates of the D5-brane in the full ten-dimensional spacetime \eqref{Eq:spacetime}, $T_{D5}$ is the tension of the D5 brane, $g_{ab}$ is the induced worldvolume metric and $F_{ab}$ the $U(1)$ world-volume field strength. In writing the above action we absorbed a factor $2 \pi \alpha'$, where $\alpha'$ is the string tension, into the field strength to make it is dimensionless.}

{\ To generate a non-zero charge density we turn on a component of the gauge field $A_v(r)$ and demand near the AdS boundary that it tends to a non-zero constant while being regular in the interior. Working in radial gauge $A_r = 0$, this is equivalent to turning on a non-zero worldvolume electric field given by $F_{rv}(r) = A'_v(r)$, where the prime denotes the derivative taken with respect to the radial coordinate, i.e. $\partial_r$. The D5-brane action then becomes 
    \begin{eqnarray} \label{eq:action}
        \mathcal{S}_{\mathrm{D5}}^{(0)} = - \mathcal{N}_5 V_{\mathbb{R}^{(2,1)}} \int_{r=0}^{r=1} dr \frac{\sqrt{1-r^4 A_v^{'2}}}{r^4}
    \end{eqnarray}
with $\mathcal{N}_5 \equiv N_f T_{D_5} V_{S^2}$, $V_{S^2}$ being the volume of the 2-dimensional unit sphere $S^2$ and $V_{\mathbb{R}^{(2,1)}}$ being the volume of $\mathbb{R}^{(2,1)}$. Note that the $D5$-brane does not occupy the $z$-direction, as shown in table \ref{tab:branembedding}, which is why we have $V_{\mathbb{R}^{(2,1)}}$ and not $V_{\mathbb{R}^{(3,1)}}$. For notational purposes we will divide \eqref{eq:action} by $V_{\mathbb{R}^{(2,1)}}$ and from now onwards will refer to the resulting action as $\mathcal{S}_{\mathrm{D5}}$, i.e. $\mathcal{S}_{\mathrm{D5}}/V_{\mathbb{R}^{(2,1)}} \rightarrow \mathcal{S}_{\mathrm{D5}}$.

{\ The charge density $\rho$ in a holographic theory is given by the variation of the on-shell action with respect to the asymptotic value of the gauge field (which is the chemical potential $\mu$ up to normalisations). For the probe brane gauge field this is given by the radially conserved current $\delta \mathcal{S}_{\mathrm{D5}}^{(0)}/\delta A_v'$, as the DBI action only depends on the $r$-derivative of the gauge field,
\begin{eqnarray}
    \langle J^t  \rangle = (2 \pi \alpha') \frac{\delta \mathcal{S}_{\mathrm{D5}}^{(0)}}{\delta A_v'} = \mathcal{N}_5 \frac{(2 \pi \alpha') A_v'}{\sqrt{1-r^4 A_v^{'2}}}.
\end{eqnarray}
Solving this background equation for the charge density gives 
\begin{eqnarray}
	\label{Eq:BackgroundGauge}
    A_v'(r) = \frac{\rho}{\sqrt{1+r^4 \rho^2}},
\end{eqnarray}
up to an irrelevant choice of sign where we have defined
\begin{eqnarray}
    \rho \equiv \frac{\langle J^t \rangle}{(2 \pi \alpha') \mathcal{N}_5} \; .
\end{eqnarray}
}

{\ We have arranged a background with non-zero charge corresponding to a non-zero gauge field \eqref{Eq:BackgroundGauge}. However, to compute the two-point functions of charge currents necessary for finding the charge conductivity, we need to solve the linearised equations of motion for fluctuations of the probe brane gauge field. In particular, we take the total gauge field to be of the form
\begin{eqnarray}
    A_{\nu}(r,x^\mu) = A_v(r) \delta^v_\nu + \delta A_\nu(r,x^\mu) \; ,
\end{eqnarray}
where $\delta A_{\mu}(r,x^{\mu})$ is a linearised fluctuation. Noticing the spatial $SO(2)$ rotation invariance of the AdS part of the background metric \eqref{Eq:spacetime}, and in the absence of any symmetry breaking fields such as an external electric field, all choices for the directions of spatial momentum are equivalent. In the following, we align the momentum along the $x$-direction and correspondingly let the fluctuations only depend on $r$, $t$ and $x$. Hence, the Fourier transformations of the fluctuations are
  \begin{eqnarray}
   \delta A_{\mu}(r,x^{\mu}) &=& \int \frac{d\omega dk}{(2 \pi)^2} \delta A_{\mu}(r,\omega,k) \exp \left( - i \omega t + i k x \right) \; ,
  \end{eqnarray}
where signs and factors match the conventions we introduced in \eqref{Eq:FourierConvention}. The equation of motion corresponding to fluctuations of the radial component of the linearised gauge field lead to a gauge constraint
\begin{eqnarray}
    \omega \, \delta A_v' + u(r)^2 k \, \delta A_x' + \frac{u(r)^2}{1-r^4} i k (k \delta A_v + \omega \delta A_x) = 0 \; ,
\end{eqnarray}
where 
\begin{eqnarray}
    u(r)^2 \equiv \frac{1-r^4}{1+r^4 \rho^2} \; . 
\end{eqnarray}
Using spatial rotation invariance as a guide we find that the following two combinations 
  \begin{eqnarray}
    \delta A_y(r,\omega,k) \; , \quad \quad 
    E_x(r,\omega,k) \equiv k \, \delta A_v(r,\omega,k) + \omega \, \delta A_x(r,\omega,k) ,
  \end{eqnarray}
are gauge invariant. Subsequently, the equations of motion for the gauge fluctuations in terms of $E_x$ and $\delta A_{y}$ are then
\begin{subequations}
\label{Eq:FluctuationEquations}
\begin{align}
\label{eq:eomEx}
E_x'' &\; +\; \frac{f(r)\!\left[\omega^2\!\left(f'(r)+2 i \omega\right)+2 \omega\, u(r)^2 \left(\rho^2 r^3 \omega - i k^2\right)- 6 k^2 \rho^2 r^3 u(r)^4 \right]}{i\omega\,u(r)^2\!\left(k^2 f'(r)+\omega\!\left(2\rho^2 r^3 \omega + i k^2\right)\right)+ u(r)^4\!\left(k^4 - 6 i k^2 \rho^2 r^3 \omega\right)} E_x' \\[6pt]
&\;+\; \frac{ i f(r)^2\,(\omega^2 - k^2 u(r)^2)}{u(r)^4\!\left(6 k^2 \rho^2 r^3 \omega + i k^4\right) -\omega\,u(r)^2\!\left(k^2 f'(r)+\omega\!\left(2\rho^2 r^3 \omega + i k^2\right)\right)} E_x =0 ,  \nonumber \\[10pt]
\label{eq:eomAy}
\delta A_y''  &\;+\; \frac{u(r)^2\!\left[\,2 r^3\!\left(-2+\rho ^2\!\left(-3 r^4+i r \omega +1\right)\right)+2 i \omega\right]}{f(r)^2}\, \delta A_y' \; \nonumber \\
& \; -\; \frac{u(r)^2\,(k^2 - 2 i \rho^2 r^3 \omega)}{f(r)^2}\, \delta A_y =0 .
\end{align}
\end{subequations}

} The bulk equations of motion for $E_x$ and $a_y$ decouple at $k = 0 $ and reduce to the same equation, i.e. a spatially homogeneous perturbation respecting $SO(2)$ spatial rotation invariance. In what follows, we will normalise our variables by temperature and denote them by $\tilde{k} =k/(\pi T)$,  $\tilde{\omega} = \omega/(\pi T)$, $\tilde{\rho} = \rho/(\pi T)^2 $ etc.}

\subsection{The holographic approximant } \label{section:holographicapproximant}

{\noindent With our fluctuation equations to hand \eqref{Eq:FluctuationEquations}, we can now discuss the procedure for deriving the holographic approximant. We will use this quantity to compute the three constraining quantities: $\sigma_{\mathrm{AC}}(i \omega)$, $\chi_{\rho \rho}$, $\chi_{\mathrm{EE}}(\vec{k}^2)$ and $\chi_{\mathrm{BB}}(\vec{k}^2)$ used to derive \eqref{Eq:TotalChargeCorrelator} in series expansions in small values of their parameters. In particular we will discuss computing $\sigma_{\mathrm{AC}}(i\omega)$ below, although the extension to the other quantities is clear.}

{\ To obtain the AC conductivity, $\sigma_{\mathrm{AC}}(i \omega)$, it will be sufficient to consider only \eqref{eq:eomAy}, for which we need to specify two boundary conditions; one at the horizon at $r = r_\mathrm{H} \equiv 1$ and one at the conformal boundary at $r=0$. The near-horizon behaviour of $\delta A_y(r)$ is given then by the series expansion in $1-r$ of the form
\begin{equation}\label{Eq:a2horizon}
    \delta A_y(r,\omega) = 
    \tilde{a}_{y}^{0}(\omega) + \tilde{a}_{y}^{1}(\omega) (1-r) + \mathcal{O}\left((1-r)^2\right).
\end{equation}
At the opposite extreme, the series expansion of $\delta A_{y}$ at the boundary $r=0$ can be written
	\begin{eqnarray}
		\label{Eq:a2boundary}
		\delta A_{y}(r,\omega) &=& a_{y}(\omega) + \left( b_{y}(\omega) - i \omega a_{y}(\omega) \right) r + \mathcal{O}(r^2) \; 
	\end{eqnarray}
where the coefficient $a_y$ corresponds to the source while $b_y$ corresponds to the vev. Note that the $\mathcal{O}(r)$ term in \eqref{Eq:a2boundary} gets shifted by the source due to working in Eddington-Finkelstein style coordinates. This is in contrast to the usual Poincar\'{e}-like coordinates where we would identify the coefficient of the $\mathcal{O}(r)$ term to be the vev.}

\begin{figure}[t!]
    \centering
    \includegraphics[width=0.5\linewidth]{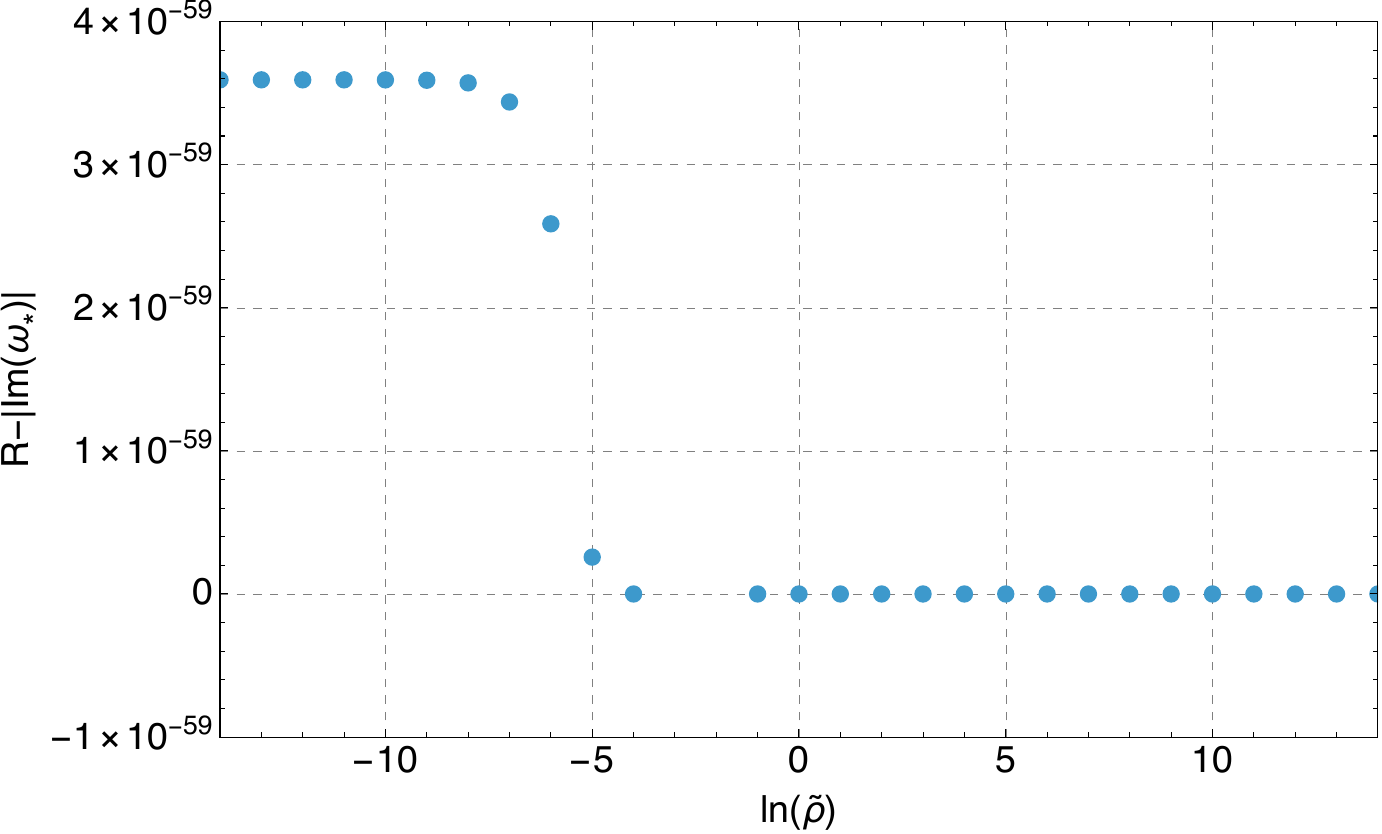}
    \caption{Difference of the position of the gapped pole obtained from the  holographic approximant and the absolute value of the imaginary non-hydrodynamic quasi-normal mode lying closest to the origin, obtained from a shooting method. The difference between the shooting method and the holographic approximant is tiny. We find excellent agreement on the order of $\sim 10^{-56}$, where the two outliers that are not captured in the plot at ln$(\tilde{\rho}) = 2,3$ are of order $\sim 10^{-58}$ and $\sim 10^{-56}$, respectively, and thus lie outside the plotting range.
    }
    \label{fig:comparison}
\end{figure}

{\ To motivate the holographic approximant, let us consider the following: suppose that we wish to extract the residue of the Green's function given by
	\begin{eqnarray}\label{Eq:GreensFct}
		G^{R}_{yy}(\omega) &=& \frac{b_{y}(\omega)}{a_{y}(\omega)} \; ,
	\end{eqnarray}
corresponding to a gapped pole at $\omega=\omega_{\mathrm{gap}}$. From the expression given in \eqref{Eq:GreensFct}, it is not enough to know $b_{y}(\omega)$ at $\omega = \omega_{\mathrm{gap}}$. In particular, at the gapped pole, $a_y(\omega)$ has a simple zero and can generally be written as
\begin{eqnarray}
    a_y(\omega) &=& (\omega - \omega_{\mathrm{gap}}) \alpha(\omega) = (\omega - \omega_{\mathrm{gap}}) \left(\alpha_{0}(\omega_{\mathrm{gap}}) + \alpha_{1}(\omega_{\mathrm{gap}})(\omega - \omega_{\mathrm{gap}}) \right. \nonumber \\
    &\;& \left. \hphantom{} + \mathcal{O}\left(((\omega - \omega_{\mathrm{gap}} \right)^2 \right) \; . 
\end{eqnarray}
Subsequently, the Laurent expansion around a gapped pole is to lowest order
	\begin{eqnarray}
		G^{R}_{yy}(\omega) &=& \frac{b_{y}(\omega_{\mathrm{gap}})}{\alpha_0(\omega_\mathrm{gap})(\omega - \omega_{\mathrm{gap}})} + \mathcal{O}^{0}( \omega - \omega_{\mathrm{gap}}) \; .
	\end{eqnarray}
Extracting the residue then requires us to also know the derivative of $a_{y}(\omega)$ at $\omega=\omega_{\mathrm{gap}}$. Approaches to obtaining this information from \eqref{Eq:a2boundary} in the past either required us to plot several nearby points and interpolate or to perform a numerical integral around the pole \cite{Amoretti_2020}. Both of these approaches suffer from issues: the former requires fitting and can introduce numerical error by the choice of fit, the latter becomes intensive numerically as one computes higher order terms in the Laurent expansion about the pole. It would be better to have direct access to the derivative information. Moreover a precise evaluation of the residue will be necessary for determining the holomorphic parts of our conductivities \eqref{Eq:TotalChargeCorrelator}.}

{\ To obtain this derivative information we need to differentiate $a_{y}(\omega,r)$ with respect to $\omega$ and generate a new auxiliary equation for the behaviour of the derivative. In particular, we promote $\delta A_{y}(r)$ in \eqref{eq:eomAy} to an explicit function of $\omega$, differentiate the whole equation with respect to frequency, and name $\partial_{\omega} \delta A_{y}(r,\omega) \rightarrow \delta A_{y}^{(1)}(r,\omega)$. We can of course iterate this procedure which is equivalent to determining higher and higher terms in the Laurent expansion about $\omega = \omega_{\mathrm{gap}}$. We subsequently define the holographic approximant to the Green's function about any point $\omega = \omega_{*}$ to be given by
	\begin{eqnarray}
		G_{yy,N}(\omega) &=& \frac{\sum_{n=0}^{N} \frac{b_{y}^{(n)}(\omega_{*})}{n!} (\omega - \omega_{*} )^n}{\sum_{n=0}^{N} \frac{a_{y}^{(n)}(\omega_{*})}{n!} (\omega - \omega_{*} )^n} \; . 
	\end{eqnarray}
} 

\begin{figure}[t!]
    \centering
    \begin{subfigure}{0.45\textwidth}
        \centering
        \includegraphics[width=\linewidth]{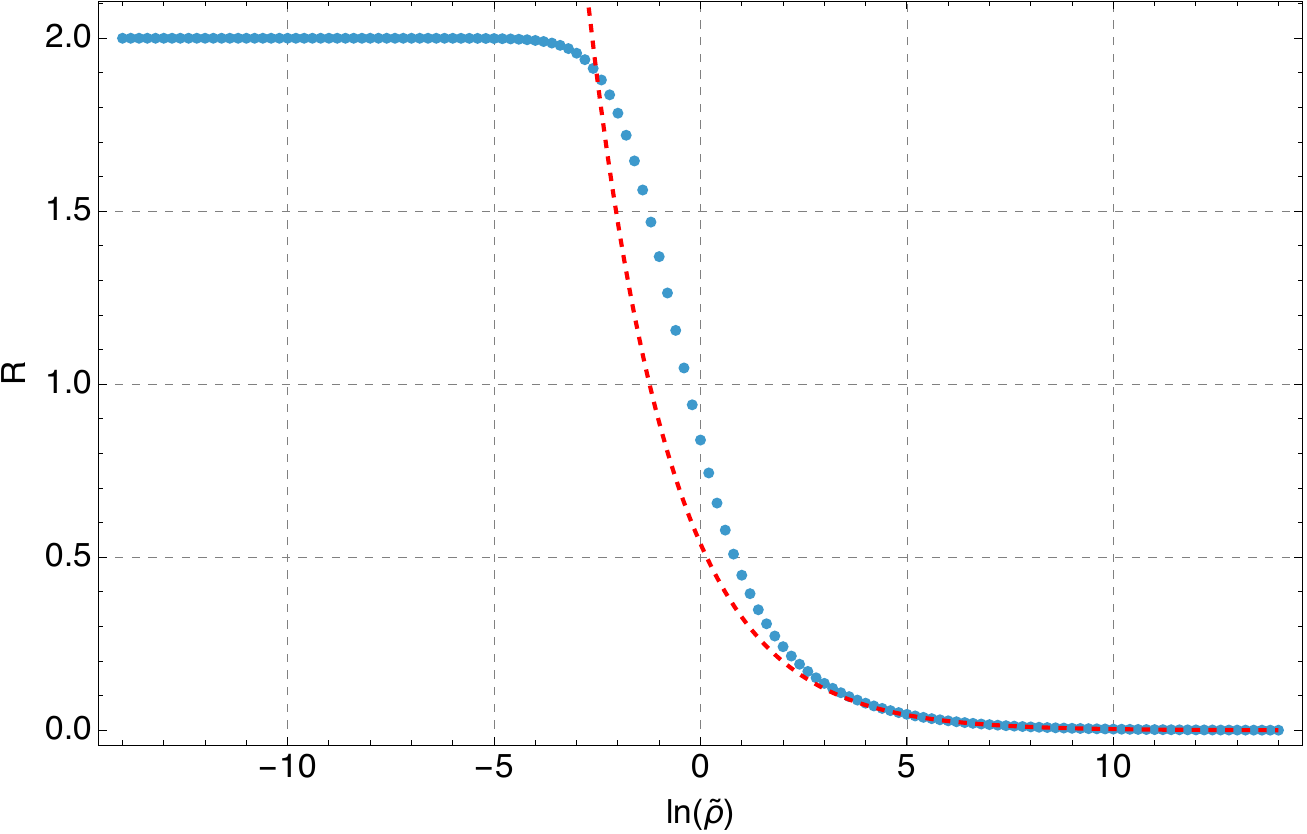}
        \caption{}
    \end{subfigure}
    \hfill
    \begin{subfigure}{0.45\textwidth}
        \centering
        \includegraphics[width=\linewidth]{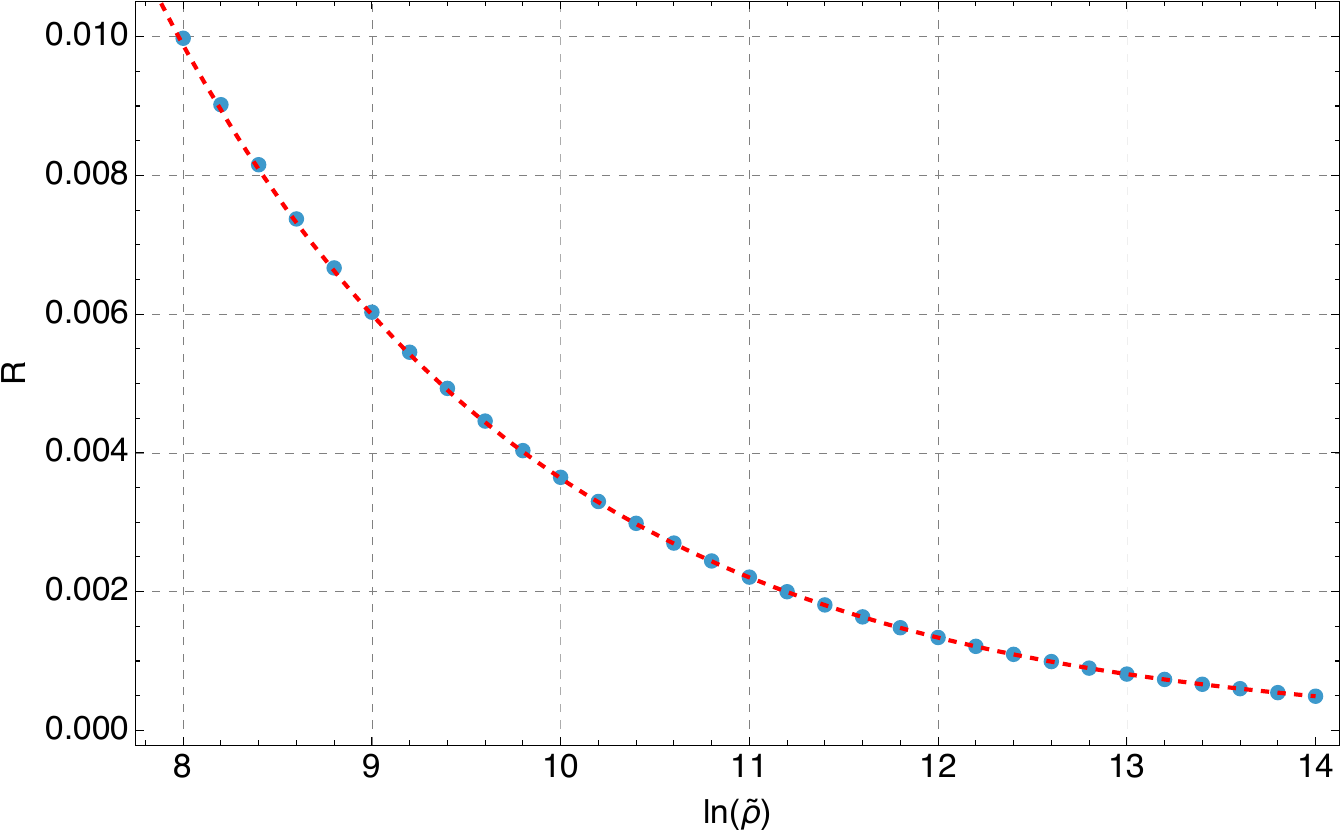}
        \caption{}
    \end{subfigure}
    \caption{Flow of the radius of convergence of the Taylor expansion of the holographic approximant around $\omega=0$ (equivalent to the gapped pole position) plotted against ln$(\tilde{\rho})$. (a) Shows the full range from $e^{-14}$ to $e^{14}$, while (b) shows a close-up for high $\tilde{\rho}$ values from $e^{8}$ to $e^{14}$.}
    \label{fig:flowradius}
\end{figure}

{\ As a benchmark, it is also natural to ask how effectively the approximant built about $\omega=0$ reproduces the quasinormal mode positions. It is well known that many probe brane models have a gapped pole that moves towards the origin at large charge density i.e. becomes long-lived \cite{Chen:2017dsy}. This is the regime of quasihydrodynamics. Given this, we used shooting on a large number of coupled equations (original + auxiliaries $\sim 200$) and compared the position of the first gapped pole given by finding zeroes of the denominator of the holographic approximant to the position of the first quasi-normal mode obtained from a traditional shooting procedure. We found agreement up to the accuracy and precision goals we requested. These are shown in Figure \ref{fig:comparison}.}

{\ Equivalently the radius of convergence of the series expansion of the approximant (a rational function) in small frequency indicates the distance from the origin to the nearest gapped pole. Figure \ref{fig:flowradius} shows the flow of the radius of convergence as a function of $\ln(\tilde{\rho})$, where $\tilde{\rho}$ ranges from $e^{-14}$ to $e^{14}$. The red dashed line denotes the analytic result for the position of this long lived mode, which is known to be $(\Gamma(1/4) \Gamma(1/4)/(4 \sqrt{\pi}))^{-1}\tilde{\rho}^{-1/2}$ for the D3/D5 system \cite{Chen:2017dsy} at leading order in large charge density. As a final check, we have determined and found agreement up to the selected precision and accuracy between the approximant evaluated about $\omega=0$ and the Green's function given by a simple evaluation of \eqref{Eq:GreensFct} over a wide range of points in a disc about $\omega=0$.}

{\ An observation we wish to record here is that for precision purposes it seems to give better results if we compute the holographic approximant away from the pole and re-expand it about a zero of the denominator; although we still get good behaviour in the coefficients about the pole to relatively high order if we evaluate there. The reason for this is natural, small errors in shooting for the position of the pole are magnified if we directly compute the approximant there. We note that the holographic approximant can also be evaluated as a rational function in frequency at fixed $\vec{k} \neq \vec{0}$, but this has proved slow. We believe this is due to using shooting, as opposed to a more efficient integration method. Nevertheless, we shall compute a small $\vec{k}$ expansion of the approximant about the pole in section \ref{sec:omegandk} to obtain the residue at non-zero $\vec{k}$.}

{\ Finally, we note that while the holographic approximant bears some similarities to the Pad\'{e} approximant, it is distinct in a quite important respect - namely the holographic approximant, unlike the Pad\'{e}, is not constructed to match the Taylor expansion of the Green’s function about $\omega_{*}$. Instead it is the ratio of two Taylor expansions - the horizon to boundary propagator for the source and vev respectively. We discuss various ways in which the analysis we use here can be extended and improved upon in the discussion section.}

\subsection{Emergence of quasihydrodynamics in the probe brane at large charge density}

{\noindent In the previous sections we developed an effective theory for an arbitrary number of poles.  For illustrative purpose, figure \ref{fig:qnm_evolution} shows the approximate movement of the quasinormal modes of the D3/D5 brane system with increasing charge density, in the regime of vanishing wavevector but finite frequency. In what follows, we shall specialise our treatment to one gapped pole in addition to the usual diffusive pole.

\begin{figure}[!t]
    \centering
    \begin{subfigure}[t]{0.32\textwidth}
        \centering
        \includegraphics[width=\textwidth]{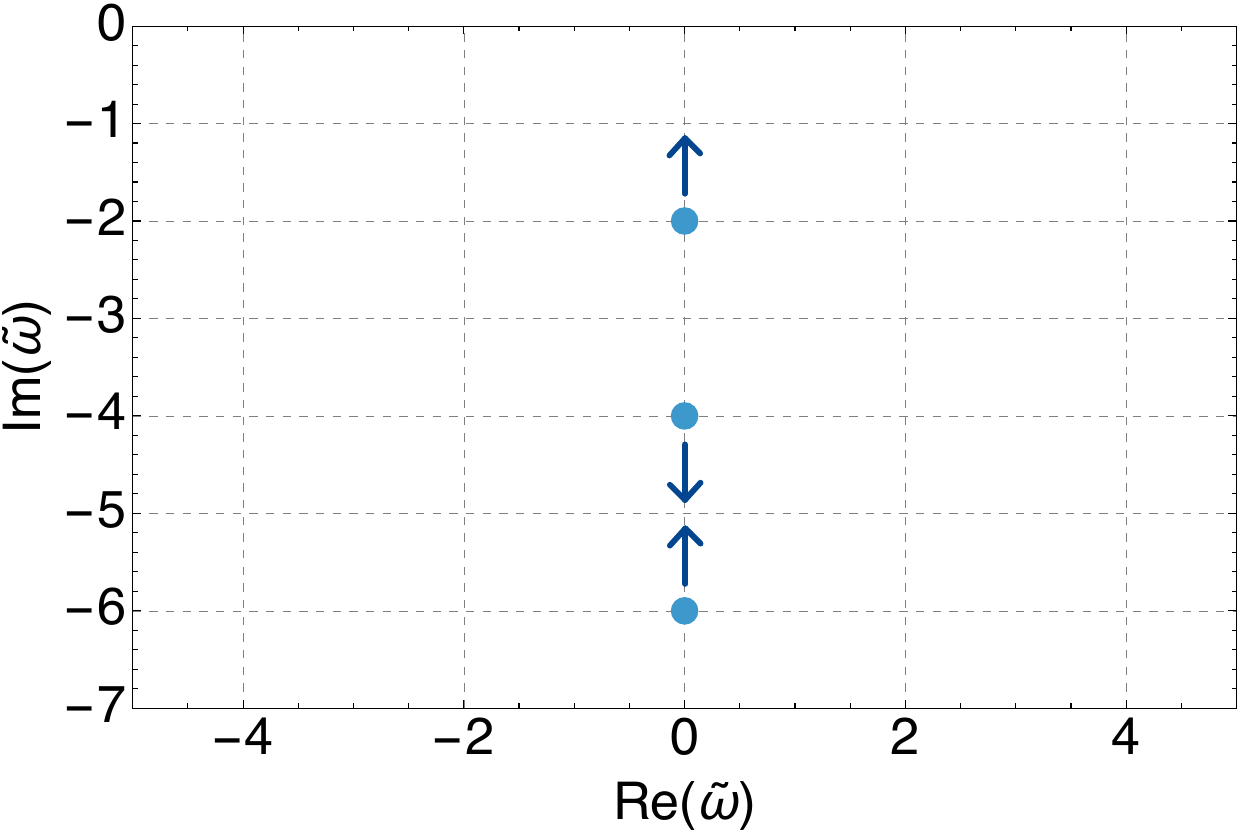}
        \caption{$\tilde{\rho}= e^{- 14}$}
        \label{fig:qnm1}
    \end{subfigure}
    \hfill
    \begin{subfigure}[t]{0.32\textwidth}
        \centering
        \includegraphics[width=\textwidth]{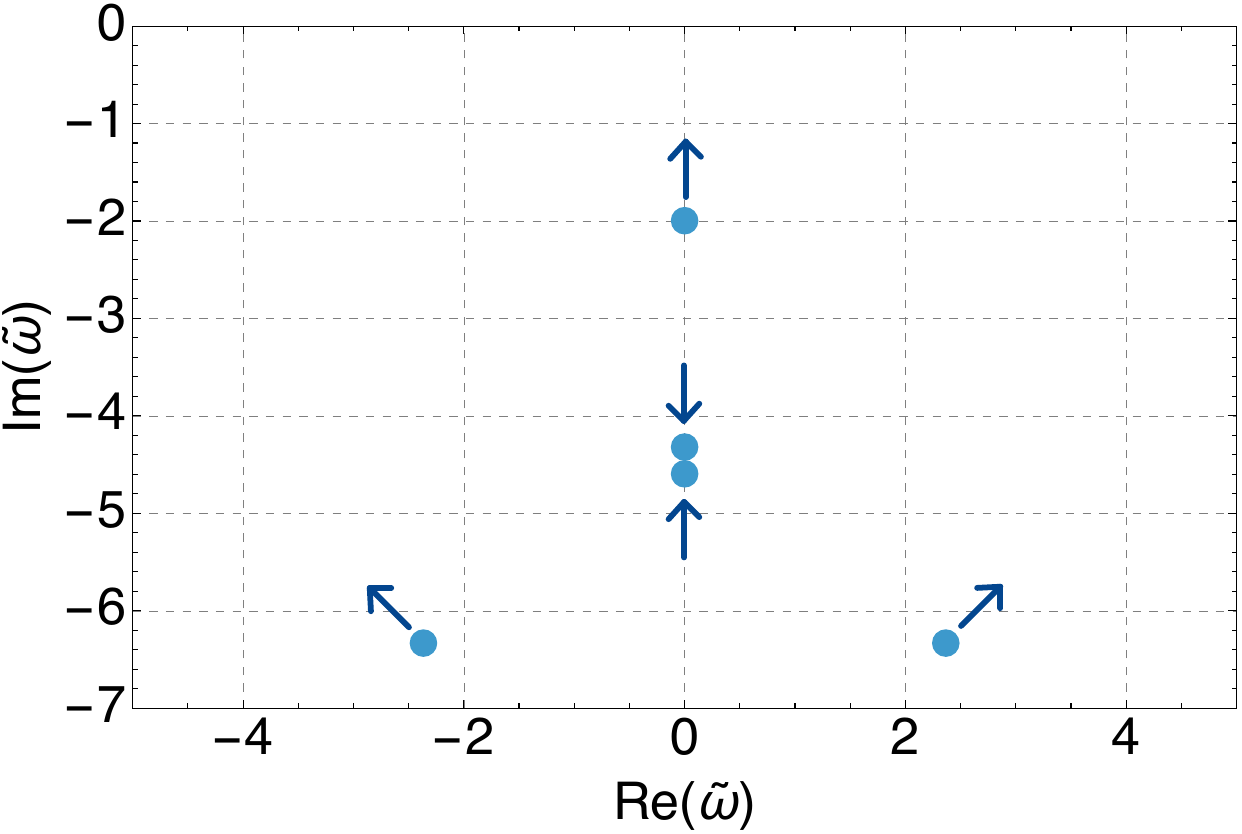}
        \caption{$\tilde{\rho}= e^{- 4.4}$}
        \label{fig:qnm47}
    \end{subfigure}
    \hfill
    \begin{subfigure}[t]{0.32\textwidth}
        \centering
        \includegraphics[width=\textwidth]{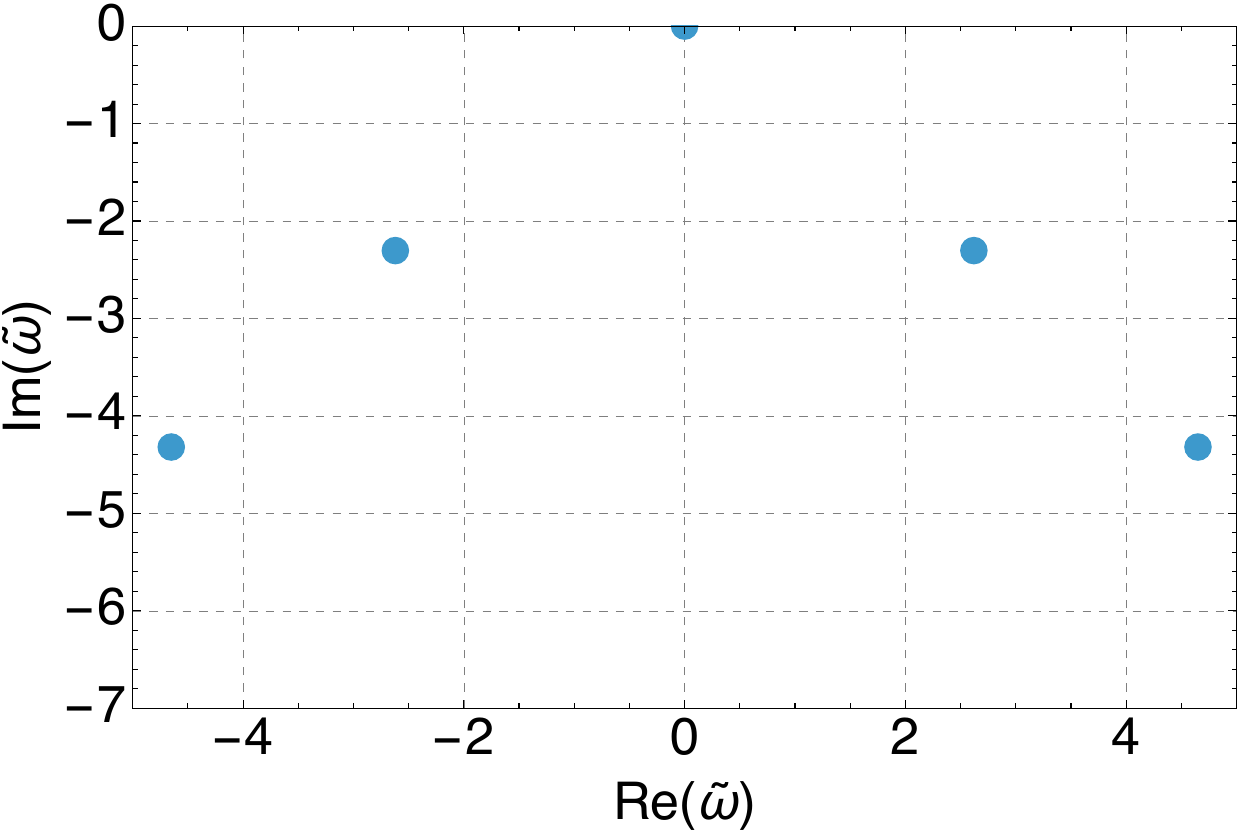}
        \caption{$\tilde{\rho}= e^{14}$}
        \label{fig:qnm135}
    \end{subfigure}

    \caption{Movement of quasi-normal modes with increasing charge density at vanishing wave vector and non-vanishing frequency.}
    \label{fig:qnm_evolution}
\end{figure}

{We recall some known results in the literature \cite{Karch:2002sh,Brattan:2012nb,Brattan:2013wya,Brattan:2014moa,Jokela_2015} - the charge susceptibility $\chi_{\rho \rho}$, DC conductivity $\sigma_{\mathrm{DC}}$ and magnetisation susceptibilities $\chi_{\mathrm{BB}}$ of the D3/D5 probe brane are given by 
	\begin{subequations}
	\label{Eq:KnownResults}
	\begin{eqnarray}
		\label{Eq:AnalyticExpressionChi}
		\chi_{\rho \rho} &=& \frac{1}{2} \left(\, _2F_1\left(\frac{1}{4},\frac{1}{2};\frac{5}{4};-\tilde{\rho} ^2\right)+\frac{1}{\sqrt{\tilde{\rho} ^2+1}}\right)^{-1}\nonumber \\ &\overset{\tilde{\rho} \gg 1}{=}& \sqrt{\tilde{\rho}} \left( \frac{2 \sqrt{\pi }}{\Gamma \left(\frac{1}{4}\right) \Gamma \left(\frac{5}{4}\right)} +  \frac{64 \pi }{5 \Gamma \left(\frac{1}{4}\right)^4 \tilde{\rho}^{5/2}} + \mathcal{O}\left( \frac{1}{\tilde{\rho}^{9/2}} \right) \right) \; , \\
		\label{Eq:MagSusc}
		\chi_{\mathrm{BB}} &=& \, _2F_1\left(\frac{1}{4},\frac{1}{2};\frac{5}{4};-\tilde{\rho}^2 \right) \overset{\tilde{\rho} \gg 1}{=} \tilde{\rho}^{-1/2} \left( \frac{\Gamma \left(\frac{1}{4}\right) \Gamma \left(\frac{5}{4}\right)}{\sqrt{\pi }}-\frac{1}{\sqrt{\tilde{\rho}} }+\mathcal{O} \left(\frac{1}{\tilde{\rho}^2 }\right) \right)\; , \qquad \\
		\label{Eq:DCanalytic}
		\sigma_{\mathrm{DC}} &=&\sqrt{1+\tilde{\rho}^2} \overset{\tilde{\rho} \gg 1}{=} \tilde{\rho} \left( 1 + \frac{1}{2 \tilde{\rho}^2} + \mathcal{O}\left( \tilde{\rho}^{-4} \right) \right) \; .
	\end{eqnarray}
for arbitrary values of $\tilde{\rho}$. From the Einstein relation \eqref{Eq:EinsteinRelation} one then finds the diffusion constant $\mathfrak{D}$ to be given by
	\begin{eqnarray}
				\label{Eq:D3D5diffusion}
		\mathfrak{D} &=& \frac{\sigma_{\mathrm{DC}}}{\chi_{\rho \rho}} =  \frac{\sqrt{1+\tilde{\rho}^2}}{2} \left(\, _2F_1\left(\frac{1}{4},\frac{1}{2};\frac{5}{4};-\tilde{\rho} ^2\right)+\frac{1}{\sqrt{\tilde{\rho} ^2+1}}\right) \nonumber \\ &\overset{\tilde{\rho} \gg 1}{=}& \frac{\sqrt{\tilde{\rho} }}{2 \sqrt{\pi }} \left( \Gamma \left(\frac{1}{4}\right) \Gamma \left(\frac{5}{4}\right) + \frac{\Gamma \left(\frac{1}{4}\right) \Gamma \left(\frac{5}{4}\right)}{2 \tilde{\rho}^2} + \mathcal{O}\left( \frac{1}{\tilde{\rho}^{\frac{5}{2}}} \right) \right) \;  ,
	\end{eqnarray}
	\end{subequations}
which is again an expression which is valid for all values of $\tilde{\rho}$.}

{\ Given our discussion of the effective theory in section \ref{sec:manypoles}, we know that we can reproduce from it conductivities of the form 
 \begin{subequations}
\begin{eqnarray}
\label{Eq:CompleteChargeCorrelatorLD3D5}
\sigma_{(\mathrm{L})}(\omega,\vec{k}^2) 
&=& 
- \frac{i \vec{k}^2 \left( R_{\mathfrak{D}}(0) + R_{\mathfrak{D}}'(0) \vec{k}^2 + \tilde{R}_{(\mathrm{L}),\mathfrak{D}}(\vec{k}^2) \vec{k}^4 \right)}
       {\omega + i \left( \frac{\sigma_{\mathrm{DC}}}{\chi_{\rho \rho}} + \tilde{\mathfrak{D}}(\vec{k}^2)\, \vec{k}^2 \right)} 
+ \frac{i \left( R_{(0)} - \vec{k}^2 R_{(\mathrm{L})}(\vec{k}^2) \right)}
       {\omega + i \left( \tau_{(0)}^{-1} - \vec{k}^2 \tau_{(\mathrm{L})}^{-1}(\vec{k}^2) \right)} 
\nonumber \\
&\;&  
+ \sum_{n=0}^{n=N_{D}} \frac{(i\omega)^n}{n!}  
\left[  \sigma_{\mathrm{AC}}^{(n)}(0) -  \frac{ n! R_{(0)}}{( \tau_{(0)}^{-1} )^n} \right] 
\nonumber \\
&\;&  
- \vec{k}^2 \left( f_{(\mathrm{L}),0}(\vec{k}^2) 
+ i \omega f_{(\mathrm{L}),1}(\vec{k}^2)  
+ \sum_{n=2}^{\infty} \frac{(i\omega)^n}{n!} f_{(\mathrm{L}),n}(\vec{k}^2)  \right)  
\; ,  
\\[4pt]
\label{Eq:CompleteChargeCorrelatorTD3D5}
\sigma_{(\mathrm{T})}(\omega,\vec{k}^2) 
&=& 
\frac{i \left( R_{(0)} - \vec{k}^2 R_{(\mathrm{T})}(\vec{k}^2) \right)}
     {\omega + i \left( \tau_{(0)}^{-1} - \vec{k}^2 \tau_{(\mathrm{T})}^{-1}(\vec{k}^2) \right)}  
\nonumber  \\
&\;& 
+ \sum_{n=0}^{n=N_{D}} \frac{(i\omega)^n}{n!}  
\left[  \sigma_{\mathrm{AC}}^{(n)}(0) 
- n! \frac{R_{(0)}}{( \tau_{(0)}^{-1} )^n} 
- \vec{k}^2 f_{(\mathrm{T}),n}(\vec{k}^2)  
\right] 
\; ,
\end{eqnarray}
\end{subequations}
where we have dropped the indices on our gapped pole terms as there is only one and we have truncated to order $N_{D}$ in frequency derivatives. We will use our formulation to discuss and parametrise the behaviour of the probe brane quasihydrodynamics of the D3/D5 probe brane system at large charge density by computing these conductivities numerically. Where relevant we will compare to the corresponding quantities in the effective theory. In the large charge density regime a gapped pole becomes parametrically close to the real frequency access and, by increasing $\tilde{\rho}$, can be made arbitrarily long-lived.}

\subsubsection{The charge and magnetisation susceptibility at non-zero wavevector }

{\noindent Let us first extract the susceptibilities $\chi_{\rho \rho}$, $\chi_{\mathrm{EE}}(\vec{k})$ and $\chi_{\mathrm{BB}}(\vec{k})$ from the holographic approximant and thus $F_{1}$, $F_{2}[-\vec{k}^2]$ and $F_{4}[-\vec{k}^2]$ of the hydrostatic generating functional \eqref{Eq:GeneratingFunctional} through the relations \eqref{Eq:SusceptibilityMatch}. Figure \ref{fig:susceptibility} shows the first coefficients of the small $\tilde{k}^2$ expansion of the charge susceptibility against the charge density. At large charge density, the coefficients grow as $\sim \tilde{\rho}^{1/2-n}$. Thus, higher orders in $\tilde{k}^2$ are suppressed by the charge density in the limit that quasihydrodynamics emerges. The exception is the thermodynamic charge susceptibility itself, $\chi_{\rho \rho}$, which has the analytic expression \eqref{Eq:AnalyticExpressionChi} for all value of $\tilde{\rho}$ and in fact  grows as $\sqrt{\tilde{\rho}}$. Meanwhile, figure \ref{fig:magnetic susceptibility} shows the first five coefficients of the small $k^2$ expansion of the magnetisation susceptibility. The leading term is suppressed as charge density grows, unlike the former situation. In particular, the coefficients behave $\sim \tilde{\rho}^{-1/2-n}$.}

\begin{figure}[t!]
  \centering

  \begin{subfigure}{0.45\textwidth}
    \centering
    \includegraphics[width=\linewidth]{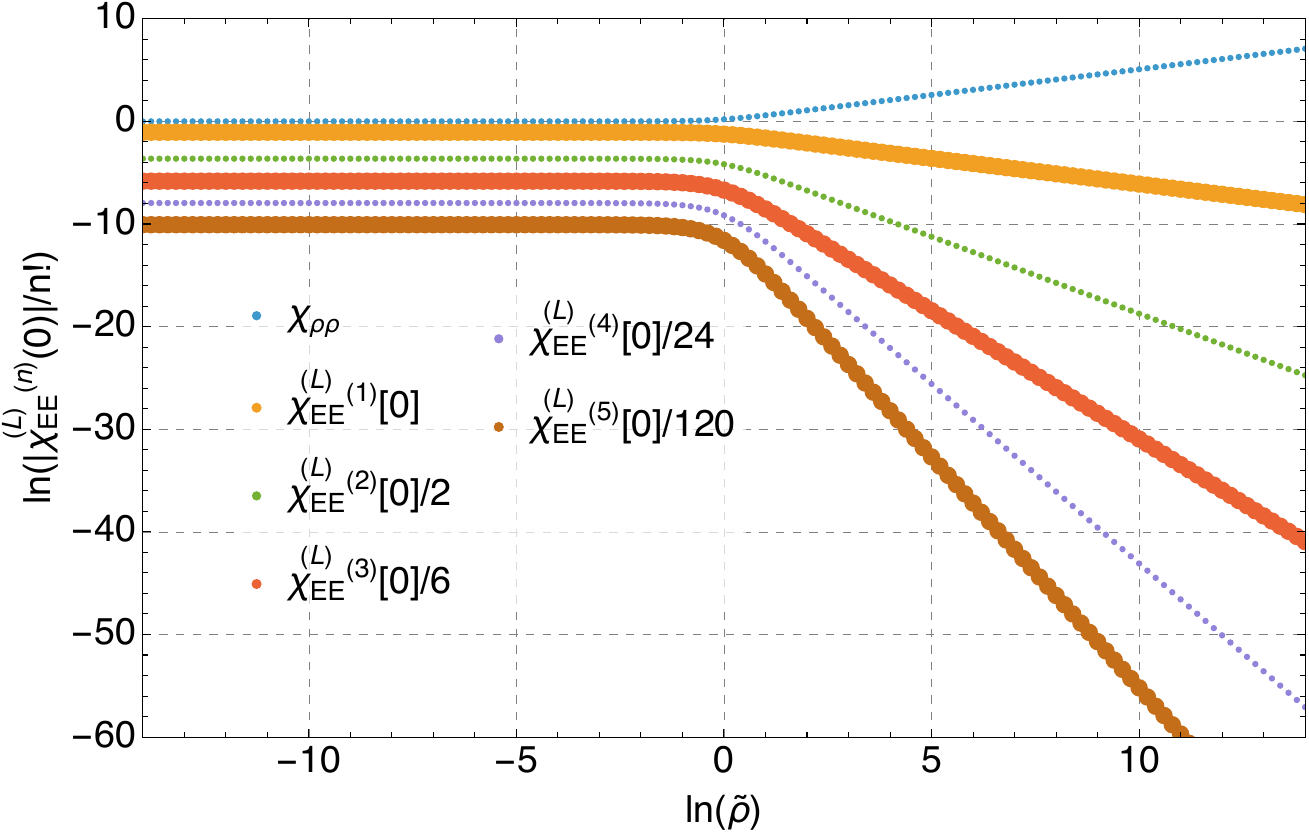}
    \caption{}
    \label{fig:susceptibility}
  \end{subfigure}
  \hfill
  \begin{subfigure}{0.45\textwidth}
    \centering
    \includegraphics[width=\linewidth]{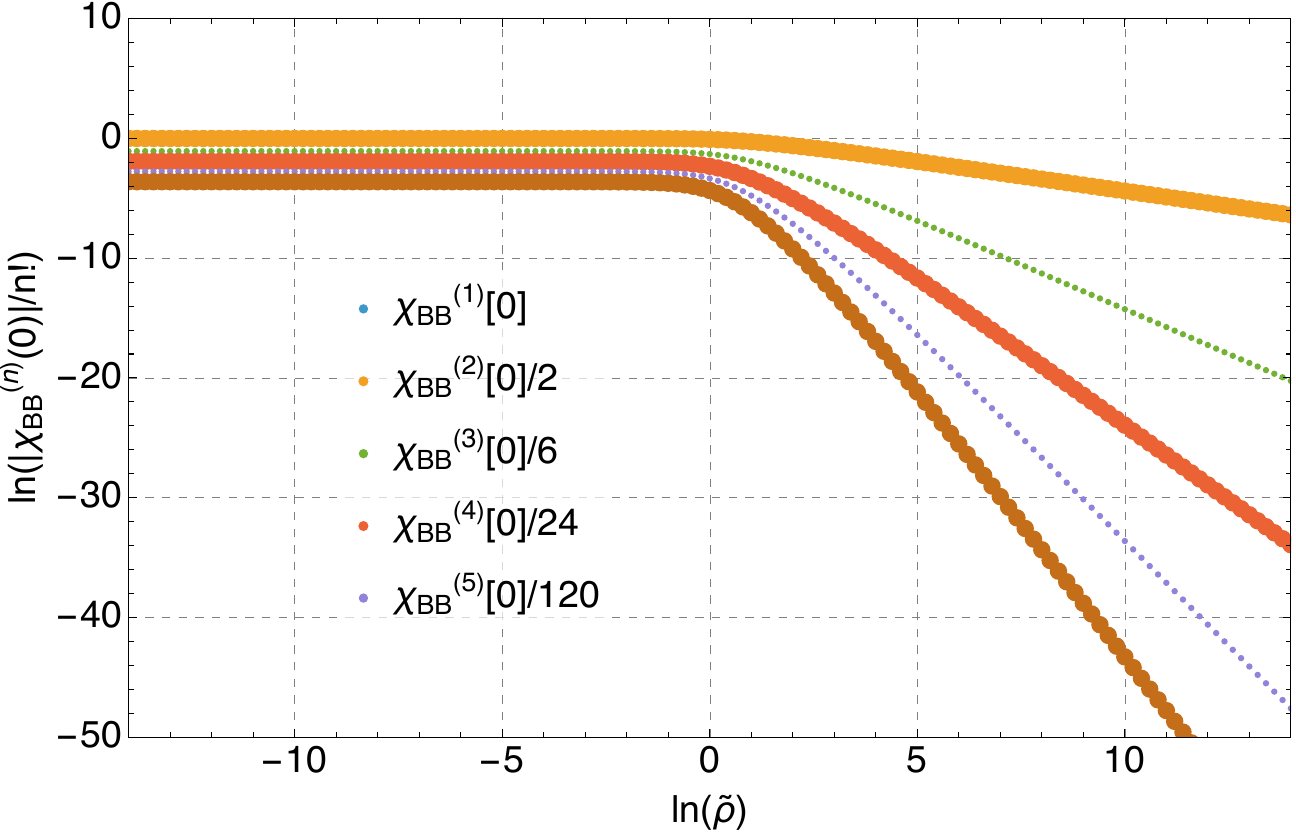}
    \caption{}
    \label{fig:magnetic susceptibility}
  \end{subfigure}
  \caption{First coefficients of the small $\vec{k}$ expansion of (a) the charge susceptibility and (b) the magnetic susceptibility against the charge density. We find that the coefficients of the charge susceptibility grow like $\sim \tilde{\rho}^{1/2-n}$ for large charge density, while the magnetisation susceptibilities grow like $\sim \tilde{\rho}^{-1/2-n}$.
  }
  \label{fig:finitekf0}
\end{figure}

{\ In our effective framework \eqref{Eq:EffectiveEquations}, we have assumed that we can make a series expansion in powers of small $\tilde{k}^2$ and matched the observed behaviour of the full theory. As per usual this is only valid up to the radius of convergence of the series. We can extract such information from the holographic approximants of $\chi_{\mathrm{EE}}^{(\mathrm{L})}(\tilde{k}^2)$ and $\chi_{\mathrm{BB}}(\tilde{k}^2)$. We have confirmed that the radius of convergence for these series corresponds to the distance to the first pole encountered as one moves away from $\tilde{k}^2=0$. In figure \ref{fig:critical} we show the critical radius of the wavevector, in other words the distance to first pole from the origin, against the charge density for a small $\tilde{k}$ expansion of the charge susceptibility $\chi_{\mathrm{EE}}^{(\mathrm{L})}(\tilde{k}^2)$ and the magnetisation susceptibility $\chi_{\mathrm{BB}}(\tilde{k}^2)$. We see that at large $\tilde{\rho}$ both the radii of the susceptibility and the magnetisation grow as $\sqrt{\tilde{\rho}}$ ensuring that the complex $\tilde{k}$ series converges for a much larger range of values as we enter the quasihydrodynamic regime.}

{\ There is another, now classic, convergence radius that has gained a lot of traction in recent years - the radius of convergence for the hydrodynamic mode series \cite{Withers_2018}. In the case of the probe brane, unlike some models, there is no need to complexify the spatial momentum to determine the collision point of the hydrodynamic mode and the first non-hydrodynamic mode. Indeed, the phenomenon of zero sound in probe branes is well known \cite{Karch:2008fa, Chen:2017dsy,Davison:2022vqh}. Moreover, given our holographic approximant formulation, it is easy to find the critical wavevector for the emergence of the zero sound regime by solving the original equation \eqref{eq:eomEx} and its first auxiliary equation (the first frequency derivative) looking for AdS boundary values where both the leading boundary terms are zero. In other words we search for a double zero of the source by scanning $\tilde{\omega}$ and $\tilde{k}$.}

{\ The zero sound critical radius is also displayed in fig. \ref{fig:critical} and we can see that it is the dominant defining radius for the failure of the small $\tilde{k}^2$ expansion to converge, and presumably for pure hydrodynamics to no longer apply. As the charge density increases the regime in which we can expand around $\tilde{k}^2=0$ becomes even smaller as $k_{\mathrm{sound}} \sim 0.382 \tilde{\rho}^{-\frac{1}{2}}$. To the authors knowledge, it is an open question as to whether the first collision of the hydrodynamic mode with another non-hydrodynamic mode always occurs at a smaller $\tilde{k}^2$ than that corresponding to radius of convergence of the series expansions of $\chi_{\mathrm{EE}}^{(\mathrm{L})}(\vec{k}^2)$ and $\chi_{\mathrm{BB}}(\vec{k}^2)$ .}

\subsubsection{AC conductivity }

\begin{figure}[t!]
    \centering
    \includegraphics[width=0.5\linewidth]{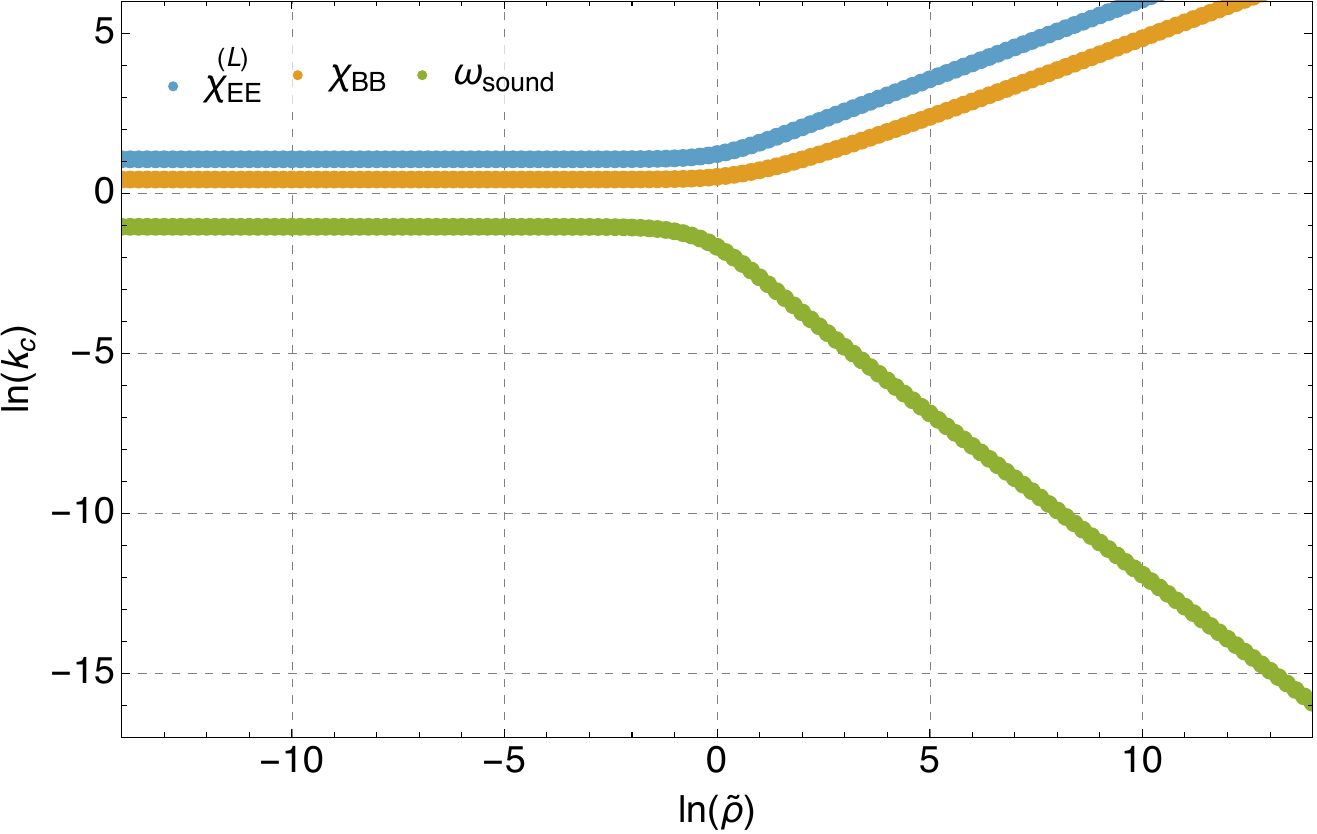}
    \caption{Critical radius of the wavevector against the charge density for a small $\tilde{k}$ expansion of the charge susceptibility $\chi_{\mathrm{EE}}^{(\mathrm{L})}(\tilde{k}^2)$ and the magnetisation susceptibility $\chi_{\mathrm{BB}}(\tilde{k}^2)$. The radii grow as $\sqrt{\tilde{\rho}}$. The green line shows the critical radius for the zero sound mode.}
    \label{fig:critical}
\end{figure}

{\noindent One of the key questions we wish to address in the application of our formalism is how the quasihydrodynamic regime emerges at large charge density. One often finds the Drude form for the conductivity of the D3/D5 probe brane in the literature (see e.g. \cite{Chen:2017dsy}):
	\begin{eqnarray}
		\label{Eq:sigmaDrude}
		\sigma_{\mathrm{Drude}}(\omega) &=& \frac{\chi_{\rho \rho} \tau}{1 - i \omega \tau} \stackrel{\omega \rightarrow 0}{\longrightarrow} \chi_{\rho \rho} \tau \; . 
	\end{eqnarray}
At any finite $\tilde{\rho}$ this expression is generally corrected - not only with frequency dependent terms which one might ignore as contact terms, but a departure of the residue from the simplistic expression given above. Our framework allows us to construct all available terms necessary to correct this expression, while being consistent with the usual hydrostaticity conditions. In particular, our AC conductivity taken from \eqref{Eq:CompleteChargeCorrelatorLD3D5} in the limit $\vec{k} \rightarrow \vec{0}$ has the form
	\begin{eqnarray}
		\label{Eq:sigmaEffective}
		\sigma_{\mathrm{AC}}(i \omega) 
        &=& \frac{i R_{(0)}}{ \omega + i\Gamma_{(0)}} + \sum_{n=0}^{N_{D}} c_{n} (i\omega)^n + \mathcal{O}(\omega^{N_{D}+1})  \; , \nonumber \\
         c_{n} &=&  \left( \frac{1}{n!} \sigma_{\mathrm{AC}}^{(n)}(0) -  \frac{R_{(0)}}{(\Gamma_{(0)})^{n+1}} \right) \; , \qquad
	\end{eqnarray}
where we have truncated the series expansion to some $N_{D}$ in frequency derivatives and $\sigma_{\mathrm{AC}}^{(n)}(0)$ is the $n^{\mathrm{th}}$ derivative of the AC conductivity evaluated at $\omega=0$. Comparing these two expressions, \eqref{Eq:sigmaDrude} and \eqref{Eq:sigmaEffective}, we soon determine that if the two expressions for the conductivity are to match it must be the case that
	\begin{eqnarray} 
		\tau = \frac{1}{\Gamma_{(0)}} \; , \qquad \chi_{\rho \rho} = R_{(0)} \; , \qquad c_{n} = 0 \; . 
	\end{eqnarray}
For the D3/D5 probe brane, this will only be the case in the limit of infinite $\tilde{\rho}$. Our goal then is to calculate subleading corrections and thus improve \eqref{Eq:sigmaDrude}.}

\begin{figure}[t!]
  \centering
  \begin{subfigure}{0.45\textwidth}
    \centering
    \includegraphics[width=\linewidth]{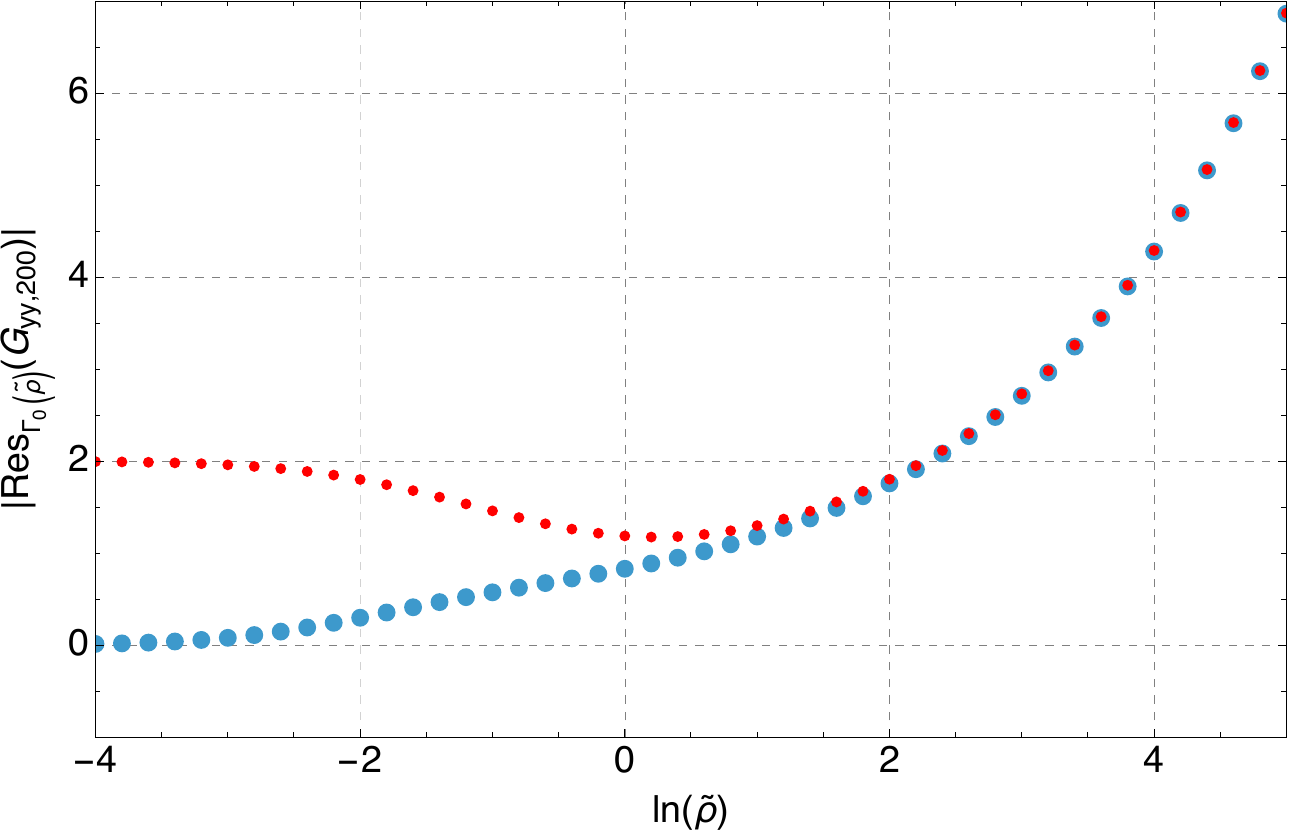}
    \caption{}
    \label{fig:residue low}
  \end{subfigure}
  \hfill
  \begin{subfigure}{0.45\textwidth}
    \centering
    \includegraphics[width=\linewidth]{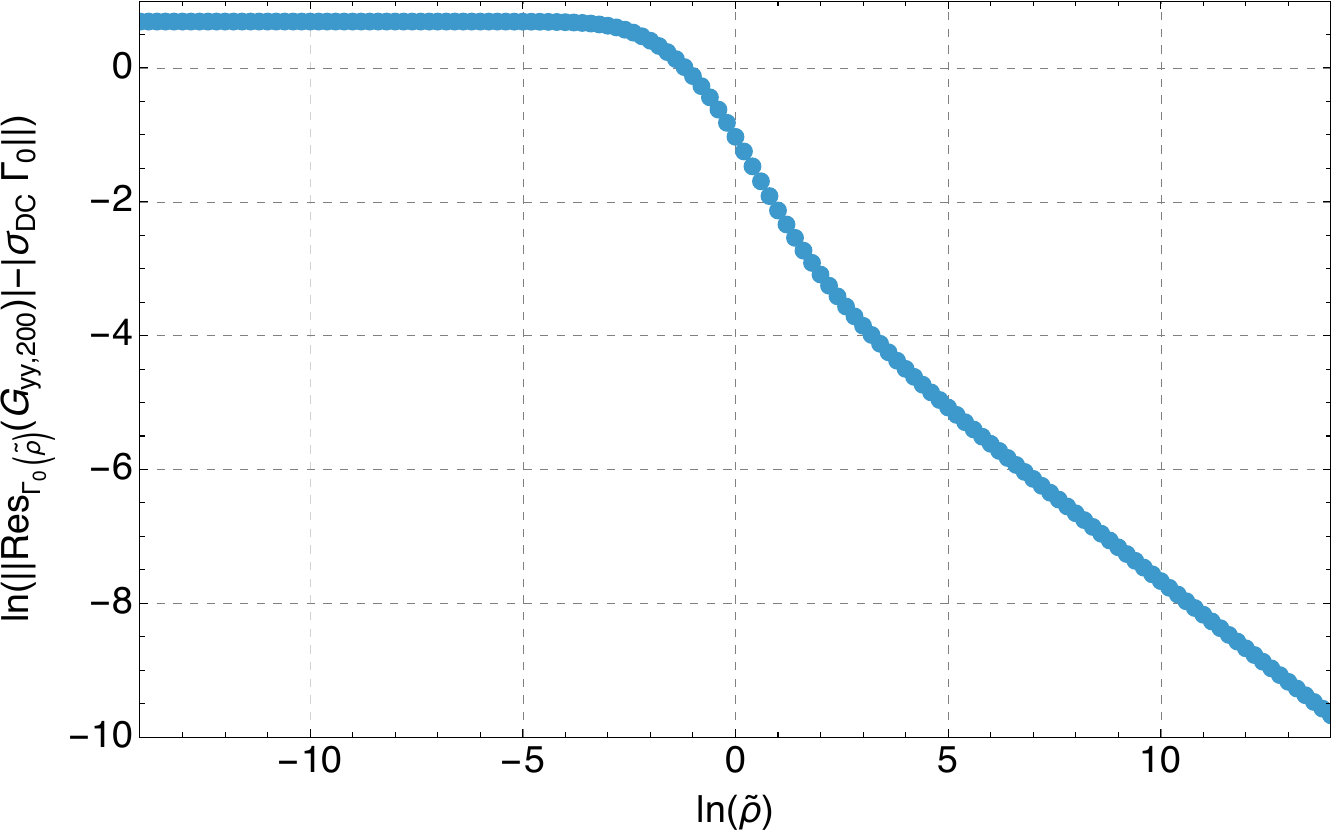}
    \caption{}
    \label{fig:residue low2}
  \end{subfigure}
  \caption{(a) Blue dots show the absolute value of the residue of the Laurent expansion of the holographic approximant expanded around the gapped pole $\Gamma_0$ with respect to the charge density. Red dots show $\sigma_{\mathrm{DC}}(\tilde{\rho}) \Gamma_0(\tilde{\rho})$. For large charge density the expressions agree. The blue dotted line shows that for large charge density the residue goes like  $(\Gamma(1/4)^2/(4 \sqrt{\pi}))^{-1}\tilde{\rho}^{1/2}$, plotted for $\mathrm{ln}(\tilde{\rho})>0$. (b) Shows the logarithm of the absolute value of the difference of the residue and $\sigma_{\mathrm{DC}} \Gamma_0$. For large $\tilde{\rho}$ the subleading behaviour of the difference goes like $\sim 0.07 \tilde{\rho}^{-1/2}$.
  }
  \label{fig:residueplots}
\end{figure}

{\ Let us first consider the DC conductivity. The value is known for all $\tilde{\rho}$ in the D3/D5 brane and it is given by \eqref{Eq:DCanalytic} in units where $\pi T=1$. In the simple Drude approximation of \eqref{Eq:sigmaDrude} one finds that charge susceptibility, DC conductivity and relaxation time are related to each other
	\begin{eqnarray}
		\sigma_{\mathrm{DC}} &\overset{\eqref{Eq:AnalyticExpressionChi}}{=}& \chi_{\rho \rho} \tau \; . 
	\end{eqnarray}
However, in our expression for the conductivity, the residue of the gapped pole is completely decoupled from the value of the DC conductivity. In particular 
	\begin{eqnarray}
		\label{Eq:CompleteACConduct}
		\sigma_{\mathrm{AC}}(\omega) &=& \sigma_{\mathrm{DC}} -  \frac{\omega R_{(0)}}{\Gamma_{(0)}\left( \omega + i\Gamma_{(0)}\right)} + \sum_{n=1}^{N_{D}} c_{n} (i\omega)^n + \mathcal{O}(\omega^{N_{D}+1}) \; , 
	\end{eqnarray}
so that the second term drops from the expression as $\tilde{\omega} \rightarrow 0$. By examining the residue of the gapped pole at large charge density using the holographic approximant, we find
	\begin{subequations}
	\begin{eqnarray}
\Gamma_{(0)} &=& \frac{1}{\sqrt{\tilde{\rho}}} \left[
\frac{4\pi}{\Gamma(\frac{1}{4})^2}
+ \frac{0.297}{\sqrt{\tilde{\rho}}}
+ \frac{0.088}{\tilde{\rho}}
+ \mathcal{O}(\tilde{\rho}^{-\frac{3}{2}})
\right] \; , \\
R_{(0)} &=& \sqrt{\tilde{\rho}} \left[
\frac{4\pi}{\Gamma(\frac{1}{4})^2}
+ \frac{0.297}{\sqrt{\tilde{\rho}}}
+ \frac{0.019}{\tilde{\rho}}
+ \mathcal{O}(\tilde{\rho}^{-\frac{3}{2}})
\right] \; .
\end{eqnarray}
and thus
	\begin{eqnarray} 
		R_{(0)} - \sigma_{\mathrm{DC}} \Gamma_{(0)}
		&=& - \frac{0.069}{\sqrt{\tilde{\rho}}} - \frac{ 0.075}{\tilde{\rho}} + \mathcal{O}(\tilde{\rho}^{-\frac{3}{2}}) \; . 
	\end{eqnarray}
Meanwhile the expansion of the pole near $\omega=0$ gives the following contribution to the DC conductivity	
	\begin{eqnarray}
		\label{Eq:DCfailure}
		\left. \frac{i R_{(0)}}{\omega + i \Gamma_{(0)}} \right|_{\omega=0} &=& \tilde{\rho} \left[ 1+ \frac{0.128}{\tilde{\rho}} + \mathcal{O}(\tilde{\rho}^{-\frac{3}{2}}) \right] \; . 
	\end{eqnarray}
	\end{subequations}
Notice that while the first term in \eqref{Eq:DCfailure} is precisely the DC conductivity at leading order in large $\tilde{\rho}$, the subleading term has entirely the wrong power behaviour. Moreover, we demonstrate the departure in figure \ref{fig:residue low} at lower values of $\tilde{\rho}$. We extract the subleading behaviour for the residue in fig. \ref{fig:residue low2}. It is a little mysterious as to why the leading term of $R_{(0)}$ has precisely the form necessary to give the Drude result, after all, from our approach it is entirely a coincidence. A potential answer comes from the emergence of a higher form symmetry (see appendix \ref{appendix:higherform}). However, the higher form approach as it stands in the literature does not seem to be able to capture the transverse behaviour, nor corrections departing from infinite $\tilde{\rho}$ and zero $\vec{k}$.} 

\begin{figure}[t!]
  \centering
  \begin{subfigure}{0.45\textwidth}
    \centering
    \includegraphics[width=\linewidth]{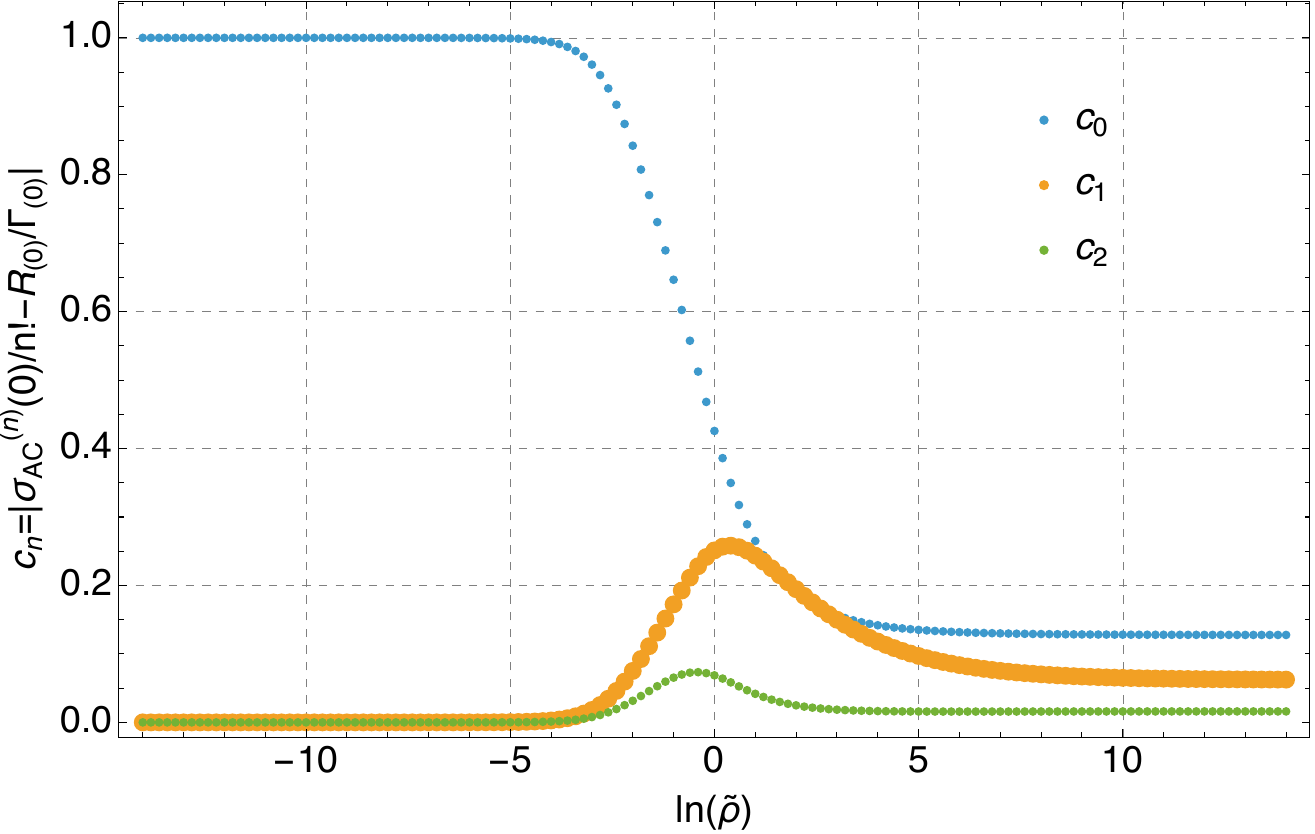}
    \caption{}
    \label{fig:coefficients}
  \end{subfigure}
  \hfill
  \begin{subfigure}{0.45\textwidth}
    \centering
    \includegraphics[width=\linewidth]{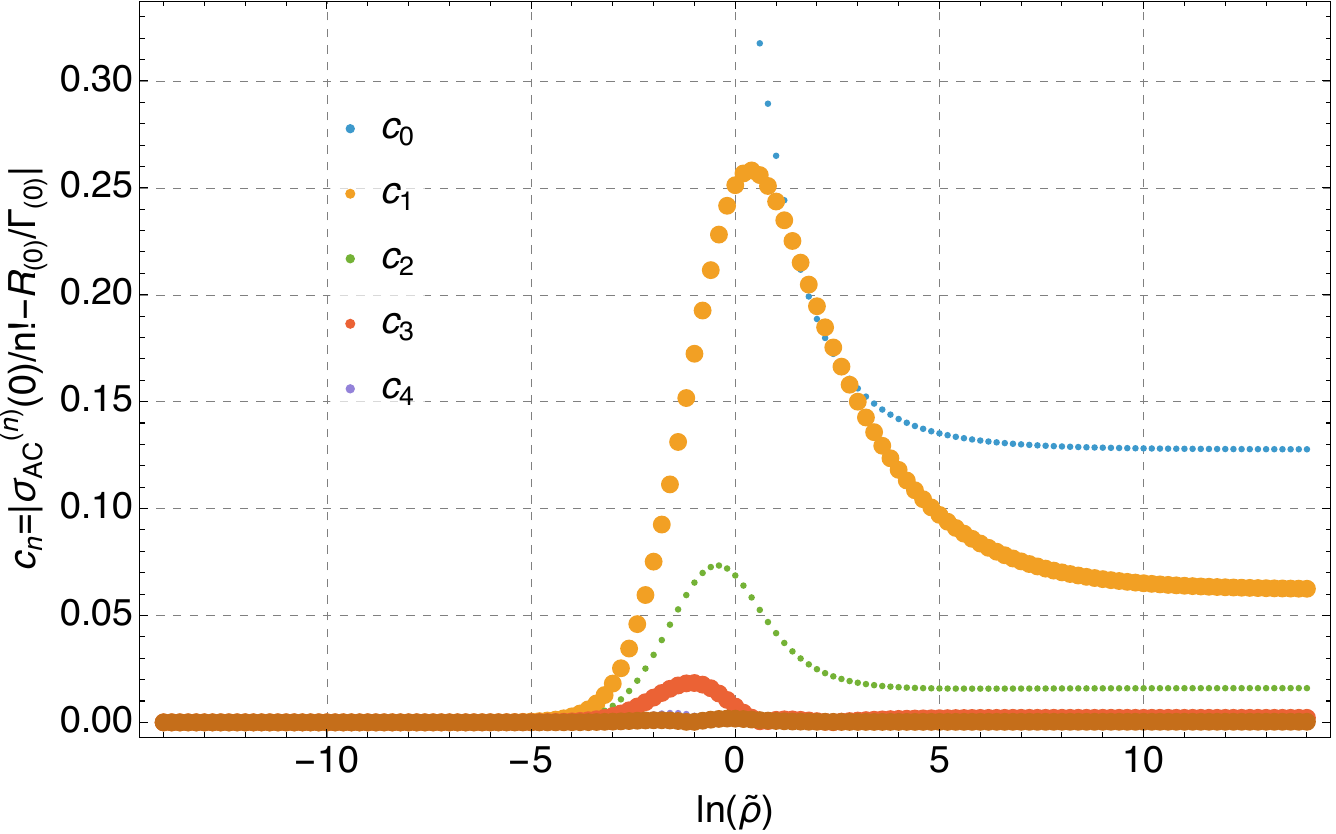}
    \caption{}
    \label{fig:coefficients_higher}
  \end{subfigure}
  \caption{The first five coefficients $c_n$ of the Series  expansion around $\omega = 0$ of the holomorphic part of $\sigma_{\mathrm{AC}}$ against the charge density, which for large charge density approach fixed values. (b) Shows the zoomed in version of the left plot. 
  }
\end{figure}

{\ For completeness, let us also consider the first few $c_{n}$ of \eqref{Eq:CompleteACConduct}. We utilise the holographic approximant to compute the complex conductivity around $\omega=0$. Subsequently, we determine the location of the gapped pole from the denominator of the approximant and expand about this pole to determine the residue. We can then remove the resultant pole from the approximant i.e.
	\begin{eqnarray}
		\sum_{n=0}^{N_{D}} c_{n} (i\omega)^n &=& \left[ G_{yy,200}(\omega) -  \frac{i R_{(0)}}{ \omega + i\Gamma_{(0)}} \right] 
	\end{eqnarray}
and order by order expand in $\tilde{\omega}$. We have confirmed that the $c_{n}$ have a disc of convergence upon whose boundary lies the next gapped pole (i.e. $|\omega_{*}|>\Gamma_{(0)}$). Figure \ref{fig:coefficients} displays the first few coefficients $c_n$ against charge density. We observe that the coefficients $c_{n}$ converge to some fixed value as $\tilde{\rho}$ gets large - and thus each term in the holomorphic function given by the sum over the $c_{n}$ is equally relevant to the behaviour of the system as opposed to being suppressed by charge density (although they do decrease in absolute value with $n$). This is because the other gapped poles at large $\tilde{\rho}$ sit at fixed positions in the infinite charge density limit. Moreover, these coefficients can in principle be determined from the AC conductivity on the real frequency axis once we have subtracted what we believe to be the pole behaviour.}

\subsubsection{Non-zero frequency and wavevector; quasinormal modes}\label{sec:omegandk}

{\noindent Finally, to construct our effective theory it was necessary to have a proper basis for the source terms. This requires that we have determined the modes (see \eqref{eq:ZerokNumerator} and \eqref{eq:FinitekNumerator} plus the following discussion). Let us report the relevant modes for the D3/D5 probe brane up to $\mathcal{O}(\vec{k}^6)$ at large charge density. We find
	\begin{subequations}
		\label{Eq:ModeSolns}
		\begin{eqnarray}
			\label{Eq:GappedTexpression}
			\omega _{\mathrm{gap},(\mathrm{T})}
		&=& - \frac{i}{\sqrt{\tilde{\rho}}}  \left[  \left( 0.539 + \frac{0.291}{\sqrt{\tilde{\rho}}} + \mathcal{O}(\tilde{\rho}^{-1}) \right) + \left( 0.618 - \frac{0.333}{\sqrt{\tilde{\rho}}} + \mathcal{O}(\tilde{\rho}^{-2}) \right) \left(\frac{\tilde{k}^2}{\tilde{\rho}}\right) \right. \nonumber \\
		&\;& \left. \hphantom{  - \frac{i}{\sqrt{\tilde{\rho}}}  \left[ \right.}  - \left( 0.142 - \frac{0.229}{\sqrt{\tilde{\rho}}} + \mathcal{O}(\tilde{\rho}^{-\frac{3}{2}}) \right) \left(\frac{\tilde{k}^2}{\tilde{\rho}}\right)^{2}   \right. \nonumber \\
		&\;& \left. \hphantom{  - \frac{i}{\sqrt{\tilde{\rho}}}  \left[ \right.} + \left( 0.046 - \frac{0.125}{\sqrt{\tilde{\rho}}} + \mathcal{O}(\tilde{\rho}^{-1}) \right) \left(\frac{\tilde{k}^2}{\tilde{\rho}}\right)^3 + \mathcal{O}^{4}\left(\frac{\tilde{k}^2}{\tilde{\rho}}\right)  \right]  \; , \\
		\label{Eq:GappedLexpression}
		\omega _{\mathrm{gap},(\mathrm{L})} 
		&=& - \frac{i}{\sqrt{\tilde{\rho}}}  \left[  \left( 0.539 + \frac{0.291}{\sqrt{\tilde{\rho}}} + \mathcal{O}(\tilde{\rho}^{-1}) \right)  - \left( 0.927 - \frac{0.118}{\tilde{\rho}} + \mathcal{O}(\tilde{\rho}^{-\frac{3}{2}}) \right) (\tilde{\rho} \tilde{k}^2) \right. \nonumber \\
		&\;& \left. \hphantom{  - \frac{i}{\sqrt{\tilde{\rho}}}  \left[ \right.}  - \left( 3.187 - \frac{1.719}{\sqrt{\tilde{\rho}}} + \mathcal{O}(\tilde{\rho}^{-\frac{3}{2}}) \right) (\tilde{\rho} k)^2 \right. \nonumber \\
		&\;& \left. \hphantom{  - \frac{i}{\sqrt{\tilde{\rho}}} \left[ \right.}  - \left( 32.865 - \frac{35.451}{\sqrt{\tilde{\rho}}} + \mathcal{O}(\tilde{\rho}^{-1}) \right) (\tilde{\rho} \tilde{k}^2)^3  + \mathcal{O}^{4}(\tilde{\rho} \tilde{k}^2)  \right] \;, \\
		\label{Eq:Dexpression}
		\omega _{\mathfrak{D}} 
		&=& - \frac{i}{\sqrt{\tilde{\rho}}} \left[ \left( 0.927 + \frac{0.464}{\tilde{\rho}^2} + \mathcal{O}(\tilde{\rho}^{-\frac{5}{2}}) \right) (\tilde{\rho} \tilde{k}^2) \right. \nonumber \\
		&\;& \left. \hphantom{  - \frac{i}{\sqrt{\tilde{\rho}}}  \left[ \right.} + \left( 1.593 - \frac{0.859}{\sqrt{\tilde{\rho}}} + \mathcal{O}(\tilde{\rho}^{-2}) \right) (\tilde{\rho} \tilde{k}^2)^{2}  \right. \nonumber \\
		&\;& \left. \hphantom{  - \frac{i}{\sqrt{\tilde{\rho}}}  \left[ \right.}   - \left( 5.477 - \frac{5.909}{\sqrt{\tilde{\rho}}} + \mathcal{O}(\tilde{\rho}^{-1}) \right) (\tilde{\rho} \tilde{k}^2)^3 + \mathcal{O}^{4}(\tilde{\rho} \tilde{k}^2)  \right] \; .
		\end{eqnarray}	
	\end{subequations}
Some comments on these expressions are appropriate; we obtained the large $\tilde{\rho}$ fall-offs by fitting, however it is in principle possible to modify the holographic approximant procedure to order by order obtain the large $\tilde{\rho}$ behaviour. In fact such an approach is favourable to this undertaking as the strict $\tilde{\rho} \rightarrow \infty$ limit of the probe brane fluctuation equations is ill-defined (there is no asymptotically AdS region at leading order in that limit). With this said, notice that the leading term in small $\vec{k}$ in \eqref{Eq:GappedTexpression} and \eqref{Eq:GappedLexpression} match as is required by spatial rotation invariance. Also, the $\mathcal{O}(\vec{k}^2)$ terms in \eqref{Eq:GappedLexpression} and \eqref{Eq:Dexpression} match at leading order in $\tilde{\rho}$ as one finds in \cite{Chen:2017dsy} - thus their trajectories will form almost perfect semi-circles in the complex frequency plane at large enough $\tilde{\rho}$. However, they differ at subleading order in $\tilde{\rho}$ and $\vec{k}$.}

{\ We have claimed in our derivation of the effective theory that there are undetermined holomorphic corrections to our correlators, i.e. $f_{(\mathrm{L}),n}(\vec{k}^2)$ and $f_{(\mathrm{T}),n}(\vec{k}^2)$, and such corrections begin at $\mathcal{O}(\vec{k}^2)$. Let us first show how one may determine such terms by examining the transverse sector conductivity. At low frequency and large charge density this conductivity $\sigma_{(\mathrm{T})} $ takes the form \small{
	\begin{align}
		&\; \left[ \highlight{\left( 1 + \frac{1}{2\tilde{\rho}^2} + \mathcal{O}(\tilde{\rho}^{-4}) \right)} - \left( 3.438 - \frac{3.708}{\sqrt{\tilde{\rho}}} + \mathcal{O}(\tilde{\rho}^{-1}) \right) \frac{\tilde{k}^2}{\tilde{\rho}} + \mathcal{O}(\vec{k}^4)  \right] \tilde{\rho} \nonumber \\
		&\; +  \left[ \highlight{\left( 1.854 - \frac{1}{\sqrt{\tilde{\rho}}} + \mathcal{O}(\tilde{\rho}^{-\frac{5}{2}}) \right)} - \left( 8.498 - \frac{13.750}{\sqrt{\tilde{\rho}}} + \mathcal{O}(\tilde{\rho}^{-1}) \right) \frac{\tilde{k}^2}{\tilde{\rho}} + \mathcal{O}(\vec{k}^4) \right] \left( i \tilde{\rho}^{\frac{3}{2}} \tilde{\omega}  \right)  \nonumber \\ 
		&\; -  \left[ \highlight{\left( 3.438 - \frac{3.708}{\sqrt{\tilde{\rho}}} + \mathcal{O}(\tilde{\rho}^{-1}) \right)} - \left( 19.695 - \frac{42.490}{\sqrt{\tilde{\rho}}} + \mathcal{O}(\tilde{\rho}^{-1}) \right) \frac{\tilde{k}^2}{\tilde{\rho}} + \mathcal{O}(\vec{k}^4) \right] \left( \tilde{\rho}^{2} \tilde{\omega}  ^2 \right)  \nonumber \\
		&\; - \left[ \highlight{\left( 6.374 - \frac{10.313}{\sqrt{\tilde{\rho}}} + \mathcal{O}(\tilde{\rho}^{-1}) \right)} - \left(43.819 - \frac{118.170}{\sqrt{\tilde{\rho}}} + \mathcal{O}(\tilde{\rho}^{-1}) \right)\frac{\tilde{k}^2}{\tilde{\rho}} + \mathcal{O}(\vec{k}^4)\right] \left( i \tilde{\rho}^{\frac{5}{2}} \tilde{\omega}  ^3 \right) \nonumber \\
		&\; + \mathcal{O}(\tilde{\omega} ^4) \; .
	\end{align} }
The red terms are nothing more than leading approximations at large charge density to the usual AC conductivity ($\sigma_{\mathrm{AC}}(i\omega)$). To extract $f_{(\mathrm{T}),n}(\vec{k}^2)$ we require the residue of the gapped pole $\omega_{\mathrm{gap},(\mathrm{T})}$. In the same limits as the mode solutions of \eqref{Eq:ModeSolns} we find
	\begin{subequations}
	\begin{eqnarray}
			\frac{R_{\mathrm{gap},(0)}}{\sqrt{\tilde{\rho}}}
		&=&  i \ \left( 0.539 + \frac{0.291}{\sqrt{\tilde{\rho}}} + \mathcal{O}(\tilde{\rho}^{-1}) \right)  \; , \\
			\sqrt{\tilde{\rho}} R_{\mathrm{gap},(\mathrm{T})}(\vec{k}^2)
		&=& i  \left(  1.236 - \frac{0.0667}{\sqrt{\tilde{\rho}}} + \mathcal{O}(\tilde{\rho}^{-\frac{3}{2}}) \right) \nonumber \\
		&\;&  - i \left( 2.054 - \frac{3.064}{\sqrt{\tilde{\rho}}} + \mathcal{O}(\tilde{\rho}^{-1}) \right) \left(\frac{\tilde{k}^2}{\tilde{\rho}} \right) \nonumber \\
		&\;&+ i \left( 3.182 - \frac{7.632}{\sqrt{\tilde{\rho}}} + \mathcal{O}(\tilde{\rho}^{-1}) \right) \left(\frac{\tilde{k}^2}{\tilde{\rho}} \right)^2 
			+ \mathcal{O}^{4}\left(\frac{\tilde{k}^2}{\tilde{\rho}}\right) \; . \qquad
	\end{eqnarray}
	\end{subequations}
Naturally the expressions we use to confirm our results contain more orders in $\tilde{\rho}$ to greater precision than is displayed above, but we have only displayed leading and subleading terms in the name of legibility. With this information, to obtain $f_{(\mathrm{T}),n}(\vec{k}^2)$, one must subtract the gapped pole term. We thus find
	\begin{eqnarray}
		&\;& \left[ \sigma_{(\mathrm{T})}(\omega ,\vec{k}^2) - \sum_{n=0}^{2} \frac{1}{n!} \sigma_{\mathrm{AC}}^{(n)}(0) (i \omega )^n \right] \nonumber \\
		&\;&- \left[ \frac{R_{\mathrm{gap},(\mathrm{T})}(\vec{k}^2)}{\omega  - \omega _{\mathrm{gap},(\mathrm{T})}(\vec{k}^2)} + \frac{R_{\mathrm{gap},(\mathrm{T})}(0)}{\omega_{\mathrm{gap}}(0)} \sum_{n=0}^{2}  \left(\frac{\omega}{\omega _{\mathrm{gap}}(0)} \right)^n \right] \nonumber \\
		&=& - \left[ \left( 0.146 + \mathcal{O}(\tilde{\rho}^{-\frac{1}{2}}) \right) \left(\frac{\tilde{k}^2}{\tilde{\rho}}\right) + \mathcal{O}^2\left(\frac{\tilde{k}^2}{\tilde{\rho}}\right)  \right]  \nonumber \\
		&\;&	+ \left[ \left( 0.071 + \mathcal{O}(\tilde{\rho}^{-\frac{1}{2}}) \right) \left(\frac{\tilde{k}^2}{\tilde{\rho}}\right) + \mathcal{O}\left(\frac{\tilde{k}^2}{\tilde{\rho}}\right) \right] \left( i  \tilde{\omega}  \right)  \nonumber \\
		&\;& + \left[ \left(  0.141 + \mathcal{O}(\tilde{\rho}^{-\frac{1}{2}}) \right) \left(\frac{\tilde{k}^2}{\tilde{\rho}}\right) + \mathcal{O}^2\left(\frac{\tilde{k}^2}{\tilde{\rho}}\right)  \right]  	\tilde{\omega} ^2	+ \mathcal{O}^{3}(\tilde{\omega} ) \; . 
	\end{eqnarray}
Notice that each of the terms on the right hand side of the equality, corresponding to the first few $f_{(\mathrm{T}),n}(\vec{k}^2)$, appear at $\mathcal{O}(k^2)$ as was argued.}

{\ In principle the same procedure can be applied to the longitudinal sector to extract $f_{(\mathrm{L}),n}(\vec{k}^2)$ with the important caveat that two poles must be subtracted - the longitudinal gapped pole and the diffusive pole. It is the latter however that causes additional problems. In particular, notice that in a small frequency expansion of the diffusive pole given by
		\begin{align}
		 - \frac{i \vec{k}^2 \mathfrak{R}_{\mathfrak{D}}(\vec{k}^2)}{\tilde{\omega}  + i \mathfrak{D}(\vec{k}^2) \vec{k}^2}						   &= \left( \frac{\mathfrak{R}_{\mathfrak{D}}(\vec{k}^2)}{\mathfrak{D}(\vec{k}^2)} \right) + \left( \frac{\mathfrak{R}_{\mathfrak{D}}(\vec{k}^2)}{\vec{k}^{2} \mathfrak{D}(\vec{k}^2)^2} \right) \omega  - \left( \frac{\mathfrak{R}_{\mathfrak{D}}(\vec{k}^2)}{\vec{k}^{4} \mathfrak{D}(\vec{k}^2)^3}  \right) \omega ^2 + \mathcal{O}(\omega^3) \, \\
	 &= - \left( \sigma_{\mathrm{DC}} + \mathcal{O}(\vec{k}^2) \right) - \left( \frac{\chi_{\rho \rho}}{\vec{k^2}} + \mathcal{O}(\|\vec{k}\|^0) \right) ( i \omega ) \nonumber \\
	 &\qquad  + \left( \frac{\chi_{\rho \rho}^2}{\sigma_{\mathrm{DC}} \vec{k}^4} + \mathcal{O}(\|\vec{k}\|^{-2})  \right) \omega^2  + \mathcal{O}(\omega ^3) 
	\end{align}
there are inverse powers of $\vec{k}^2$. In writing the above we have defined
	\begin{eqnarray}
		\mathfrak{R}_{\mathfrak{D}}(\vec{k}^2) &=& R_{\mathfrak{D}}(0) + R_{\mathfrak{D}}'(0) \vec{k}^2 + \tilde{R}_{\mathfrak{D}}(\vec{k}^2) \vec{k}^4 \; ,
	\end{eqnarray}
to simplify the notation in comparison to \eqref{Eq:TotalChargeCorrelator} . Our earlier numerical procedure cannot capture these; to include such non-analytic powers of $\vec{k}$ in our small $\vec{k}$ expansion we must rescale so that
	\begin{align}
		\label{Eq:DiffusiveEexpansion}
		E_{x}(r;\omega,\vec{k}^2) \rightarrow \sum_{n=0}^{\infty} \left(\frac{\omega}{|\vec{k}|}\right)^{n} E_{n}(r,\vec{k}^2) = \sum_{n=0}^{\infty} \sum_{m=0}^{\infty} \left(\frac{\omega}{|\vec{k}|}\right)^{n} E_{n,m}(r) \vec{k}^{2m} \;  .
	\end{align}
This introduces the expected divergences in small $\vec{k}$ into the low frequency Green's function associated with the expansion of the diffusive pole in small $\tilde{\omega}$.}

{\ With this said, these leading small $\vec{k}$ divergences of the longitudinal conductivity $\sigma_{(\mathrm{L})}(\omega,\vec{k}^2)$ take the form
	\begin{align}
\label{Eq:LowFrequencyKsigmaL}
&\; - \left[ \left( 1.079 + \frac{0.233}{\tilde{\rho}^\frac{5}{2}} + \mathcal{O}(\tilde{\rho}^{-\frac{9}{2}}) \right) + \mathcal{O}(\tilde{k}^2) \right] \frac{i \sqrt{\tilde{\rho}} \tilde{\omega}}{\tilde{k}^2} \nonumber \\
&\; + \left[ \left( 1.164 - \frac{0.582}{\tilde{\rho}^2} + \mathcal{O}(\tilde{\rho}^{-3}) \right) - \left( 0.622 + \frac{0.131}{\tilde{\rho}} + \mathcal{O}(\tilde{\rho}^{-2}) \right) \frac{\tilde{k}^2}{\tilde{\rho}} + \mathcal{O}(\vec{k}^4) \right] \left(\frac{\tilde{\omega}}{\tilde{k}^2} \right)^2 \nonumber \\
&\; + \left[ \left( 1.255 - \frac{1.255}{\tilde{\rho}^2} + \mathcal{O}(\tilde{\rho}^{-\frac{5}{2}}) \right) - \left( 2.157 - \frac{1.164}{\sqrt{\tilde{\rho}}} + \mathcal{O}(\tilde{\rho}^{-2}) \right) (\tilde{\rho} \tilde{k}^2) \right. \nonumber \\
&\; \left. \hphantom{+ \left[  \right.} + \left( 0.423 - \frac{0.622}{\sqrt{\tilde{\rho}}} + \mathcal{O}(\tilde{\rho}^{-\frac{3}{2}}) \right) (\tilde{\rho} \tilde{k}^2)^2 + \mathcal{O}^{3}(\tilde{\rho} \tilde{k}^2) \right] \frac{i}{\sqrt{\tilde{\rho}}} \left(\frac{\tilde{\omega}}{\tilde{k}^2}\right)^3
+ \mathcal{O}^{4}(\tilde{\omega}) \; .
\end{align}
Meanwhile, the leading and sub-leading terms in the diffusive pole residue at large charge density are
	\begin{eqnarray}
			\label{Eq:DiffusiveResidue}
			\frac{\mathfrak{R}_{\mathfrak{D}}(\vec{k}^2)}{\sqrt{\tilde{\rho}}}
		&=& \left( 0.927 + \frac{0.927}{\tilde{\rho}^2} + \mathcal{O}(\tilde{\rho}^{-\frac{5}{2}}) \right) \left( \tilde{\rho} \tilde{k}^2 \right)
			+  \left( 4.780 - \frac{2.578}{\sqrt{\tilde{\rho}}} + \mathcal{O}(\tilde{\rho}^{-2}) \right)  \left(\tilde{\rho} \tilde{k}^2 \right)^2 \nonumber \\
		&\;& + \left( 27.387 - \frac{29.543}{\sqrt{\tilde{\rho}}} + \mathcal{O}(\tilde{\rho}^{-1}) \right) \left(\tilde{\rho} \tilde{k}^2 \right)^3
			 + \mathcal{O}^{4}\left(\tilde{\rho} \tilde{k}^2\right) \; .
	\end{eqnarray}
To reproduce all the powers in $\tilde{\rho}$ displayed in \eqref{Eq:LowFrequencyKsigmaL}, it is necessary to work to more orders in $\tilde{\rho}$ than we display in the residue and diffusive mode positions, although what we have supplied is sufficient to cancel the leading terms. Nevertheless we have  checked that we can reproduce the divergences in small $\vec{k}$ that are presented in \eqref{Eq:LowFrequencyKsigmaL}. Consequently, it is possible to match our effective theory in the longitudinal sector to the observed behaviour of the correlator as claimed.}

\section{Discussion and outlook}\label{sec:Discussion}

In this paper we have constructed an effective linearised theory for a conserved $U(1)$ charge current that incorporates an arbitrary but finite number of isolated poles under mild and physically natural assumptions. The central conceptual ingredient of our approach is the explicit separation between time and spatial derivatives. This allows poles to be treated as non-perturbative objects in frequency, while retaining a controlled derivative expansion in momentum. As a result, the theory is fully compatible with standard hydrostaticity conditions and does not require any modification of thermodynamics.

Applying this framework to the D3/D5 probe brane system, we were able to track how transport coefficients reorganise as additional long-lived modes approach the origin of the complex frequency plane at large charge density. In this sense, the emergence of quasihydrodynamics in this model admits a precise and quantitative interpretation: it reflects a truncation of the Mittag-Leffler representation of the current correlator, rather than the appearance of parametrically small gaps. The effective theory makes this reorganisation manifest and allows it to be matched systematically to microscopic data.

We emphasise that the construction itself is not restricted to large charge density. For any value of $\rho$, the effective theory reproduces the conductivity within the disc of analyticity set by the next excluded pole. That hydrodynamics continues to perform remarkably well even when additional long-lived gapped modes become relevant is therefore best understood as a statement about the robustness of linearised hydrodynamics once its non-perturbative content is properly identified. An important open question is to what extent this robustness extends to the non-linear regime, where mode coupling and genuinely dynamical effects become important.  We expect to be able to derive in the linearised approximation generalisations of the work of \cite{Haehl_2015} to allows us to incorporate gapped poles into the small frequency effective theory, as we have achieved here. Some results along this line are discussed in \cite{Jain:2023obu} with regard to the Maxwell-Cattaneo model; although in that paper the relaxation term is constant and holomorphic terms are not included. However, a serious obstacle exists at the non-linear level, because transport coefficients become full functions of the hydrodynamic variables. Unless we are perturbing around a constant background, which is in effect linearisation, it is difficult to see how to proceed. Moreover, even in the case of linearising a constant background, incorporating non-linearity requires that we introduce multiple integrals over products of the Green's functions we have derived here.

Although our explicit large density analysis focused on leading and subleading orders, the auxiliary-equation techniques we employed admit a natural extension. In particular, they can be used to compute derivatives of both sources and expectation values with respect to the charge density. This is particularly important as the equations for the modes do not have a well-defined limit as $\tilde{\rho} \rightarrow \infty$. This allows a systematic exploration of higher-order density corrections to pole positions and residues, providing a more complete picture of how quasihydrodynamic behaviour develops across parameter space.

We also discussed the construction of a holographic approximant for current correlators. Several improvements suggest themselves. One natural refinement is to replace Taylor expansions of the source and vev by Pad\'{e} approximants before taking their ratio, which is particularly advantageous when the correlator exhibits known power-law behaviour at large frequency. Another possibility is to supplement the near-origin data with information about the next quasinormal mode, obtained for instance via standard shooting methods, thereby constraining higher-order coefficients in the frequency expansion.

From a numerical perspective, we observed that at non-zero frequency and momentum the computation of higher-order frequency coefficients becomes increasingly costly. A practical alternative is to construct multi-point Pad\'{e} approximants for the source and vev propagators directly. This strategy, which has been employed previously in the context of analytic continuation of Green's functions \cite{Witczak-Krempa:2013nua}, has the advantage of being readily parallelisable and of incorporating global information more efficiently. While the optimal choice of sampling points is non-trivial, established techniques such as adaptive grid methods can be used to refine the approximant in regions where it deviates from the exact correlator.

Our results also clarify a long-standing issue in holographic magnetohydrodynamics. Early studies found that hydrodynamic predictions for thermo-electric conductivities in the presence of external magnetic fields agreed with holography only at leading order in the field, despite the fact that Ward identities must hold exactly. In earlier work \cite{Amoretti_2020,Amoretti_2021,Amoretti_2021_2,Amoretti_2022, Amoretti_2024_relaxation} some of the present authors proposed resolving this tension by enforcing the Ward identities at the outset, which necessitated the inclusion of an incoherent Hall conductivity. While this correctly captured the Laurent expansion around the cyclotron pole, a small mismatch in pole positions persisted at higher magnetic field.

The resolution is now clear. The effective theories considered previously implicitly truncated background couplings at too low an order in time derivatives. Once one recognises that poles must be treated non-perturbatively, it becomes natural to retain arbitrarily high time-derivative couplings to background fields while truncating only the thermodynamic sector. With these terms included, it is possible to respect all Ward identities and exactly reproduce the holographic pole structure. Linearised hydrodynamics in external magnetic or electric fields therefore does not require these fields to be small in derivatives, contrary to some claims in the literature.

The effective theory we have derived in this work applies to conservative (in the sense of following from an action principle), non-driven models. However a core feature of the formalism is the inclusion of decaying modes manifesting as non-conservation of the spatial charge current which is quite reminiscent of non-conservative systems. Moreover, it is well-known that probe branes can form driven steady states \cite{Karch:2007pd,Mas_2009,Brattan_2024} which, to the author's knowledge, are not necessarily solutions to the hydrostaticity conditions we have discussed. Understanding how the formalism can be modified to include steady driven states and truly dissipative effects \cite{Amoretti_2023,Amoretti_2023_2,Amoretti_2023_3,Amoretti_2024,Amoretti_2024_2} is a core goal for future work. In this respect, we may need to pass beyond the limitations of holography and consider more generalised systems as test-beds such as: flocking matter \cite{RevModPhys.85.1143,Amoretti_2024,Armas_2025} and numerical simulations of exotic particles \cite{Amoretti_2025}.}

 A natural question is whether the pole-based reorganisation of linear response developed here is related to the hydrodynamic attractor programme (see \cite{Romatschke:2017ejr} for a review). Hydrodynamic attractors arise from a resummation of the gradient expansion in real-time evolution. In contrast, our construction reorganises the theory in frequency space, treating poles as non-perturbative input data and expanding only the holomorphic sector. It is plausible that both structures reflect the same underlying analytic properties of the theory: in particular, the finite radius of convergence of the gradient expansion is controlled by the location of non-hydrodynamic singularities in the complex frequency plane. From this viewpoint, attractors might be understood as the time-domain manifestation of the same analytic structure that governs the Mittag–Leffler expansion. Establishing this connection more precisely, especially beyond linear response, would be an interesting direction for future work.

 It is worth noting that other approaches to extend the validity of hydrodynamics such as the ``Hydro+'' framework proposed in \cite{Stephanov:2017ghc} have been applied to various systems. Hydro+ in particular has been motivated experimentally by beam energy scans which probe the Quark Gluon Plasma (QGP) near its critical point. At this critical point, non-linear fluctuations subside as slow as hydrodynamic fluctuations. This allows the inclusion of slow modes into the model, effectively producing a quasi-hydrodynamic theory. Besides the QGP, the Hydro+ framework has additionally been applied to describe spin within hydrodynamics as an almost conserved quantity. These examples give potential areas for the application of our effective framework.

{Finally, we comment on a general feature of holographic correlators that places intrinsic constraints on effective descriptions. Extending the Mittag-Leffler representation to cover a larger portion of the real frequency axis requires the inclusion of poles deeper in the complex plane (see fig. \ref{fig:christmastree})}. However, such poles tend to collide at smaller values of the spatial momentum. When pole collisions occur, the small-momentum expansion necessarily breaks down \cite{Amoretti:2013xya,Withers_2018,Grozdanov:2019kge,Grozdanov:2019uhi,Heller:2020uuy,Abbasi:2020ykq,Grozdanov:2020koi,Jansen:2020hfd,Heller:2020hnq,Abbasi:2020xli,Liu_2022,Grozdanov_2023,Grozdanov_2023_2,cartwright2024exampleconvergencehydrodynamicsstrong,grozdanov2025algebraicstructuregradientexpansion}. As a consequence, enlarging the disc of convergence in frequency generically reduces the critical wavevector below which the effective theory remains valid. This behaviour appears to be universal in holographic systems and provides a sharp criterion for the regime of applicability of quasihydrodynamic effective theories.

\section*{Acknowledgements}

{\noindent A.A. \& D.B. have received support from the project PRIN 2022A8CJP3 by the Italian Ministry of University and Research (MUR). D.B. is currently funded by PNRR GIOVANI RICERCATORI, CUP D33C25000470006. This project has also received funding from the European Union’s Horizon 2020 research and innovation programme under the Marie-Sk\l{}odowska-Curie grant agreement No. 101030915.}

\appendix

\section{Refining the conductivities of interest }
\label{appendix:refining}

{\noindent At any fixed frequency where the poles of interest remain isolated we can expect to be able to write down a Mittag-Leffler expansion for the conductivity of the following form
	\begin{eqnarray}
		\sigma^{ij}(\omega) &=& f^{ij}(\omega) + \sum_{n=1}^{N-1} \frac{R_{n}^{ij}}{\omega - \omega_{n}}
	\end{eqnarray}
where the first term is a set of holomorphic functions on some disc containing the origin, $R^{ij}$ are pole residues and $\omega_{n}$ the $N-1$ additional poles we wish to include in our analysis. Among the constraints mentioned in section \ref{section:single_u1} is spatial rotation invariance. This symmetry implies that all two-tensor terms can be decomposed into the following forms
    \begin{subequations}
    \label{Eq:RotationIdentifications}
    \begin{eqnarray}
        \Xi^{ij}(\vec{k}) &=& \Xi_{(0)} \delta^{ij} - \Xi_{(\mathrm{L})}(\vec{k}^2) k^{i} k^{j} - \Xi_{(\mathrm{T})}(\vec{k}^2) \left( \vec{k}^2 \delta^{ij} - k^{i} k^{j} \right) \; , \\
        \Xi^{ij}(\omega,\vec{k}) &=& \Xi_{(0)}(\omega) \delta^{ij} - \Xi_{(\mathrm{L})}(\omega,\vec{k}^2) k^{i} k^{j} - \Xi_{(\mathrm{T})}(\omega,\vec{k}^2) \left( \vec{k}^2 \delta^{ij} - k^{i} k^{j} \right) \; .
    \end{eqnarray}
    \end{subequations} 
Hence the Mittag-Leffler expansions for the conductivities of interest for generic $\vec{k}$ will have the form
    \begin{subequations}
    \label{Eq:GenericConductivity}
	\begin{eqnarray}
		\sigma^{ij}(\omega,\vec{k}) &=& \sigma_{(\mathrm{L})}(\omega,\vec{k}^2) \left(  \frac{k^{i} k^{j}}{\vec{k}^2} \right)  + \sigma_{(\mathrm{T})}(\omega,\vec{k}^2) \left( \delta^{ij} - \frac{k^{i} k^{j}}{\vec{k}^2} \right) \; , \\
            \sigma_{(\mathrm{L})}(\omega,\vec{k}^2) 
        &=& - \frac{i \vec{k}^2 R_{\mathfrak{D}}(\vec{k})}{\omega + i \mathfrak{D}(\vec{k}^2) \vec{k}^2} + i \sum_{m=1}^{N-1} \frac{R_{(0),m} - \vec{k}^2 R_{(\mathrm{L}),m}(\vec{k})}{\omega + i (\tau_{(0),m}^{-1} - \vec{k}^2 \tau_{(\mathrm{L}),m}^{-1}(\vec{k}^2))} \nonumber \\
        &\;& + \left( f_{(0)}(\omega) - \vec{k}^2 f_{(\mathrm{L})}(\omega,\vec{k}^2) \right) \; ,  \\
            \sigma_{(\mathrm{T})}(\omega,\vec{k}^2) 
        &=& i \sum_{m=1}^{N-1} \frac{R_{(0),m} - \vec{k}^2 R_{(\mathrm{T}),m}(\vec{k})}{\omega + i (\tau_{(0),m}^{-1} - \vec{k}^2 \tau_{(\mathrm{T}),m}^{-1}(\vec{k}^2))} \nonumber \\
        &\;& + \left( f_{(0)}(\omega) - \vec{k}^2 f_{(\mathrm{T})}(\omega,\vec{k}^2) \right) \; , 
	\end{eqnarray}
    \end{subequations}
where signs and factors of the imaginary number on residues have been chosen for stylistic reasons, $f_{(\mathrm{L}),(\mathrm{T})}(\omega,\vec{k})$ are holomorphic functions in a disc on the complex frequency plane containing the origin, and we have explicitly constructed our conductivity so that the diffusive pole is only present in the longitudinal sector. As the disc where the Mittag-Leffler expansion applies contains the origin we can expand holomorphic function in small frequency 
    \begin{eqnarray}
      f_{(\mathrm{L}),(\mathrm{T})}(\omega,\vec{k}) =  \sum_{n=0}^{\infty} \frac{f_{(\mathrm{L}),(\mathrm{T}),n}(\vec{k})}{n!} (i \omega)^{n} \; ,
    \end{eqnarray}
and eventually truncate to some order $N_{D}$ in the frequency expansion.

Let us now further identify which parts of this \eqref{Eq:GenericConductivity} correspond to the (complex) AC conductivity\footnote{Here we consider the AC conductivity to be a complex function of complex frequency. If the system is time reversal invariant, this means the real part must be even in frequency while the imaginary part must be odd. This is the origin of the argument $i \omega$}. Taking $\vec{k} \rightarrow \vec{0}$ we see that
    \begin{eqnarray}
            \sigma_{\mathrm{AC}}(i \omega)
        &=& \sum_{n=1}^{N-1} \frac{i R_{(0),n}}{\omega + i \tau_{(0),n}^{-1}} + \sum_{n=0}^{\infty} \frac{f_{(0),n}}{n!} (i \omega)^{n} \; . 
    \end{eqnarray}
A balance between the non-hydrodynamic poles and the holomorphic function is necessary to match the AC conductivity at low frequencies. In particular, we can identify
    \begin{eqnarray}
            \label{Eq:ConductConstraint}
            f_{(0),n} 
            &=& \partial_{\omega}^{n} \left[  \sigma_{\mathrm{AC}}(i \omega) - \sum_{m=1}^{N-1} \frac{i R_{(0),m}}{\omega + i \tau_{(0),m}^{-1}} \right]_{\omega=0} \nonumber \\
            &=&  i^{n} \left[ \sigma_{\mathrm{AC}}^{(n)}(0) - n! \sum_{m=1}^{N-1} \frac{R_{(0),m}}{ (\tau_{(0),m}^{-1})^{n+1}} \right]
    \end{eqnarray}
as a power series about $\omega =0$ and thus
    \begin{eqnarray}
            \sigma_{\mathrm{AC}}(i \omega)
        &=& \sum_{m=1}^{N-1} \frac{i R_{(0),m}}{\omega + i \tau_{(0),m}^{-1}} + \sum_{n=0}^{\infty}  \left( \frac{1}{n!}  \sigma_{\mathrm{AC}}^{(n)}(0) - \sum_{m=1}^{N-1} \frac{R_{(0),m}}{ (\tau_{(0),m}^{-1})^{n+1}} \right) (i \omega)^{n} \; .  \qquad
    \end{eqnarray}
Care must be taken with this expression to understand that the second term is a series expansion of a holomorphic function; in general it has a finite radius of convergence up to the next pole not included in the summation. Moreover, when we work to a finite number of time derivatives the above expression will be replaced by
	    \begin{eqnarray}
            \sigma_{\mathrm{AC}}(i \omega)
        &=& \sum_{m=1}^{N-1} \frac{i R_{(0),m}}{\omega + i \tau_{(0),m}^{-1}} + \sum_{n=0}^{N_{D}}  \left( \frac{1}{n!} \sigma_{\mathrm{AC}}^{(n)}(0) - \sum_{m=1}^{N-1} \frac{R_{(0),m}}{ (\tau_{(0),m}^{-1})^{n+1}} \right) (i \omega)^{n} \nonumber \\
        &\;& + \mathcal{O}(\omega^{N_{D}+1}) \; .   \qquad
    \end{eqnarray}
   }

{\ To generate further constraints on the conductivities of interest \eqref{Eq:GenericConductivity}, we can take this expression and convert it to the equivalent expression for the spatial charge current correlator i.e.
    \begin{eqnarray}
        \label{Eq:SpatialCorrelatorfromsigma}
        \langle J^{i} J^{j} \rangle_{\mathrm{R}}(\omega,\vec{k})
        &=& - i \omega \sigma^{ij}(\omega,\vec{k}) + \eta^{ij}(\vec{k}) \; , 
    \end{eqnarray}
where $\eta^{ij}(\vec{k})$ is some function we must identify. Importantly, this spatial current correlator must satisfy the $U(1)$ Ward identity which, allowing for generic but analytic contact terms compatible with spatial rotation invariance, takes the form
    \begin{subequations}
    \label{Eq:U1Ward}
    \begin{eqnarray}
        \label{Eq:WardI}
        - i \omega \langle J^{t} J^{t} \rangle_{\mathrm{R}}(\omega,\vec{k}) &=& i k_{i} \langle J^{i} J^{t} \rangle_{\mathrm{R}}(\omega,\vec{k}) + C_{(1)}(\vec{k}^2) \; , \\
        \label{Eq:WardII}
        - i \omega \langle J^{i} J^{t} \rangle_{\mathrm{R}}(\omega,\vec{k}) &=& i k_{j} \langle J^{i} J^{j} \rangle_{\mathrm{R}}(\omega,\vec{k}) + k^{i} C_{(2)}(\vec{k}^2) \; . 
    \end{eqnarray}
    \end{subequations}
where $C_{(1)}$ and $C_{(2)}$ are arbitrary functions of $\vec{k}^2$ as required by spatial rotation invariance. From this expression it follows that
    \begin{eqnarray}
        \label{Eq:GenericJtJt}
        \langle J^{t} J^{t} \rangle_{\mathrm{R}}(\omega,\vec{k})
        &=& \frac{i C_{(1)}(0)}{\omega} + \frac{\vec{k}^2}{\omega^2}\left[ - i \omega ( \sigma_{(0)}(\omega) - C_{(1)}'(0))  + ( \eta_{(0)} - i C_{(2)}(0) ) \right] \nonumber \\
        &\;& - \frac{\vec{k}^4}{\omega^2} \left[ - i \omega \left( \sigma_{(\mathrm{L})}(\omega,\vec{k}^2) + \sum_{n=2}^{\infty} \frac{\vec{k}^{2(n-2)}}{n!} \frac{\partial^{n}}{\partial (\vec{k}^2)^{n}} C_{(1)}(\vec{k}^2) \right) \right. \nonumber \\
        &\;& \left. \hphantom{+ \frac{\vec{k}^4}{\omega^2} \left[ \right.} + \left( \eta_{(\mathrm{L})}(\vec{k}^2) + i \sum_{n=1}^{\infty} \frac{\vec{k}^{2(n-1)}}{n!} \frac{\partial^{n}}{\partial (\vec{k}^2)^{n}} C_{(2)}(\vec{k}^2) \right) \right]  \; . 
    \end{eqnarray}
Our next constraint involves ensuring that \eqref{Eq:GenericJtJt} remains finite in the $\omega \rightarrow 0$ limit so that the charge susceptibility remains finite. We see one set of constraints immediately, namely
    \begin{subequations}
    \begin{eqnarray}
        \label{Eq:FirstContact}
        C_{(1)}(0) = 0 \; ,
    \end{eqnarray}
and
    \begin{eqnarray}
        \label{Eq:SecondContact}
        &\;& \eta_{(0)} = i C_{(2)}(0) \; , \qquad \eta_{(\mathrm{L})}(\vec{k}^2) = - i \sum_{n=1}^{\infty} \frac{\vec{k}^{2(n-1)}}{n!} \frac{\partial^{n}}{\partial (\vec{k}^2)^{n}} C_{(2)}(\vec{k}^2) \; . 
    \end{eqnarray}
    \end{subequations}
The presence of contact term $C_{(2)}$ in \eqref{Eq:WardII} precisely corresponds to the longitudinal, frequency independent part of the spatial current-current correlator given in \eqref{Eq:SpatialCorrelatorfromsigma}.}

{\ The first constraint \eqref{Eq:FirstContact} is particularly useful as it allows us to eliminate the $C_{(1)}(\vec{k}^2)$ contact term by redefining $\langle J^{i} J^{t} \rangle_{\mathrm{R}}(\omega,\vec{k})$ and $\langle J^{i} J^{j} \rangle_{\mathrm{R}}(\omega,\vec{k})$. In particular, we can set $C_{(1)}(\vec{k}^2) =0$ in \eqref{Eq:U1Ward} if we make the redefinition
    \begin{subequations}
    \begin{eqnarray}
        C_{(1)}(\vec{k}^2) = \vec{k}^2 \tilde{C}_{(1)}(\vec{k}^2)
    \end{eqnarray}
and define new correlation functions by
    \begin{eqnarray}
        \tilde{\langle J^{i} J^{t} \rangle}_{\mathrm{R}}(\omega,\vec{k}) &=& \langle J^{i} J^{t} \rangle_{\mathrm{R}}(\omega,\vec{k}) - i k^{i} C'_{(1)}(\vec{k}^2) \; , \\
        \tilde{\langle J^{i} J^{j} \rangle}_{\mathrm{R}}(\omega,\vec{k}) &=& \langle J^{i} J^{j} \rangle_{\mathrm{R}}(\omega,\vec{k}) - i \frac{k^{i} k^{j}}{\vec{k}^2} \left( C_{2}(\vec{k}^2) - \omega  \tilde{C}_{1}(\vec{k}^2) \right)
    \end{eqnarray}
    \end{subequations}
These correlation functions satisfy the Ward identities \eqref{Eq:U1Ward} without contact terms and with the $\eta_{(0)}$ and $\eta_{(\mathrm{L})}(k^2)$ in \eqref{Eq:SpatialCorrelatorfromsigma} equal to zero i.e.
       \begin{eqnarray}
        \label{Eq:SpatialCorrelatorfromsigmareduced}
        \langle J^{i} J^{j} \rangle_{\mathrm{R}}(\omega,\vec{k})
        &=& - i \omega \sigma^{ij}(\omega,\vec{k}) + \eta_{(\mathrm{T})}(\vec{k}) \left( \vec{k}^2 \delta^{ij} - k^{i} k^{j} \right) \; , 
    \end{eqnarray}
The transverse component $\eta_{\mathrm{(T)}}$ corresponds to the magnetisation response and will not be zero. We shall only consider such correlation functions from this point onward.}

\subsection{Only the diffusive pole }

{\noindent Restricting our Mittag-Leffler \eqref{Eq:GenericConductivity} to contain only the diffusive pole gives
    \begin{eqnarray}
        \label{Eq:DiffusionConductivity}
        \sigma^{ij}(\omega,\vec{k}) &=& \left( \sigma_{\mathrm{AC}}(i \omega)- \frac{i \vec{k}^2 R_{\mathfrak{D}}(\vec{k}^2)}{\omega + i \mathfrak{D}(\vec{k}^2) \vec{k}^2} - \vec{k}^2 \sum_{n=0}^{N_{D}} \frac{f_{(\mathrm{L}),n}(\vec{k}^2)}{n!} (i \omega)^n \right) \frac{k^{i} k^{j}}{\vec{k}^2} \nonumber \\
        &\;& + \left( \sigma_{\mathrm{AC}}(i \omega) - \vec{k}^2 \sum_{n=0}^{N_{D}} \frac{f_{(\mathrm{T}),n}(\vec{k}^2)}{n!} (i \omega)^n \right) \left( \delta^{ij} - \frac{k^i k^{j}}{\vec{k}^2} \right) \nonumber \\
        &\;& + \mathcal{O}(\omega^{N_{D}+1}) \; , 
    \end{eqnarray}
where we have imposed \eqref{Eq:ConductConstraint} and truncated the frequency expansion to order $N_{D}$. Notice that the residue of the pole vanishes in the $\vec{k} \rightarrow \vec{0}$ limit as required for a finite conductivity. Meanwhile, from the contact-term-free Ward identities \eqref{Eq:U1Ward} and the conductivity given in \eqref{Eq:DiffusionConductivity} one finds from \eqref{Eq:GenericJtJt}, again in the absence of contact terms, the following expression:
    \begin{eqnarray}
        \langle J^{t} J^{t} \rangle_{\mathrm{R}}(\omega,\vec{k})
        &=& - i \frac{\vec{k}^2}{\omega} \sigma_{\mathrm{AC}}(i \omega)
        - \frac{\vec{k}^4}{\omega} \left( \frac{R_{\mathfrak{D}}(\vec{k}^2)}{\omega + i \mathfrak{D}(\vec{k}^2) \vec{k}^2} - i \sum_{n=0}^{N_{D}} \frac{f_{(\mathrm{L}),n}(\vec{k}^2)}{n!} (i\omega)^n  \right) \; . \qquad 
    \end{eqnarray}
Taking $\omega \rightarrow 0$ with fixed $\vec{k}$ we find
    \begin{eqnarray}
            \label{Eq:Lowomegaexpansion}
            \langle J^{t} J^{t} \rangle_{\mathrm{R}}(\omega,\vec{k})
        &=&  \frac{i \vec{k}^2}{\omega} \left( \frac{R_{\mathfrak{D}}(0)}{\mathfrak{D}(0)} - \sigma_{\mathrm{DC}} \right) \nonumber \\
        &\;&  + \frac{i \vec{k}^4}{\omega} \left( \frac{1}{\vec{k}^2} \left( \frac{R_{\mathfrak{D}}(\vec{k}^2)}{\mathfrak{D}(\vec{k}^2)} - \frac{R_{\mathfrak{D}}(0)}{\mathfrak{D}(0)} \right)  +  f_{(\mathrm{L}),0}(\vec{k}^2) \right) \nonumber \\
        &\;& + (\mathrm{finite \; as \;}\omega \rightarrow 0) \; .
    \end{eqnarray}
In the $\omega \rightarrow 0$ limit of \eqref{Eq:Lowomegaexpansion}, for the result to be finite as required by our constraints, it must be the case that
    \begin{subequations}
    \label{Eq:JtJtconstr}
    \begin{eqnarray}
        \label{Eq:JtJtconst1}
        \sigma_{\mathrm{DC}} &=& \frac{R_{\mathfrak{D}}(0)}{\mathfrak{D}(0)} \; , \\
        \label{Eq:JtJtconst2}
        f_{(\mathrm{L}),0}(\vec{k}^2) &=& - \frac{1}{\vec{k}^2} \left( \frac{R_{\mathfrak{D}}(\vec{k}^2)}{\mathfrak{D}(\vec{k}^2)} - \sigma_{\mathrm{DC}} \right) \; . 
    \end{eqnarray}
    \end{subequations}
More than being finite, we require that $\langle J^{t} J^{t} \rangle_{\mathrm{R}}(0,\vec{k})$ reproduces the charge conductivity and polarisation effects at non-zero $\vec{k}$. These latter quantities can be derived from the hydrostatic generating functional which we discuss in \ref{sec:Hydrostatic}. In particular, we wish to understand what is necessary for
    \begin{eqnarray}
        \lim_{\omega \rightarrow 0} \langle J^{t} J^{t} \rangle_{\mathrm{R}}(\omega,\vec{k}) &=& - \left( \chi_{\rho \rho} + \vec{k}^2 \left( \chi_{\mathrm{EE}}^{(0)} + k^2 \chi_{\mathrm{EE}}^{(\mathrm{L})}(\vec{k}^2) \right) \right) \; , 
    \end{eqnarray}
to hold. Computing the $\mathcal{O}(\omega^{0})$ term in $\langle J^{t} J^{t} \rangle_{\mathrm{R}}(\omega,\vec{k})$ we find
    \begin{eqnarray}
        \langle J^{t} J^{t} \rangle_{\mathrm{R}}(\omega,\vec{k}) &=& - \frac{R_{\mathfrak{D}}(\vec{k}^2)}{\mathfrak{D}^2(k^2)} + \vec{k}^2 \sigma_{\mathrm{AC}}'(0) - \vec{k}^4 f_{(\mathrm{L}),1}(\vec{k}^2) + \mathcal{O}(\omega) \nonumber \\
        &=& - \frac{R_{\mathfrak{D}}(0)}{\mathfrak{D}^2(0)}
            + \vec{k}^2 \left( f_{(0),1} - \left. \partial_{\vec{k}^2} \left( \frac{R_{\mathfrak{D}}(\vec{k}^2)}{\mathfrak{D}^2(\vec{k}^2)} \right) \right|_{\vec{k}^2=0} \right)  \nonumber \\
        &\;& - \vec{k}^4 \left( f_{(\mathrm{L}),1}(\vec{k}^2) + \frac{1}{\vec{k}^4} \left( \frac{R_{\mathfrak{D}}(\vec{k}^2)}{\mathfrak{D}^2(\vec{k}^2)} - \frac{R_{\mathfrak{D}}(0)}{\mathfrak{D}^2(0)} - \vec{k}^2 \left. \partial_{\vec{k}^2} \left( \frac{R_{\mathfrak{D}}(\vec{k}^2)}{\mathfrak{D}^2(\vec{k}^2)} \right) \right|_{\vec{k}^2=0} \right) \right) \nonumber \\
        &\;& + \mathcal{O}(\omega) \; , 
    \end{eqnarray}
which means we must identify
    \begin{subequations}
    \begin{eqnarray}
        \label{Eq:EinsteinConstraint}
        R_{\mathfrak{D}}(0) &=& \chi_{\rho \rho} \mathfrak{D}^2(0) \; , \\
        \label{Eq:EinsteinConstraint2}
        \sigma_{\mathrm{AC}}'(0) &=& \frac{
   R_{\mathfrak{D}}'(0)}{\mathfrak{D}^2(0)} -\frac{2 \chi_{\rho \rho} \mathfrak{D}'(0)}{\mathfrak{D}(0)}- \chi_{\mathrm{EE}}^{(0)}\; , \\
        \label{Eq:fLfunction}
        f_{(\mathrm{L}),1}(\vec{k}^2) &=& - \left( \chi_{\rho \rho} + \vec{k}^4 \chi_{\mathrm{EE}}^{(\mathrm{L})}(\vec{k}^2) \right)  + \chi_{\rho \rho} \left( 1 + \frac{1}{\vec{k}^4} \right)  \nonumber \\
        &\;& - \frac{\chi_{\rho \rho}^2}{\sigma_{\mathrm{DC}} \vec{k}^2} \left( 2 \mathfrak{D}'(0) - \frac{R_{\mathfrak{D}}'(0)}{\sigma_{\mathrm{DC}}} \right) - \frac{R_{\mathfrak{D}}(\vec{k}^2)}{\vec{k}^4 \mathfrak{D}^2(\vec{k}^2)} \; , 
    \end{eqnarray}
    \end{subequations}
We can now see that, by comparing \eqref{Eq:EinsteinConstraint} to \eqref{Eq:JtJtconst1}, the famous Einstein relation is satisfied
    \begin{eqnarray}
        \label{Eq:EinsteinRelation2}
        \mathfrak{D}(0) &=& \frac{\sigma_{\mathrm{DC}}}{\chi_{\rho \rho}} 
    \end{eqnarray}
as a consequence of the constraints we have imposed. This fixes the diffusion constant at $\vec{k}=\vec{0}$ in terms of the DC conductivity and thermodynamic charge susceptibility. A second relation, in the same spirit, follows from \eqref{Eq:EinsteinConstraint2}. It relates the derivative of the AC conductivity (the imaginary part of said conductivity if the system is time reversal invariant) and a term in the small $\vec{k}^2$ expansion on the residue. One finds
    \begin{subequations}
    \begin{eqnarray}
           \sigma_{\mathrm{AC}}'(0) &=& \frac{
   R_{\mathfrak{D}}'(0)}{\mathfrak{D}^2(0)} -\frac{2 \chi_{\rho \rho} (0) \mathfrak{D}'(0)}{\mathfrak{D}(0)}- \chi_{\mathrm{EE}}^{(0)} \; , \\
            \label{Eq:ResidueConstraint}
            \Rightarrow  R_{\mathfrak{D}}'(0) 
        &=& \left( \frac{\sigma_{\mathrm{DC}}}{\chi_{\rho \rho}(0)} \right)^2 ( \sigma_{\mathrm{AC}}'(0) + \chi_{\mathrm{EE}}^{(0)} ) + 2 \sigma_{\mathrm{DC}} \mathfrak{D}'(0) \; .
    \end{eqnarray}
    \end{subequations}
This relation is an unescapable consequence of the constraints we have imposed, however we shall check both it and the Einstein relation \eqref{Eq:EinsteinRelation2} when we consider the D3-D5 probe brane system. Regardless, we can substitute \eqref{Eq:ResidueConstraint} into \eqref{Eq:fLfunction} so that
    \begin{eqnarray}
         f_{(\mathrm{L}),1}(\vec{k}^2) &=& \frac{\chi_{\rho \rho} - \frac{R_{\mathfrak{D}}(\vec{k}^2)}{\mathfrak{D}^2(\vec{k}^2)}}{\vec{k}^4} + \frac{\sigma_{\mathrm{AC}}'(0) + \chi_{\mathrm{EE}}^{(0)}}{\vec{k}^2} - \vec{k}^2 \chi_{\mathrm{EE}}^{(\mathrm{L})}(\vec{k}^2)  \; , \qquad
    \end{eqnarray}
which expresses the derivative of the longitudinal residue in terms of other quantities.

{\ Finally, we consider the transverse part of the correlator and match it to the magnetisation susceptibility. The transverse part of the the correlator has the form
    \begin{eqnarray}
            \label{Eq:MagnetisationSusceptibility}
            - \vec{k}^2 \chi_{\mathrm{BB}}(\vec{k}^2)
        &=& \left( \delta_{ij} - \frac{k_{i} k_{j}}{\vec{k}^2} \right) \langle J^{i} J^{j} \rangle_{\mathrm{R}}(0,\vec{k}) \nonumber \\
        &=& \vec{k}^2 \chi_{(\mathrm{T})}(\vec{k}^2) \; ,
    \end{eqnarray}
and thus we fix $\chi_{(\mathrm{T})}$ appearing in \eqref{Eq:SpatialCorrelatorfromsigmareduced} in terms of the magnetisation susceptibility. Thus, no parts in the transverse conductivity are constrained, but to extract it from the current-current correlator we need knowledge of the magnetisation susceptibility.}

{\ Compiling these constraints, and setting $\mathfrak{D} = \mathfrak{D}(0)$, the conductivities of interest truncated to containing only a single diffusive pole have the form
    \begin{subequations}
    \begin{eqnarray}
            \sigma^{ij}(\omega,\vec{k})
        &=& \sigma_{\mathrm{(L)}}(\omega,\vec{k}) \frac{k^{i} k^{j}}{\vec{k}^2} 
            + \sigma_{\mathrm{(T)}}(\omega,\vec{k}) \left( \delta^{ij} - \frac{k^{i} k^{j}}{\vec{k}^2} \right) \; , \\
            \sigma_{(\mathrm{L})}(\omega,\vec{k})
        &=& \sigma_{\mathrm{AC}}(i \omega) - \frac{ i \vec{k}^2 ( R_{\mathfrak{D}}(0) + \vec{k}^2 R_{\mathfrak{D}}'(0)  + \vec{k}^4 {\color{red} \tilde{R}_{(\mathrm{L}),\mathfrak{D}}(\vec{k}^2) }  )}{\omega + i \left( \mathfrak{D} \vec{k}^2 + {\color{red} \tilde{\mathfrak{D}}(\vec{k}^2)} \vec{k}^4 \right)} \nonumber \\
        &\;& - \vec{k}^2 \left[ f_{(\mathrm{L}),0}(\vec{k}^2)  + i \omega f_{(\mathrm{L}),1}(\vec{k}^2) 
        + \sum_{n=2}^{N_{D}} \frac{1}{n!} {\color{red} f_{(\mathrm{L}),n}(\vec{k}^2)} (i \omega)^n \right] \nonumber \\
        &\;& + \mathcal{O}(\omega^{N_{D}+1}) \; , \\
            \sigma_{(\mathrm{T})}(\omega,\vec{k})
        &=& \sigma_{\mathrm{AC}}(i \omega) + \vec{k}^2 \sum_{n=0}^{N_{D}}  {\color{red} f_{(\mathrm{T}),n}(\vec{k}^2)} (i \omega)^n  + \mathcal{O}(\omega^{N_{D}+1}) \; . 
    \end{eqnarray}
    \end{subequations}
where terms not in red have values dictated by either $\sigma_{\mathrm{AC}}(\omega)$ and/or $\chi_{\rho \rho}(\vec{k})$ and, $\tilde{R}_{(\mathrm{L}),\mathfrak{D}}(\vec{k}^2)$ and $\tilde{\mathfrak{D}}(\vec{k}^2)$ are arbitrary functions of $\vec{k}^2$. We shall show in our effective hydrodynamic theory that there are exactly enough transport coefficients to both satisfy the supplied relations and to be fixed by the red terms.}

\subsection{Multiple poles}

{\noindent Now that the procedure for implementing the constraints is set, the generalisation to the effect of multiple poles is relatively straightforward. After imposing \eqref{Eq:ConductConstraint} our conductivities take the form:
	\begin{subequations}
        \label{Eq:MultipoleconductivityAppendix}
	\begin{eqnarray}
		\sigma^{ij}(\omega,\vec{k}) &=& \sigma_{(\mathrm{L})}(\omega,\vec{k}^2) \left(  \frac{k^{i} k^{j}}{\vec{k}^2} \right)  + \sigma_{(\mathrm{T})}(\omega,\vec{k}^2) \left( \delta^{ij} - \frac{k^{i} k^{j}}{\vec{k}^2} \right) \; , \\
            \sigma_{(\mathrm{L})}(\omega,\vec{k}^2) 
        &=& - \frac{i \vec{k}^2 R_{\mathfrak{D}}(\vec{k}^2)}{\omega + i \mathfrak{D}(\vec{k}^2) \vec{k}^2} \nonumber \\
        &\;&  + \sum_{m=1}^{N-1} \frac{i \left( R_{(0),m} - \vec{k}^2 R_{(\mathrm{L}),m}(\vec{k}^2) \right)}{\omega + i (\tau_{(0),m}^{-1} - \vec{k}^2 \tau_{(\mathrm{L}),m}^{-1}(\vec{k}^2))}\nonumber \\
        &\;&  + \sum_{n=0}^{n=N_{D}} \frac{(i\omega)^n}{n!}  \left[  \sigma_{\mathrm{AC}}^{(n)}(0)   - n! \left( \sum_{m=1}^{N-1}  \frac{R_{(0),m}}{( \tau_{(0),m}^{-1} )^{n+1}} \right)  \right] \nonumber \\
        &\;&  - \vec{k}^2 \sum_{n=0}^{N_{D}} \frac{(i\omega)^n}{n!}  f_{(\mathrm{L}),n}(\vec{k}^2) + \mathcal{O}(\omega^{N_{D}+1}) \; ,  \\
            \sigma_{(\mathrm{T})}(\omega,\vec{k}^2) 
        &=& \sum_{m=1}^{N-1} \frac{i \left( R_{(0),m} - \vec{k}^2 R_{(\mathrm{T}),m}(\vec{k}^2) \right)}{\omega + i (\tau_{(0),m}^{-1} - \vec{k}^2 \tau_{(\mathrm{T}),m}^{-1}(\vec{k}^2))}  \nonumber  \\
        &\;& + \sum_{n=0}^{n=N_{D}} \frac{(i\omega)^n}{n!}  \left[  \sigma_{\mathrm{AC}}^{(n)}(0)   - n! \left( \sum_{m=1}^{N-1}  \frac{R_{(0),m}}{( \tau_{(0),m}^{-1} )^{n+1}} \right) - \vec{k}^2  f_{(\mathrm{T}),n}(\vec{k}^2)  \right]  \nonumber \\
        &\;& + \mathcal{O}(\omega^{N_{D}+1})  \; . \; \;
	\end{eqnarray}
    \end{subequations}
Again we derive $\langle J^{t} J^{t} \rangle_{\mathrm{R}}(\omega,\vec{k})$ as in \eqref{Eq:Lowomegaexpansion} where the divergences as $\omega \rightarrow 0$ are given by
    \begin{eqnarray}
            \label{Eq:Lowomegaexpansion2}
            \langle J^{t} J^{t} \rangle_{\mathrm{R}}(\omega,\vec{k})
        &=&  \frac{i \vec{k}^2}{\omega} \left( \frac{R_{\mathfrak{D}}(0)}{\mathfrak{D}(0)} - \sigma_{\mathrm{DC}} \right) \nonumber \\
        &\;&  + \frac{i \vec{k}^4}{\omega} \left[ \frac{1}{\vec{k}^2} \left( \frac{R_{\mathfrak{D}}(\vec{k}^2)}{\mathfrak{D}(\vec{k}^2)} - \frac{R_{\mathfrak{D}}(0)}{\mathfrak{D}(0)} \right)  +  f_{(\mathrm{L}),0}(\vec{k}^2) \right. \nonumber \\
        &\;& \left. \hphantom{+ \frac{i \vec{k}^4}{\omega} \left( \right.} + \sum_{m=1}^{N-1} \frac{R_{(\mathrm{L}),m}(\vec{k}^2)}{\tau_{(0),m}^{-1} - \vec{k}^2 \tau_{(\mathrm{L}),m}^{-1}(\vec{k}^2)} \right.  \nonumber \\
        &\;& \left. \hphantom{+ \frac{i \vec{k}^4}{\omega} \left( \right.} - \sum_{m=1}^{N-1} \frac{R_{(0),m}}{\vec{k}^2} \left( \frac{1}{\tau_{(0),m}^{-1} - \vec{k}^2 \tau_{(\mathrm{L}),m}^{-1}(\vec{k}^2)} - \frac{1}{\tau_{(0),m}^{-1}} \right) \right] \nonumber \\
        &\;& + (\mathrm{finite \; as \;}\omega \rightarrow 0) \; .
    \end{eqnarray}
Thus we find the following coefficient is modified:
    \begin{eqnarray}
        f_{(\mathrm{L}),0}(\vec{k}^2) &=& - \frac{1}{\vec{k}^2} \left( \frac{R_{\mathfrak{D}}(\vec{k}^2)}{\mathfrak{D}(\vec{k}^2)} - \sigma_{\mathrm{DC}} \right) + \sum_{m=1}^{N-1} \frac{R_{(0),m}}{\vec{k}^2} \left( \frac{1}{\tau_{(0),m}^{-1} - \vec{k}^2 \tau_{(\mathrm{L}),m}^{-1}(\vec{k}^2)} - \frac{1}{\tau_{(0),m}^{-1}} \right) \nonumber \\
        &\;& - \sum_{m=1}^{N-1} \frac{R_{(\mathrm{L}),m}(\vec{k}^2)}{\tau_{(0),m}^{-1} - \vec{k}^2 \tau_{(\mathrm{L}),m}^{-1}(\vec{k}^2)} \; . \qquad \;
    \end{eqnarray}
The $\mathcal{O}(\omega)$ in $\langle J^{t} J^{t} \rangle_{\mathrm{R}}(\omega,\vec{k})$ will also receive contributions from the additional poles to the relevant identifications. The resultant expressions are relatively long and straightforwardly obtained following the path outlined above. Regardless, the same components of the correlator are constrained even in the presence of additional poles. Consequently, our conductivities take the from given in \eqref{Eq:TotalChargeCorrelator}.}

\section{Frame transformations}
\label{appendix:frametrans}

{\noindent With fixed $u^\mu=(1,\vec{0})$ and $T$,  the linearised current takes the form \eqref{Eq:ChargeCurrentConstitutive}
\begin{equation}
\delta J^\mu = \delta \mathcal{N}\, u^\mu + \delta \mathcal{J}^\mu,
\end{equation}
with
\begin{equation}
\delta \mathcal{N} = \chi_{\rho\rho}\, \delta \mu - \nabla_\nu^\perp \delta p^\nu, \qquad
\delta \mathcal{J}^\mu = \delta \bar{J}^\mu - \Pi\indices{^{\mu}_{\nu}} \nabla_\rho \delta M^{\rho\nu} \; . 
\end{equation}
Note that the linearised current fluctuation written in this way resembles the standard decomposition of the fluctuation of the charge current into longitudinal and transverse components with respect to the fluid velocity.}

{\ We now investigate allowed frame transformations of the chemical potential
\begin{equation}
    \delta \mu \rightarrow \delta \mu' = \delta \mu + \Delta \delta \mu
\end{equation}
that leave the charge current invariant. Take the general transformation 
\begin{equation}\label{eq:chemicalFrameTransf}
\begin{aligned}
 \delta \mu \rightarrow \delta \mu' = & \delta \mu + \Delta \delta \mu =  \delta \mu + C^{\mu \nu} [\nabla_\lambda] \nabla_\mu \nabla_\nu \delta \mu + \nonumber \\
 & D^\mu [\nabla_\lambda] \delta E_\mu + F [\nabla_\lambda] \nabla_{\mu} \delta E^\mu + G^{\mu\nu} [\nabla_\lambda] \nabla_\mu \delta E_\nu \; . 
\end{aligned}
\end{equation}
The above tensor structures respect spatial rotation and parity invariance.  Applying this transformation schematically to the charge current gives
\begin{equation}
    \delta J'^\mu = (\delta \mathcal{N} + \Delta \delta \mathcal{N}) u^\mu + (\delta j^\mu + \Delta \delta j^\mu)
\end{equation}
such that in order to not alter the physical current, i.e. to have $\delta J^\mu \stackrel{!}{=} \delta J'^\mu$, the frame transformation of the transverse part of the fluctuation of the charge current $\Delta \delta j^\mu$ needs to be
\begin{equation}
    \Delta \delta \mathcal{J}^\mu = - \Delta \delta \mathcal{N} u^\mu.
\end{equation}
This dictates the transformation of $ \delta \bar{J}^\mu $. In particular, it has the form
\begin{equation}
\begin{aligned}
    \Delta \delta \bar{J}^\mu = & \chi_{\rho \rho} \Delta \delta \mu u^\mu \nonumber \\ = & \chi_{\rho \rho} \bigl[C^{\mu \nu} [\nabla_\lambda] \nabla_\mu \nabla_\nu \delta \mu +D^\mu [\nabla_\lambda] \delta E_\mu + F [\nabla_\lambda] \nabla^{\mu} \delta E_\mu + G^{\mu\nu} [\nabla_\lambda] \nabla_\mu \delta E_\nu  \bigr] u^\mu \; .  
\end{aligned}
\end{equation}
In this way we have added a contribution of the dissipative part of the charge current that is along the fluid velocity $u^\mu$, where before it was completely orthogonal to it ($u_\mu \bar{J}^\mu = 0$). It is clear from the construction of $\delta \mathcal{J}^\mu$ that the charge conservation equation is satisfied. But we can confirm this using that $u^\mu = u^\mu_0$ is a constant background field that does not depend on time and $\nabla_\mu = \nabla^\perp_\mu - u_\mu D$, as well as the normalisation $u_\mu u^\mu = -1$ i.e.
\begin{align}
    \nabla_{\mu}(\delta J^{\mu} +\Delta \delta J^{\mu}) &= \nabla_{\mu} \delta J^{\mu} - \nabla_\mu( \chi_{\rho \rho} \Delta \delta \mu u^\mu) + \nabla_{\mu}( \Delta  \delta \bar{J}^{\mu}) \nonumber \\
    & =  \chi_{\rho \rho} D \Delta \delta \mu + \nabla_{\mu}( \Delta  \delta \bar{J}^{\mu}) = 0.
\end{align}
}

{\ Next we consider if there are restrictions on frame transformations that follow from the relaxed equation by examining the mode spectrum. For this we consider the source-free case of \eqref{Eq:Nonconserv}, i.e. we set $\delta E_\mu= 0$,
\begin{align}
    \left[ \Pi_{n=1}^{N-1} \left( \Pi^{\mu \nu} D + \Gamma_{n}^{\mu \nu}[\nabla_{\perp}^2] \right) \right] \delta \bar{J}_{\nu} 
	= - \bar{\sigma}^{\mu \nu}\left[ \nabla_{\perp}^{2} \right]  \nabla^{\perp}_{\nu} \delta \mu   \; . \qquad
\end{align}
After the frame transformation we have 
\begin{align}
    \left[ \Pi_{n=1}^{N-1} \left( \Pi^{\mu \nu} D + \Gamma_{n}^{\mu \nu}[\nabla_{\perp}^2] \right) \right] (\delta \bar{J}_{\nu} + \Delta \delta \bar{J}_{\nu})
	= - \bar{\sigma}^{\mu \nu}\left[ \nabla_{\perp}^{2} \right]  \nabla^{\perp}_{\nu} (\delta \mu + \Delta \delta \mu)   \; . \qquad
\end{align}
We immediately notice that after Fourier transforming, the differential operator $\nabla_\mu = \nabla^\perp_\mu - u_\mu D$ becomes $i k_\mu^\perp + i \omega u_\mu $. Correspondingly, in order to not add new modes to the spectrum, we restrict ourselves by setting the time derivatives in \eqref{eq:chemicalFrameTransf} to zero. Thus we replace all $\nabla_\mu \rightarrow \nabla_\mu^\perp$, restricting the frame transformations to be of the form
\begin{equation}\label{eq:chemicalFrameTransf2}
\begin{aligned}
 \delta \mu \rightarrow \delta \mu' = & \delta \mu + \Delta \delta \mu =  \delta \mu + C^{\mu \nu} [\nabla_\lambda^\perp] \nabla_\mu^\perp \nabla_\nu^\perp \delta \mu + \nonumber \\
 & D^\mu [\nabla_\lambda^\perp] \delta E_\mu + F [\nabla_\lambda^ \perp] \nabla_{\mu}^\perp \delta E^\mu + G^{\mu\nu} [\nabla_\lambda^\perp] \nabla_\mu^\perp \delta E_\nu.
\end{aligned}
\end{equation}
With that restriction being made, to compute the resulting modes after the frame transformation we split the tensors into longitudinal and transversal components with respect to the wavevector and for reasons of clarity refrain from writing down the dependency of the frame transformation functions on the spatial derivatives
\begin{equation} \label{eq:relFrameTransf}
    \begin{aligned} 
    & \left[  \Pi_{n=1}^{N-1} \left(- \Pi^{\mu \nu} i \omega + \Gamma_{(0),n} \delta^{\mu \nu}- \Gamma_{(\mathrm{T}),n} (k_\perp^2 \delta^{\mu \nu} - k_\perp^\mu k_\perp^\nu) - \Gamma_{(\mathrm{L}),n} k_\perp^\mu k_\perp^\nu \right) \right] \nonumber \\
    & \times (\delta \bar{J}_{\nu} - \chi_{\rho \rho} \Delta \delta \mu u_\nu) \nonumber \\
    =& - \left( \bar{\sigma}_{(0)} \delta^{\mu \nu}- \bar{\sigma}_{(\mathrm{T})}(k_\perp^2 \delta^{\mu \nu} - k_\perp^\mu k_\perp^\nu) - \bar{\sigma}_{(\mathrm{L})} k_\perp^\mu k_\perp^\nu  \right) i k^\perp_\nu \left( \delta \mu + \Delta \delta \mu \right). \; \qquad
    \end{aligned}
\end{equation}
We can set the electric field again to be zero such that we have
\begin{equation}
    \Delta \delta \mu =  - C^{\lambda \tau} (ik^\perp_\nu, - i \omega)  k_\lambda^\perp k_\tau^\perp \delta \mu.
\end{equation}
From the charge conservation equation we have 
\begin{equation}
    \Delta \delta \mu = \frac{k_\mu^\perp \Delta  \delta \bar{J}^{\mu}}{\omega \chi_{\rho \rho}}
\end{equation}
and use it to replace the corresponding terms in \eqref{eq:relFrameTransf}. The modes in the transversal sector do not change compared to the ones obtained from the expression of the untransformed chemical potential and thus do not put restrictions on the allowed transformations. For the longitudinal sector, the modes are obtained  from the zeroes of the following polynomial
\begin{align} 
    & \omega \left[  \Pi_{n=1}^{N-1} \left(- i \omega + \Gamma_{(0),n} [i k_\lambda^\perp]  - \Gamma_{(\mathrm{L}),n} [i k_\lambda^\perp] k^2  \right) \right]  \\
    & = - \left( \bar{\sigma}_{(0)} [i k_\lambda^\perp] - \bar{\sigma}_{(\mathrm{L})} [i k_\lambda^\perp] k^2 \right) \frac{i k^2}{\chi_{\rho\rho}} 
    \left( 1 - C^{\rho \tau} [i k_\lambda^\perp]  k_\rho^\perp k_\tau^\perp  \right).  \; \qquad
\end{align}
We immediately see that at $\vec{k}=\vec{0}$ this equation gives exactly the same solutions as the frame we started with, namely
    \begin{eqnarray}
        \omega_{\mathfrak{D}} = 0 \; , \qquad \omega_{n=1,\ldots,N-1} = - i \Gamma_{(0),n} 
    \end{eqnarray}
and thus match again the $\vec{k}=\vec{0}$ modes in \eqref{Eq:TransverseModes}, up to the presence of the diffusive pole $\omega_{\mathfrak{D}}$. Additionally, at leading non-zero order in $\vec{k}$, we have the same equation as in the untransformed case, since the $C^{\rho \tau}$ terms start only contributing at $\mathcal{O}(k^4)$. Thus the corresponding transport coefficients related to the dispersion relations such as the diffusion constant do not get altered from the frame transformation. The frame transformation as written above thus does not receive further constraints from the analysis of the mode spectrum.}

{\ In this appendix we only focused on frame transformations in terms of fluctuations of the chemical potential. It is a matter to debate whether one should also allow frame transformations dependent on $\delta \bar{J}^{\mu}$.}

\section{Comparison to higher form formulations of probe brane quasihydrodynamics }\label{appendix:higherform}

{\noindent In \cite{Davison:2022vqh} the equation of motion for the quasihydrodynamics of the probe brane is reformulated as an expression for the almost conservation of a $d$-form. Let us determine whether we can straightforwardly rewrite our prescription in terms of non-conservation of a higher form current.}

{\ In \cite{Davison:2022vqh}, it is claimed that there is an almost conserved $d$-dimensional $K$ form that describes the dynamics of our system at large charge density. As we are working in $(2+1)$-dimensions a general antisymmetric $2$-form can be decomposed with respect to the fluid velocity $u^{\mu}$ as
	\begin{subequations}
	\label{Eq:GenericKformdecomp}
	\begin{eqnarray}
		\delta K&=& \frac{1}{2} \left[ \Sigma_{\mu \nu} \delta \theta +  \left( u_{\mu} \Sigma_{\nu \rho} - u_{\nu} \Sigma_{\mu \rho} \right) \delta \bar{\theta}^{\rho} \right] \mathrm{d}x^{\mu} \wedge \mathrm{d}x^{\nu} \; , 
	\end{eqnarray}
	\end{subequations}
where $\delta \bar{\theta}^{\sigma}$ is a transverse vector field. The Hodge dual of $\delta K$, denoted $* \delta K$ is given by
	\begin{eqnarray}
		(* \delta K) &=& \frac{1}{2} \epsilon\indices{^{\mu \nu}_{\sigma}} \left[ \Sigma_{\mu \nu} \delta \theta +  \left( u_{\mu} \Sigma_{\nu \rho} - u_{\nu} \Sigma_{\mu \rho} \right)  \delta \bar{\theta}^{\rho} \right]  \mathrm{d}x^{\sigma} = \left[ -  u_{\sigma} \delta \theta +  \delta \bar{\theta}_{\sigma} \right] \mathrm{d}x^{\sigma} \; . \qquad
	\end{eqnarray}
Acting upon this with an exterior derivative we find
	\begin{eqnarray}
        \label{Eq:DiffdhodgeK}
			\mathrm{d} * \delta K
		&=& - \left( D \delta \bar{\theta}_{\mu} - \nabla_{\mu}^{\perp} \delta \theta \right) \mathrm{d} u \wedge dx^{\mu} + \left( \nabla_{\mu}^{\perp} \delta \bar{\theta}_{\nu} \right) \mathrm{d}x^{\mu} \wedge \mathrm{d}x^{\nu} \; . 
	\end{eqnarray}
This is the most general decomposition for the exterior derivative of a two-form.}
    
{\ Comparing to \cite{Davison:2022vqh} we now add a generic term linear in the external gauge field strength to \eqref{Eq:DiffdhodgeK}. For ease of comparison to that work, let
	\begin{eqnarray}
		\vartheta_{\mu \nu \rho \sigma} &=& \left( \eta_{\mu \rho} \eta_{\nu \sigma} - \eta_{\mu \sigma} \eta_{\nu \rho} \right) + \theta_{\mu \nu \rho \sigma} 
	\end{eqnarray}
so that the first term reproduces $F$ as found in \cite{Davison:2022vqh} and the second term $\theta_{\mu \nu \rho \sigma}$ is at least order one in derivatives. Equation \eqref{Eq:DiffdhodgeK} then becomes
	\begin{eqnarray}
        \label{Eq:DiffdhodgeK2}
		&\;& \mathrm{d} * \delta K + \frac{1}{2} \vartheta_{\mu \nu \rho \sigma}  \delta F^{\rho \sigma} \mathrm{d}x^{\mu} \wedge \mathrm{d} x^{\nu} \nonumber \\
		&=&  - \left( D \delta \bar{\theta}_{\mu}  +  (\delta E_{\mu} - \nabla_{\mu}^{\perp} \delta \theta) + u^{\alpha} \theta_{\alpha \mu \rho \sigma} \delta F^{\rho \sigma} \right) \mathrm{d} u \wedge dx^{\mu} \nonumber \\
		&\;& + \left[ \nabla_{\mu}^{\perp} \left( \delta \bar{\theta}_{\nu} + \delta A_{\nu}^{\perp} \right)  + \frac{1}{2} \Pi\indices{_{\mu}^{\alpha}} \Pi\indices{_{\nu}^{\beta}}\theta_{\alpha \beta \rho \sigma} \delta F^{\rho \sigma} \right] \mathrm{d}x^{\mu} \wedge \mathrm{d}x^{\nu}  \; , 
	\end{eqnarray}
We note that the second line in \eqref{Eq:DiffdhodgeK2} is a two-form equation, but as our effective theory consists of only a current and a scalar equation, we will need this second term to be identically satisfied.}

{\ Allowing for a more general coupling to the external gauge field, $\theta_{\mu \nu \rho \sigma}$, introduces some ambiguities into the equations which we can use to our advantage. In particular, the first term in \eqref{Eq:DiffdhodgeK2} looks suspiciously like the conservation equation for the non-hydrostatic spatial charge current. Let us then identify
	\begin{subequations}
	\begin{eqnarray}
		 \delta \bar{\theta}_{\mu} &:=& - (\bar{\sigma}^{-1})_{\mu \nu}[\nabla_{\perp}^2] \delta \bar{J}^{\nu}  \;  , \\
		 \delta \theta &:=& \delta \mu \; , \\
		  u^{\alpha} \Pi\indices{_{\mu}^{\nu}} \theta_{\alpha \mu \rho \sigma} \delta F^{\rho \sigma} &:=& (\bar{\sigma}^{-1})_{\mu \nu}[\nabla_{\perp}^2]  \times \nonumber \\
		  		   &\;& \hphantom{} \left(   r\indices{^{\nu}_{\sigma}}[D,\nabla_{\perp}^2] \delta E^{\sigma} + \chi_{B}^{\sigma \nu}[D,\nabla_{\perp}^{2}]  \Sigma\indices{_{\nu}^{\rho}} \nabla_{\rho}^{\perp} \delta B \right) \; .
	\end{eqnarray}
	\end{subequations}
Using our equation of motion at $N=2$ we find
	\begin{eqnarray}
			&\;&  D \delta \bar{\theta}_{\mu}  +  (\delta E_{\mu} - \nabla_{\mu}^{\perp} \delta \theta) + u^{\alpha} \theta_{\alpha \mu \rho \sigma} \delta F^{\rho \sigma}  \nonumber \\
			&=& - (\bar{\sigma}^{-1})_{\mu \nu} \left( D \delta \bar{J}^{\nu} -  \bar{\sigma}^{\nu \beta} (\delta E_{\beta} - \nabla_{\beta}^{\perp} \delta \mu)  
				 -  r\indices{^{\nu}_{\sigma}}[D,\nabla_{\perp}^2] \delta E^{\sigma} -   \chi_{B}^{\nu \sigma }[D,\nabla_{\perp}^{2}]  \Sigma\indices{_{\sigma}^{\rho}} \nabla_{\rho}^{\perp} \delta B \right) \nonumber \\
			&=& (\bar{\sigma}^{-1})_{\mu \nu} \Gamma^{\nu \rho} \delta \bar{J}_{\rho} = (\bar{\sigma}^{-1})_{\mu \nu} \Gamma^{\nu \rho} (\bar{\sigma}^{-1})_{\rho \sigma} \delta \bar{\theta}^{\sigma} =  (\bar{\sigma}^{-1})_{\mu \nu} \Gamma^{\nu \rho} (\bar{\sigma}^{-1})_{\rho \sigma} \Sigma^{\sigma \alpha} u^{\beta} \delta K_{\alpha \beta} \nonumber \\
			&=& \Gamma\indices{_{\mu}^{\alpha}} \delta K_{\alpha \beta} u^{\beta} \; . 
	\end{eqnarray}
Hence, we arrive at an equation of the form
	\begin{eqnarray}
		\label{Eq:2formformalism}
		&\;& \mathrm{d} (* \delta K) +  \Gamma\indices{_{\mu}^{\alpha}} \delta K_{\alpha \beta} u^{\beta} u \wedge \mathrm{d}x^{\mu} + \left( F_{\mu \nu} +  \frac{1}{2} \theta_{\mu \nu \rho \sigma} \delta F^{\rho \sigma} \right) \mathrm{d}x^{\mu} \wedge \mathrm{d} x^{\nu} \nonumber \\
		&=&  \left[ \nabla_{\mu}^{\perp} \left( \delta \bar{J}_{\nu} + \delta A_{\nu}^{\perp} \right)  + \frac{1}{2} \Pi\indices{_{\mu}^{\alpha}} \Pi\indices{_{\nu}^{\beta}}\theta_{\alpha \beta \rho \sigma} \delta F^{\rho \sigma} \right] \mathrm{d}x^{\mu} \wedge \mathrm{d}x^{\nu}  \; . 
	\end{eqnarray}
}

{\ Thus far, what we have produced is an identity. Our problems will arrive when we request that the left hand side of \eqref{Eq:2formformalism} vanishes. The $u \wedge \mathrm{d}x^{\mu}$ component is then the analogue of our spatial charge non-conservation equation. However, the spatially projected part of the equation
	\begin{eqnarray}
		\label{Eq:TransverseConstraint}
		  \left[ \nabla_{\mu}^{\perp} \left( \delta \bar{J}_{\nu} + \delta A_{\nu}^{\perp} \right)  + \frac{1}{2} \Pi\indices{_{\mu}^{\alpha}} \Pi\indices{_{\nu}^{\beta}}\theta_{\alpha \beta \rho \sigma} \delta F^{\rho \sigma} \right] \mathrm{d}x^{\mu} \wedge \mathrm{d}x^{\nu}
	\end{eqnarray}
has no analogue in our formalism. In the absence of external fields, requiring that it is zero places a constraint on the transverse part of the spatial current
	\begin{eqnarray}
		\label{Eq:TransverseConstraint1}
		\Sigma^{\mu \nu} \nabla_{\mu}^{\perp} \delta \bar{J}_{\nu} &=& 0 \; . 
	\end{eqnarray}
}

{\ In \cite{Davison:2022vqh} at leading order in large charge density it is argued that the spatial current satisfies \eqref{Eq:TransverseConstraint} identically. This is done by treating \eqref{Eq:TransverseConstraint} as an equation for the exactness of $\delta \bar{J}_{\mu}$ (up to terms dependent on the external field). Let us attempt to complete our identification and assume that we can restrict $\theta_{\mu \nu \rho \sigma}$, such that it becomes
	\begin{subequations}
	\begin{eqnarray}
		 \frac{1}{2} \Pi\indices{_{[\mu}^{\alpha}} \Pi\indices{_{\nu]}^{\beta}}\theta_{\alpha \beta \rho \sigma} \delta F^{\rho \sigma} &=& \nabla_{[\mu}^{\perp} \left(  \zeta\indices{_{\nu]}^{\rho}} \delta A_{\rho}^{\perp} \right) \; ,
	\end{eqnarray}
	\end{subequations}
with $\zeta_{\mu \nu}$ some transverse tensor. It further follows that if \eqref{Eq:TransverseConstraint} is to be satisfied identically, i.e.
	\begin{eqnarray}
		\nabla_{\left[ \mu \right.}^{\perp} \left( \delta \bar{J}_{\left. \nu\right]} + \sigma_{\left. \nu \right] \alpha} (\Pi+\zeta)\indices{^{\alpha \rho}} \delta A_{\rho}^{\perp} \right) \equiv 0
	\end{eqnarray}
then
	\begin{eqnarray}
		\delta \bar{J}_{\nu}  &=& \sigma_{0}[0]  \nabla_{\nu}^{\perp} \varphi - \bar{\sigma}_{\nu \alpha}[\nabla_{\perp}^2] (\Pi+\zeta)\indices{^{\alpha \rho}} \delta A_{\rho}^{\perp} 
	\end{eqnarray}
for some scalar $\varphi$ where again we have used that our spacetime has vanishing Riemann tensor. We can in fact do further by restricting $\zeta$. In particular, at leading order in $\vec{\partial}$ we have
	\begin{eqnarray}
		\bar{\sigma}_{\mu \nu} = \sigma_{0}[0] \Pi_{\mu \nu} + \mathcal{O}(\vec{\partial}^2) \; . 
	\end{eqnarray}
and thus, if we take
	\begin{eqnarray}
		\zeta^{\mu \nu} = \frac{1}{\bar{\sigma}_{0}[0]}  \left( \bar{\sigma}^{\mu \nu}[\nabla_{\perp}^2] - \Pi^{\mu \nu} \bar{\sigma}_{0}[0]  \right) \sim \mathcal{O}(\partial^2)
	\end{eqnarray}
then we can use the small $\vec{\partial}^2$ expansion to correct $\zeta^{\mu \nu}$ order by order in derivatives to ensure that
	\begin{eqnarray}
		\label{Eq:ScalarIntroduction}
		\delta \bar{J}_{\nu} &\equiv& \sigma_{0}[0] \left( \nabla_{\nu}^{\perp} \varphi - \delta A_{\nu}^{\perp} \right)	\; . 
	\end{eqnarray}
The issue remains however, as an identity \eqref{Eq:TransverseConstraint} yields no modes and we explicitly have a gapped mode in the transverse sector of our theory due to a non-trivial $\Gamma_{(\mathrm{T})} \Pi_{(\mathrm{T})}^{\mu \nu}$ term; moreover this gapped mode in the transverse sector decays at $\vec{k}=\vec{0}$ with exactly the same rate as the one present in the longitudinal sector. Thus, without modification of the constraint \eqref{Eq:TransverseConstraint} the reformulation in terms of a higher form symmetry is not just a large $\rho$ limit, but also can only capture leading order in spatial gradients.}

\bibliographystyle{JHEP}
\bibliography{refs}

\end{document}